\documentclass[11pt]{article}

       \newcommand{\Jc}{ {\mathcal{J}} }
       
       \newcommand{\Lc}{ {\mathcal{L}} }

       \newcommand{\Pc}{ {\mathcal{P}} }

       \newcommand{\Tc}{ {\mathcal{T}} }

       \newcommand{\Zc}{ {\mathcal{Z}} }
  \newcommand{\nn}{\nonumber}
  
  \newcommand{\bsx}{\boldsymbol{x}}
  \newcommand{\bsb}{\boldsymbol{b}}
  \newcommand{\bsA}{\boldsymbol{A}}
  \newcommand{\bsB}{\boldsymbol{B}}
  \newcommand{\bsE}{\boldsymbol{E}}
  \newcommand{\bsp}{\boldsymbol{p}}
  \newcommand{\bsxi}{\boldsymbol{\xi}}
  \newcommand{\bsZ}{\boldsymbol{Z}}
  \newcommand{\bsT}{\boldsymbol{T}}

\usepackage{graphicx}
\usepackage{epstopdf, epsfig}
\usepackage[colorlinks]{hyperref}
\usepackage{amsfonts}
\usepackage{amssymb}
\usepackage[T1]{fontenc}
\usepackage{mathtools}
\usepackage{amsfonts}
\usepackage{dcolumn}
\usepackage{bm}
\usepackage[export]{adjustbox}
\usepackage{wrapfig}
\usepackage{lipsum}
\usepackage{authblk}
\usepackage[font=small,labelfont=bf]{caption}
\allowdisplaybreaks
\usepackage[titletoc,toc,title]{appendix}	
\DeclareMathOperator*{\argmin}{arg\,min}
\usepackage[style=ieee, citestyle=numeric-comp, backend=biber]{biblatex}

\usepackage{blindtext}
\usepackage[a4paper,
            bindingoffset=0.2in,
            left=1in,
            right=1in,
            top=1in,
            bottom=1in,
            footskip=.25in]{geometry}

\usepackage{hyperref}
\hypersetup{
    colorlinks=true,
    linkcolor=red,
    filecolor=magenta,      
    urlcolor=cyan,
    citecolor = cyan,
    }

\title{Constrained Hamiltonian Systems and Physics-Informed Neural Networks: Hamilton-Dirac Neural Networks} 
\author[1]{Dimitrios A. Kaltsas \thanks{kaltsas.d.a@gmail.com}}
\affil[1]{Department of Physics, University of Ioannina, Ioannina, GR 451 10, Greece}

\date{}
 
\begin{document}
\maketitle
\begin{abstract}
The effectiveness of the Physics Informed Neural Networks (PINNs) for learning the dynamics of constrained Hamiltonian systems is demonstrated using the Dirac theory of constraints for regular systems with holonomic constraints and systems with non-standard Lagrangians. By utilizing Dirac brackets, we derive the Hamilton-Dirac equations and minimize their residuals, incorporating also energy conservation and the Dirac constraints, using appropriate regularization terms in the loss function. The resulting PINNs, referred to as Hamilton-Dirac Neural Networks (HDNNs), successfully learn constrained dynamics without deviating from the constraint manifold. Two examples with holonomic constraints are presented: the nonlinear pendulum in Cartesian coordinates and a two-dimensional, elliptically restricted harmonic oscillator. In both cases, HDNNs exhibit superior performance in preserving energy and constraints compared to traditional explicit solvers. To demonstrate applicability in systems with singular Lagrangians, we computed the guiding center motion in a strong magnetic field starting from the guiding center Lagrangian. The imposition of energy conservation during the neural network training proved essential for accurately determining the orbits of the guiding center. { The HDNN architecture enables the learning of parametric dependencies in constrained dynamics by incorporating a problem-specific parameter as an input, in addition to the time variable. Additionally, an example of semi-supervised, data-driven learning of guiding center dynamics with parameter inference is presented.}
\end{abstract}


\section{Introduction}
\label{Sec_I}
The motion of mechanical systems is often restricted by several types of constraints that limit the trajectories within a specific area of the phase space, known as the constraint manifold. When numerically integrating these systems, a common issue arises: the simulated dynamics tend to drift away from the constraint manifold. As a result, the numerical solution becomes inherently inaccurate and unreliable. Addressing this challenge has been a focus for several researchers, and thus many works have been dedicated to the simulation of constrained Hamiltonian systems over the past decades, e.g. \cite{Leimkuhler1994,Jay1996,Seiler1999,Fu2022,Kinon2023}. In these studies, constraints arising due to singular Lagrangians or externally imposed holonomic constraints on regular systems, have been taken into account employing the Dirac theory of constraints \cite{Dirac1950,Dirac1958,Sundermeyer1982}, either by using extended Hamiltonians or the Dirac bracket approach. Dirac brackets have also been employed in calculating steady state vortex solutions to fluid equations via a Dirac bracket simulated annealing method \cite{Flierl2011,Flierl2019}.

In recent years there has been a surge of interest in developing numerical methods that leverage unsupervised and semi-supervised machine learning to solve initial and boundary value problems. Various neural network (NN) architectures are utilized for this purpose, with the most commonly employed being  the feed-forward deep neural networks (DNNs). This approach is based on the universal approximation theorem of artificial neural networks which asserts that an artificial NN with at least one hidden layer can approximate any continuous function with arbitrary accuracy \cite{Hornik1989}. These networks are trained so that they {learn solutions to the physical differential equations} and the underlying physical laws, and consequently, they are referred to as Physics Informed Neural Networks (PINNs) \cite{Raissi2019, Lagaris1998, Sirignano2018,Cuomo2022,Karniadakis2021}. In several works, PINNs are augmented with ground truth data and constraints corresponding to experimental measurements or observations. This allows for the determination of specific parameters in the differential equation, as demonstrated in several works  \cite{Raissi2019, Guo2023, Kaltsas2022, Regazzoni2021, Kaltsas2024}. Additionally, PINNs can be employed to discover the governing differential equations from the provided data, as shown in studies like \cite{Rudy2017, Raissi2017, Qin2020}. { In this work, we employ PINNs to resolve constrained Hamiltonian dynamics using the Dirac theory of constraints. Our approach offers several advantages over traditional numerical algorithms, including energy conservation, the ability to learn parametric dependencies, and the computation of accurate solution trajectories even in cases where traditional algorithms fail to confine the dynamics on the constraint manifold. }

\subsection{Related work}
\subsubsection{Hamiltonian neural networks}
Unsupervised PINNs have been applied to integrate both integrable and chaotic canonical Hamiltonian systems that conserve the Hamiltonian function \cite{Mattheakis2022,Gelbrecht2021}, building on previous work in data-driven learning of Hamiltonian and Lagrangian dynamics \cite{Greydanus2019,Bertalan2019,Cranmer2020,Qin2020,Choudhary2020,Desai2021}. In \cite{Mattheakis2022}, the authors employ a completely unsupervised approach to construct neural network solutions for systems like a nonlinear oscillator and the chaotic Hénon-Heiles system, demonstrating NNs' ability to learn Hamiltonian dynamics without data. Single Hamiltonian neural networks (HNNs) are used to model the entire system, capturing the interdependencies between variables. The networks are trained to satisfy both the Hamiltonian equations of motion and energy conservation, which is imposed by a regularization term introduced in the loss function. This approach ensures accurate long-term simulations, with the HNNs outperforming explicit symplectic integrators in terms of accuracy and energy conservation.

\subsubsection{Symplectic neural networks}
{
Another interesting deep learning approach to solve Hamiltonian dynamics is the construction of symplectic networks. Symplectic networks were introduced by various authors in \cite{Zhong2020,Chen2020,Jin2020} to identify Hamiltonian systems from data. In \cite{Zhong2020} the authors use a symplectic neural network (SymODEN) to infer the dynamics of Hamiltonian systems from data while in \cite{Chen2020}, symplectic recurrent neural networks (SRNNs) are used for this purpose. In \cite{Jin2020} a new network architecture, called SympNet, introduced to approximate arbitrary symplectic maps \cite{Arnold2013} using appropriate activation functions and supervised learning. Then the authors extended SympNets to learn Hamiltonian dynamics and symplectic maps in an unsupervised manner, applying this method to optimal control problems, thus introducing SympOCnets in \cite{Meng2022}. Symplectic learning for Hamiltonian systems have also been discussed more recently in \cite{David2023} where the authors explore an improved training method for Hamiltonian neural networks, exploiting the symplectic structure of Hamiltonian systems with a different loss function.

}

\subsubsection{Deep learning approaches for constrained Hamiltonian systems}
Neural networks have recently been applied also to constrained Hamiltonian systems \cite{Finzi2020,White2023}. In \cite{Finzi2020}, the authors explored the use of DNNs to learn the dynamics of chaotic systems embedded in Cartesian coordinates, explicitly imposing constraints in the Lagrangian rather than using generalized coordinates to implicitly encode holonomic constraints. This data-driven approach approximates the Hamiltonian and Lagrangian functions via neural networks trained on ground-truth data, while incorporating known constraints. They found that using Cartesian coordinates with explicit constraints significantly improves accuracy and data efficiency. It is argued that imposing implicitly holonomic constraints through the introduction of generalized coordinates can complicate the learning, while separating the constraints from the Hamiltonian and Lagrangian simplifies the process. A more recent supervised approach in \cite{White2023} introduced stabilized neural differential equations, incorporating stabilization terms to maintain dynamics on the constraint manifold, without relying on the Hamiltonian or Lagrangian formalism.

\subsection{Present work}
In the present work we revisit the Dirac theory of constraints, reviewing the Dirac algorithm to identify primary and secondary constraints \cite{Dirac1950,Dirac1958}, the distinction between first-class and second-class constraints and we describe the construction of the Dirac bracket exploiting second-class constraints. The Dirac bracket endows with a new symplectic structure the phase-space and the constraint manifold becomes a Casimir manifold, i.e. the constraints are Casimir functions of the Dirac bracket. By the Dirac bracket and the canonical Hamiltonian of the system the Hamilton-Dirac equations governing the constrained dynamics can be derived. Then, these equations can be solved using machine learning methods by approximating the solutions of the Hamilton-Dirac equations with DNNs.

\subsubsection{Contribution}
{
The main contribution of this work is the application of PINNs in simulating constrained Hamiltonian dynamics by solving the Hamilton-Dirac equations directly, rather than by just introducing regularization terms in the loss function to handle the constraints. We refer to these feed-forward deep neural networks as Hamilton-Dirac Neural Networks (HDNNs). Then, incorporating the Dirac constraints also as regularization terms can further enhance HDNN performance and improve the accuracy of the NN solutions. Specifically, we recommend using the architecture and learning method proposed in \cite{Mattheakis2022}, which treats the outputs of a single network as canonical variables, creating shared internal connectivity among them. 

We further extend this approach by considering networks that take problem-specific parameters as inputs, enabling them to learn parametric dependencies of the constrained dynamics. This allows the network to provide accurate solutions not only for a single value of the selected parameter but also for adjacent values, which would not be possible if training was restricted to just one value. This approach not only enhances the learning of variable and parameter interdependencies but also seamlessly integrates constraints, eliminating the need for generalized coordinates. As noted in \cite{Finzi2020}, introducing generalized coordinates can reduce the training efficiency of Hamiltonian or Lagrangian networks. Our approach differs from previous studies by using an unsupervised methodology, similar to \cite{Mattheakis2022}, and leveraging the Dirac bracket approach to stabilize motion on the constraint manifold. Additionally, we apply this algorithm to tackle problems involving singular Lagrangians, which commonly appear in plasma physics, such as the guiding center motion. Finally, we present an example of data-driven, semi-supervised learning of the guiding center motion, where ground truth data are used to infer the values of two problem parameters.
}

\subsubsection{Challenges and potential enhancements}
Training DNNs requires significantly more computational time than solving the differential equations with traditional numerical algorithms, often by many orders of magnitude, as shown in section \ref{sec_IV} and previous works (e.g., \cite{Cuomo2022,Grossmann2024}). However, achieving the exceptional conservation properties of the NN solutions typically requires implicit or semi-implicit symplectic integrators, which are computationally expensive due to solving nonlinear algebraic systems at each time step. This is particularly prohibitive for high-dimensional systems or constrained problems where it is difficult to reduce dimensionality. In contrast, DNNs avoid the ``curse of dimensionality'' \cite{Han2018}, making them more efficient for high-dimensional problems, and their solutions are in closed-form, continuous, and free from interpolation requirements.

{Also, the ability of DNNs to learn parametric dependencies makes them advantageous in multi-query problems. Moreover, there are cases where traditional numerical algorithms fail to keep the solution trajectories on the constraint manifold, whereas our HDNN method succeeds. Such an example is provided in subsection \ref{subsec_4.2}. Another advantage of neural network approach lies in the predictive capabilities of some network architectures which can be quite accurate for solutions  that exhibit some form of periodicity. This predictive capability can be enhanced using recurrent architectures like echo-state networks. Other promising approaches to improve the efficiency and utility of PINNs include leveraging parallel computing, which is straightforward for NNs, and transfer learning with pre-trained networks suited to specific physical problems, as, for example, the one-shot transfer learning approach in \cite{Desai2022} and the GPT-PINNs introduced in \cite{Chen2024}. These enhancements make the PINN approach more efficient when dealing with multi-query problems.}  

The rest of the paper is structured as follows: Section \ref{sec_II} provides a brief overview of the Dirac theory of constraints, introducing key concepts such as primary and secondary constraints, first-class and second-class constraints, the Dirac bracket, and the Hamilton-Dirac equations of motion. In Section \ref{sec_III}, we present the HDNN algorithm designed for solving constrained Hamiltonian systems. Section \ref{sec_IV} presents a series of numerical experiments, demonstrating the effectiveness of the HDNN approach in simulating Hamiltonian systems with holonomic constraints. Additionally, an example featuring a system with a singular Lagrangian is included. Finally, we summarize our findings in Section \ref{sec_V}.

\section{Constrained Hamiltonian systems and Dirac brackets}
\label{sec_II}
\subsection{Hamiltonian mechanics}
In this section, we briefly review the canonical Hamiltonian dynamics and the Dirac theory of constraints, {as presented in previous texts  (e.g.,  \cite{Seiler1999,Dirac1950,Dirac1958,Sundermeyer1982}). The reader who is familiar with this formulation can skip the present section and proceed directly to section \ref{sec_III}}. 

Central to the canonical Hamiltonian description of an $N$ degree-of-freedom Hamiltonian system is the canonical Hamiltonian function  defined through a Legendre transformation \cite{Goldstein2002}:
\begin{eqnarray}
    H_c = \dot{q}^i p_i - L(q, \dot{q})\,, \label{Hc}
\end{eqnarray}
where $q^i$, $i=1,...,N$,  are the canonical coordinates, $L(q,\dot{q})$ is the Lagrangian function and 
\begin{eqnarray}
    p_i = \frac{\partial L}{\partial \dot{q}^i}\,, \quad i=1,...,N\,, \label{p_def}
\end{eqnarray}
are the conjugate momenta. We can introduce a phase-space vector $\textbf{z}= (q^1,...,q^N,p_1,...,p_N)^T$ to describe the dynamics in the phase space in terms of the Hamiltonian function and the Poisson bracket, which is given by
\begin{eqnarray}
    \{F,G\} = \frac{\partial F}{\partial q^i}\frac{\partial G}{\partial p_i} - \frac{\partial G}{\partial q^i}\frac{\partial F}{\partial p_i} = \frac{\partial F}{\partial z^i}\Jc_c^{ij} \frac{\partial G}{\partial z^j} \,, \label{Poisson_bracket}
\end{eqnarray}
where $F(\textbf{z})$, $G(\textbf{z})$ are two phase space functions and $\Jc_c$ is the canonical Poisson operator defined as
\begin{eqnarray}
    \Jc_c = \begin{pmatrix}
        0_N & I_N\\
        - I_N & 0_N
    \end{pmatrix}\,,
\end{eqnarray}
with $0_N$ and $I_N$ being the N-dimensional zero and identity matrices, respectively.

The Poisson bracket endows the phase space with the structure of a symplectic manifold and together with the canonical Hamiltonian \eqref{Hc} determine the dynamics through the Hamilton's equations:
\begin{eqnarray}
    \dot{z}^i = \{z^i,H_c\} = \Jc_c^{ij} \frac{\partial H_c}{\partial z^j}\,. \label{Hamilton_eqs}
\end{eqnarray}
As $\Jc_c$ is an antisymmetric matrix, the time derivative of the Hamiltonian, is $dH_c/dt = \{H_c, H_c\}=0$. Consequently, the Hamiltonian $H_c$ is a constant of motion. Moreover, from a geometric perspective, the symplectic volume form (phase-space volume) is preserved under the Hamiltonian flow, determined by the Hamiltonian function and the above Poisson structure.

\subsection{Non-standard Lagrangians and Dirac constraints}
For non-standard Lagrangians where the Hessian $\partial^2L/\partial \dot{z}^i \partial\dot{z}^j$ is singular, some of the equations \eqref{p_def} cannot be used to solve for the corresponding $\dot{q}^i$'s, instead yielding relations of the form:
\begin{eqnarray}
    \Phi_\alpha(\textbf{z})=0\,,\quad \alpha = 1,...,K\leq N\,, \label{prim_constraints}
\end{eqnarray}
which are called primary constraints. These constraints define a constraint manifold in phase space. On this manifold the total Hamiltonian
\begin{eqnarray}
    H_t = H_c + \zeta^\alpha \Phi_\alpha\,, \label{H_t}
\end{eqnarray}
where $\zeta^\alpha$ can be functions of $q$'s and $p$'s, should produce the same dynamics as $H_c$. Substituting \eqref{H_t} into Hamilton's equations \eqref{Hamilton_eqs} leads to:
\begin{eqnarray}
    \dot{z}^i = \{z^i,H_t\} \approx \Jc_c^{ij}\frac{\partial H_c}{\partial z^j} + \zeta^\alpha \Jc_c^{ij} \frac{\partial \Phi_\alpha}{\partial z^j}\,. \label{Hamilton_eqs_constr}
\end{eqnarray}
Here, $\approx$ denotes weak equality, meaning that equality holds only after the constraints are applied. This weak equality ($\approx$) becomes strong equality ($=$) on the constraint manifold, i.e. for $\Phi_\alpha =0$, $\forall \alpha$.   Hence, the system \eqref{Hamilton_eqs_constr} differs from the Hamiltonian system $\dot{z}^i = \{z^i,H_t\}$, only by a linear combination of the constraint functions $\Phi_\alpha$.

The dynamics of the system should comply with the constraints and therefore the latter must be preserved by the dynamics. Hence, in a consistent description we should have:
\begin{eqnarray}
    \dot{\Phi}_\alpha = \{\Phi_\alpha, H_t \} \approx 0\,, \quad \forall \alpha\label{consistency}
\end{eqnarray}
EquationsS \eqref{consistency} can generally lead to a new set of equations of the form:
\begin{eqnarray}
    \Psi_\beta(\textbf{z})=0\,,\quad \beta =1,...,M\leq K\,. \label{second_constraints}
\end{eqnarray}
Equations \eqref{second_constraints} can be either used to determine the multipliers $\zeta^\alpha$ or, in cases where this is not possible, they are seen as secondary constraints, which should be preserved by the dynamics as well. The Dirac algorithm \cite{Dirac1950,Dirac1958} dictates that the consistency requirement \eqref{consistency} should be applied also to the secondary constraint functions $\Psi_\beta$ until we have determined all primary and secondary constraints.

{ For a brief presentation of the application of the Dirac algorithm in a particular class of systems with singular Lagrangians, readers are referred to Appendix \ref{appendix}.
}

\subsection{The Dirac bracket}
After identifying all primary and secondary constraints using the Dirac algorithm, we can further classify them into first class and second class constraints. Denoting all constraint functions, including primary and secondary, as $\chi_\alpha$ for $\alpha = 1, \ldots, A$, where $A$ is the total number of primary and secondary constraints, we define second class constraints as those for which the matrix $C$ of their Poisson brackets:
\begin{eqnarray}
    C_{\alpha\beta} = \{\chi_\alpha,\chi_\beta\}\,, \label{constraint_matrix}
\end{eqnarray}
is regular and thus can be inverted. First class constraints, on the other hand, do not satisfy this condition as their Poisson bracket with all other constraints vanishes weakly $\{\chi_a,\chi_\beta\}\approx 0$. First class constraints are associated with gauge symmetries of the Lagrangian, while second class constraints are associated with redundant degrees of freedom. By introducing the Dirac bracket \cite{Seiler1995}, the second class constraints become Casimir invariants of the Dirac bracket defined by \eqref{Dirac_brackt} (e.g. see \cite{Seiler1999}). In the subsequent analysis we consider cases with second class constraints only, so that the matrix \eqref{constraint_matrix} is invertible, and the following generalized Poisson bracket between two phase-space functions $F$, $G$ can be defined:
\begin{eqnarray}
  \{F,G\}_* = \{F,G\} - \{F,\chi_\alpha\}(C^{-1})^{\alpha\beta}\{\chi_\beta,G\}\,.  \label{Dirac_brackt}
\end{eqnarray}
The bracket $\{F,G\}_*$, referred to as the Dirac bracket, retains the same algebraic properties as the ordinary Poisson bracket. As such, it also imparts a symplectic structure to the phase space. It can be easily shown \cite{Seiler1999} that:
\begin{eqnarray}
    \{F,H_c\}_* \approx \{F,H_t\}\,,
\end{eqnarray}
i.e. the two brackets generate identical dynamics on the constraint manifold. Therefore, Hamilton's equations \eqref{Hamilton_eqs} can be replaced by  the equations:
\begin{eqnarray}
    \dot{z}^i = \{z^i,H_c\}_* = \{z^i,H_c\} - \{z^i,\chi_\alpha\}(C^{-1})^{\alpha\beta}\{\chi_\beta,H_c\}\,,\nn
\end{eqnarray}
or
\begin{eqnarray}
    \dot{z}^i= \Jc_c^{ij} \frac{\partial H_c}{\partial z^j} - \Jc_c^{ij} \frac{\partial \chi_\alpha}{\partial z^j} (C^{-1})^{\alpha \beta} \{\chi_\beta,H_c\}\,, \label{Hamilton_Dirac_eqs}
\end{eqnarray}
known as the Hamilton-Dirac equations. The second term on the right-hand side of \eqref{Hamilton_Dirac_eqs} can be interpreted as generalized forces that enforce the constraints on the dynamics. Also notice that \eqref{Hamilton_Dirac_eqs} can be written as: 
\begin{eqnarray}
    \dot{z}^i = \left[\delta^{i}_k-\Jc_c^{ij}\frac{\partial \chi_\alpha}{\partial z^j} (C^{-1})^{\alpha \beta}\frac{\partial \chi_\beta}{\partial z^k} \right] \Jc_c^{k\ell} \frac{\partial H_c}{\partial z^\ell} = \Pc^{i}_k \Jc_c^{k\ell} \frac{\partial H_c}{\partial z^\ell}  \,,
\end{eqnarray}
where $\Pc = I- \Jc_c\cdot (\nabla_z \chi_\alpha) (C^{-1})^{\alpha \beta} (\nabla_z \chi_\beta)$ is a projection operator. Therefore the constrained dynamics can be interpreted as projections of the original dynamics on the constraint manifold.

With this formulation, it is straightforward to show that the constraint functions $\chi_\alpha$, $\forall \alpha$ are preserved by the dynamics, as they are in fact, Casimir invariants of the Dirac bracket, i.e. they Dirac-commute with any phase space function $F(z)$:
\begin{eqnarray}
    \{\chi_\alpha,F\}_* = \{\chi_\alpha,F\} -\{\chi_\alpha,\chi_\beta\}(C^{-1})^{\beta\gamma}\{\chi_\gamma,F\} =0\,, \quad \forall F\,.
\end{eqnarray}
As a final theory note we mention that one can retrieve the Hamilton-Dirac equations \eqref{Hamilton_Dirac_eqs} without introducing the Dirac bracket \eqref{Dirac_brackt} but by constructing an extended Hamiltonian function: 
\begin{eqnarray}
    H_e = H_c + \zeta^\alpha \chi_\alpha\,,\label{extended_H}
\end{eqnarray}
where we recall that $\chi_\alpha$ denote all the constraints of the system, primary and secondary. Then, one can show \cite{Seiler1999} that the multipliers $\zeta^\alpha$ can be computed by the consistency requirement $\{\chi_\alpha,H_e\}\approx 0$, which yields:
\begin{eqnarray}
    \zeta^\alpha = - (C^{-1})^{\alpha \beta} \{\chi_\beta,H_e\}\,, \label{multiplier_det_eq_1}
\end{eqnarray}
and the following weak equality holds:
\begin{eqnarray}
    \{z^i,H_e\} \approx \{z^i,H_c\}_*\,, \quad i=1,...,2N\,,\label{equivalence_1}
\end{eqnarray}
as shown in \cite{Leimkuhler1994}. Notice that in view of \eqref{equivalence_1}, the equations of motion produced by the extended Hamiltonian \eqref{extended_H} and the ordinary Poisson bracket, are weakly  equivalent to the Hamilton-Dirac equations \eqref{Hamilton_Dirac_eqs}.

\subsection{Regular systems with holonomic constraints}
We started the presentation of the Dirac theory of constraints assuming that they arise due to singular, non-standard Lagrangians. However, the methodology presented above can be employed in order to impose holonomic constraints of the form $\Phi_\alpha(q)=0$ on a system with a regular Lagrangian $L$. Again, one can define the total Hamiltonian $H_t = H_c + \zeta^\alpha \Phi_\alpha$ and treat $\Phi_\alpha$ as primary constraints. Then the consistency condition $\dot{\Phi}_\alpha=0$ leads to secondary constraints $\Psi_\beta$. The multipliers $\zeta^\alpha$ can be determined by imposing the consistency condition to the secondary constraints, leading to:
\begin{eqnarray}
    \{\Psi_\beta,H_c\} + \zeta^\alpha\{\Psi_\beta,\Phi_\alpha\}\approx 0\,,
\end{eqnarray}
which can be solved for $\zeta^\alpha$ as follows:
\begin{eqnarray}
    \zeta^{\alpha} \approx - (\tilde{C}^{-1})^{\alpha\beta}\{\Psi_\beta,H_c\}= -(\tilde{C}^{-1})^{\alpha\beta} \frac{\partial \Psi_\beta}{\partial z^i} \Jc_c^{ij} \frac{\partial H_c}{\partial z^j}\,. \label{multiplier_det_eq_2}
\end{eqnarray}
Here, $\tilde{C}^{-1}$ is the inverse of the matrix $\tilde{C}_{\alpha\beta}=\{\Psi_\alpha,\Phi_\beta\}$. A standard approach (see \cite{Seiler1999}) is then to use the Hamilton equations substituting the total Hamiltonian with multipliers $\zeta^\alpha$ determined by Eqs.~\eqref{multiplier_det_eq_2}. Hence, after taking into account the constraints we have equations \eqref{Hamilton_eqs_constr} with multipliers \eqref{multiplier_det_eq_2}, i.e.
\begin{eqnarray}
    \dot{z}^i = \left[\delta^{i}_k - \Jc_c^{ij} \frac{\partial \Phi_\alpha}{\partial z^j}(\tilde{C}^{-1})^{\alpha\beta}\frac{\partial\Psi_\beta}{\partial z^k} \right]\Jc_c^{k \ell} \frac{\partial H_c}{\partial z^\ell}\,.
\end{eqnarray}
Alternatively, we can construct the Dirac bracket  or define the extended Hamiltonian and use the standard Poisson bracket as shown in the previous section. All formalisms are weakly equivalent, i.e. the equations of motion obtained by the three formulations are identical after imposing the constraints, but there might be significant differences regarding the stabilization of the phase space trajectories on the constraint manifold or the computational cost as shown in \cite{Seiler1999}.

\section{Approximating constrained dynamics with PINNs}
\label{sec_III}

\begin{figure}[ht!]
    \centering
    \includegraphics[scale=0.65]{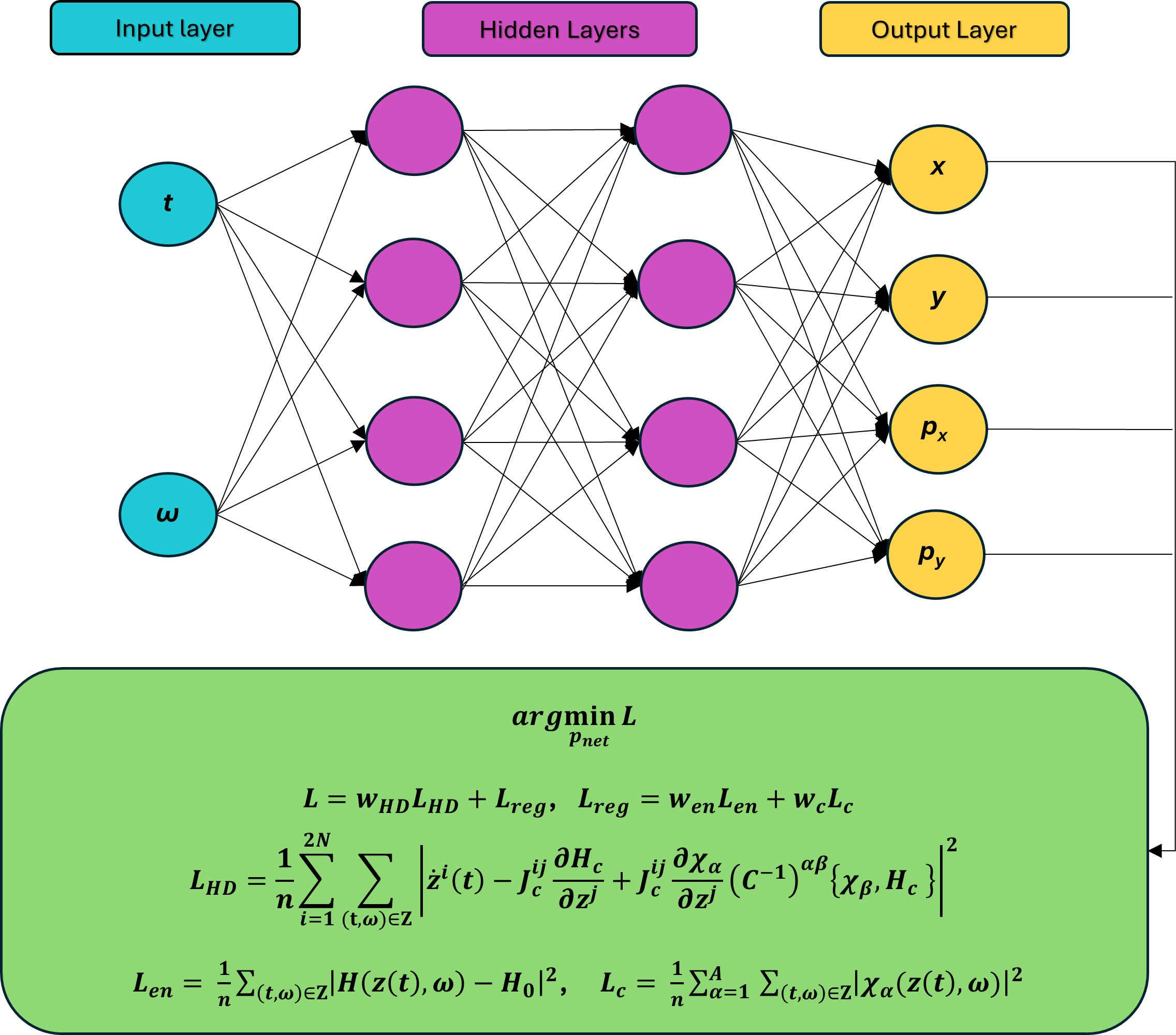}
    \caption[font=small,labelfont=bf]{The HDNN architecture and algorithm.}
    \label{fig_hdnn_architecture}
\end{figure}

{The  canonical dynamical variables $z^i$ ($i=1,...,2N$) are approximated by the outputs $Z^i_{net}$ ($i=1,...,2N$) of a single, feed-forward, fully-connected, DNN. This network has two input neurons: one for the time variable $t$ and another for a parameter $\omega$, which influences the trajectories of the system (see Fig. \ref{fig_hdnn_architecture}).  $\omega$ may represent a parameter that enters the dynamical equations, an initial condition, or a constraint-related parameter. The architecture is designed to learn the dynamics of a constrained system over a range of parameter values, enabling the neural network to function as a surrogate model for solving an entire class of equations. By varying $\omega$ within the training range, the network can provide a new solution at inference time without requiring retraining}. 

The initial conditions: 
\begin{eqnarray}
    z^i(t_0) = z^{i}_0\,, \quad i=1,...,2N\,,
\end{eqnarray}
can be enforced either during the training process by penalizing the network for violating the initial conditions or by imposing them as a hard constraint through considering solutions of the form
\begin{eqnarray}
    z^i(t) = z^i_0 + f(t)Z^{i}_{net}(t)\,, \quad i =1,...,2N\,, \label{z_net}
\end{eqnarray}
where $Z_{net}^{i}(t)$ are the outputs of the DNN {  and $f(t)$ is appropriately selected function so that $f(t_0) = 0$. This is a standard procedure for enforcing initial conditions, (e.g. see \cite{Lagaris1998,Lyu2021}) and has been exploited also in \cite{Mattheakis2022} where the authors selected $f(t) = 1 - e^{-(t-t_0)}$ and $f(t)=t-t_0$, with the second choice having the disadvantage of being unbounded from above. Alternatives, can be $f(t) = tanh(t-t_0)$ and $f(t)=\sigma(t)=1/(1+e^{-(t-t_0)})$. In this study we use $f(t) = 1 - e^{-\gamma (t-t_0)}$ where $\gamma$ can be a learnable parameter, which gives greater flexibility in the representation of the solution. In the subsequent experiments we have  observed an improvement in accuracy of the neural network solutions for $\gamma>1$.}

{The independent variable $t$ is discretized so as the constrained dynamics are evaluated on a set $\Tc = \{t_1,...,t_{n_t}\}$ of $n_t$ discrete time points while $n_{\omega}$ values of the parameter $\omega$ are considered within a set $\Omega=\{\omega_1,...,\omega_{n_\omega}\}$. We then form the following loss function:
\begin{eqnarray}
    \Lc = \frac{1}{n}\sum_{i=1}^{2N}\sum_{(t,\omega)\in \Zc}\Big|  \dot{z}^i(t) - \Jc_c^{ij}\frac{\partial H_c(\textbf{z}(t))}{\partial z^j} + \Jc_c^{ij}\frac{\partial \chi_\alpha(\textbf{z}(t))}{\partial z^j} (C^{-1})^{\alpha \beta}(\textbf{z}(t))\{\chi_\beta(\textbf{z}(t)),H_c(\textbf{z}(t))\} \Big|^2\nn\\ + L_{reg}\,,\label{loss_gen}
\end{eqnarray}
where $\Zc = \Tc\times\Omega$, $n$ is the number of discrete points considered in $
\Zc$ and
\begin{eqnarray}
    \Lc_{reg} = w_{en}\Lc_{en} + w_{c} \Lc_{c}\,, \label{loss_reg}
\end{eqnarray}
with $w_{en}$ and $w_c$ being weight parameters and 
\begin{eqnarray}
    \Lc_{en} = \frac{1}{n} \sum_{(t,\omega)\in \Zc} \big|H_t(\textbf{z}(t)) - H_{0} \big|^2\,, \label{energy_loss}\\ 
    \Lc_c = \frac{1}{n} \sum_{\alpha=1}^{A} \sum_{(t,\omega)\in \Zc} \big|\chi_\alpha(\textbf{z}(t))\big|^2\,, \label{constr_loss}
\end{eqnarray}}
are two regularization terms which penalize the network whenever the dynamics do not satisfy, respectively, the conservation of the total Hamiltonian and the Dirac constraints. Note that  Eq. \eqref{loss_gen} involves the Hamilton-Dirac equations \eqref{Hamilton_Dirac_eqs} rather than the standard Hamilton equations \eqref{Hamilton_eqs} and evidently the quantity $H_{0} = H_{t}(\textbf{z}_0)$ is the initial value of the total Hamiltonian.

{In cases where the parameter $\omega$ enters the dynamics through the constraints, then it enters $\Lc$ through $\chi_\alpha$, $\chi_\beta$ and $(C^{-1})^{\alpha\beta}$ in \eqref{loss_gen} and $\chi_\alpha$ in \eqref{constr_loss}. On the other hand, if the parameter $\omega$ is associated with the unconstrained Hamiltonian dynamics, it  enters $\Lc$ through $H_c$ in \eqref{loss_gen} and $H_t$ in \eqref{energy_loss}. If $\omega$ instead relates to the initial conditions, then we have $z^i(t_0,\omega) = z_{0}^i(\omega)$ and it enters the loss function through  \eqref{z_net}.}

The internal parameters $\boldsymbol{p}_{net}$ of the neural network (i.e. the weights and biases of the connections) are trained so that the loss function \eqref{loss_gen} is minimized, i.e. the training of the HDNN amounts to the optimization problem
\begin{eqnarray}
    \argmin_{\boldsymbol{p}_{net}} \Lc\,,
\end{eqnarray}
with $\Lc$ given by \eqref{loss_gen}. The time derivative in \eqref{loss_gen} is computed using the PyTorch automatic differentiation engine that utilizes the backpropagation algorithm. The discretization method of the variable $t$ is the same with the method described in \cite{Mattheakis2022} which utilizes stochastic perturbation of the grid points in each training epoch. Regarding the activation function, various options can be used for the neurons in the HDNN, e.g. the sigmoid, tanh, SiLU, $sin(x)$ are common candidates. For systems exhibiting periodic motions, the most natural choice is a trigonometric activation function, i.e. $sin(\delta x)$, {and $\delta$ can be a trainable parameter. Letting $\delta$ to be trainable changes the behavior of the sine activation function for a properly normalized input vector $x$, from linear to oscillatory.

In this work, we introduce a novel activation function defined as the product of a sine and a sigmoid function:
\begin{eqnarray}
sinegmoid(x) = \frac{\sin(\delta_1 x)}{1 - e^{-\delta_2 x}}.\label{sinegmoid}
\end{eqnarray}
 We refer to \eqref{sinegmoid} as the \textit{Sinegmoid} activation function and $\delta_1$ and $\delta_2$ are trainable parameters. This is essentially an adaptive activation function approach. It has been demonstrated that adaptive activation functions can accelerate convergence in deep neural networks (e.g., \cite{Jagtap2020}). Preliminary results from our numerical experiments indicate that using the sinegmoid with trainable parameters improves training efficiency across all three cases studied in this work. However, a systematic study to quantify this effect is beyond the scope of the present work and is left for future research.}

\section{Numerical experiments}
\label{sec_IV}
\subsection{Regular Lagrangians and holonomic constraints: Planar pendulum in Cartesian coordinates}

Let us consider a nonlinear pendulum described in Cartesian coordinates $(x, y)$, moving in a gravitational acceleration field $\boldsymbol{g} = -g\hat{y}$ and restricted to the plane $(x, y)$ at a constant distance $\ell=\sqrt{x_0^2+y_0^2}$ from the origin $(0, 0)$. Here, $x_0$ and $y_0$ are initial conditions. The Lagrangian is given by:
\begin{eqnarray}
    L= \frac{m}{2}(\dot{x}^2+\dot{y}^2) - mgy\,,
\end{eqnarray}
and the primary constraint is
\begin{eqnarray}
    \Phi = \frac{1}{2}(x^2 + y^2 -\ell^2)\,.\label{prim_constr_1}
\end{eqnarray}
 The canonically conjugate momenta, the canonical and the total Hamiltonian functions are, respectively,
\begin{eqnarray}
    p_x &=& \frac{\partial L}{\partial \dot{x}} = m\dot{x}\,, \quad p_y = \frac{\partial L}{\partial \dot{y}} = m\dot{y}\,, \\
    H_c &=& \frac{1}{2m} (p_x^2 + p_y^2)+ mgy\,, \\ 
    H_t &=& \frac{1}{2m} (p_x^2+p_y^2) + mgy - \frac{\lambda}{2}\left(x^2+y^2-\ell^2\right)\,, \label{Ht_pendulum}
\end{eqnarray}
where $\lambda$ is a Lagrangian multiplier. The primary constraint \eqref{prim_constr_1} should be preserved by the dynamics, thus
\begin{eqnarray}
    \{\Phi,H_t\} = m^{-1}(x p_x + yp_y)\approx 0\,,
\end{eqnarray}
that leads to the secondary constraint
\begin{eqnarray}
    \Psi = xp_x +y p_y\,, \label{secondary_constr_1}
\end{eqnarray}
which is a Pfaffian constraint \cite{Ardema2005}. Demanding the weak conservation of $\Psi$ we find an explicit relation for the multiplier $\lambda$:
\begin{eqnarray}
    \lambda = \frac{1}{x^2+y^2}\left( mgy - \frac{p_x^2 + p_y^2}{m} \right)\,, \label{lambda_pendulum}
\end{eqnarray}
and thus the Dirac algorithm concludes at this point with the identification of the constraints $\Phi$ and $\Psi$ and the determination of the Lagrangian multiplier. We can now define the matrix $C$ as follows:
\begin{eqnarray}
    C = \begin{pmatrix}
        \{\Phi,\Phi\}& \{\Phi,\Psi\}\\
        \{\Psi,\Phi\} & \{\Psi,\Psi\}
    \end{pmatrix}= \begin{pmatrix}
        0 & x^2 +y^2\\
        -x^2 -y^2 &0
    \end{pmatrix}\,,
\end{eqnarray}
whose inverse is
\begin{eqnarray}
    C^{-1} = \begin{pmatrix}
        0 & -\frac{1}{x^2 +y^2}\\
        \frac{1}{x^2 +y^2} &0
    \end{pmatrix}\,.
\end{eqnarray}
Knowing the matrix $C^{-1}$ we can construct the corresponding Dirac bracket between two phase-space functions $F$ and $G$ using \eqref{Dirac_brackt}:
\begin{eqnarray}
    \{F,G\}_* = \{F,G\} - \frac{1}{x^2+y^2}\left(\{F,\Psi\}\{\Phi,G\}-\{F,\Phi\}\{\Psi,G\}\right)\,.
\end{eqnarray}
Consequently, the Hamilton-Dirac equations of motion are: 
\begin{eqnarray}
    \dot{x} &=& \{x,H_c\}_*= \frac{p_x}{m} + \mu x\,, \label{dx_pendulum}\\
    \dot{y} &=& \{y,H_c\}_*= \frac{p_y}{m} +\mu y\,,\\
    \dot{p}_x &=& \{p_x,H_c\}_*= \lambda x -\mu p_x\,,\\
    \dot{p}_y &=& \{p_y,H_c\}_*=\lambda y -\mu p_y - mg\,, \label{dpy_pednulum}
\end{eqnarray}
where
\begin{eqnarray}
    \mu =- \frac{xp_x + y p_y}{m(x^2+y^2)}\,.
\end{eqnarray}
An alternative to the Dirac bracket approach is the use of the ordinary Poisson bracket along with the extended Hamiltonian, which is composed by the canonical Hamiltonian plus the linear combination of the constraints $\Phi$ and $\Psi$ with multipliers $\zeta$ determined by \eqref{multiplier_det_eq_1}. The resulting equations are weakly identical to equations \eqref{dx_pendulum}--\eqref{dpy_pednulum}, meaning they reduce to those equations after applying the constraints, as indicated by equation \eqref{equivalence_1}. Another standard approach involves using the total Hamiltonian \eqref{Ht_pendulum} substituted into the standard Hamilton equations with $\lambda$ given by equation \eqref{lambda_pendulum} (which can be calculated from equation \eqref{multiplier_det_eq_2}). In \cite{Seiler1995}, it was demonstrated that among these three approaches, the Dirac bracket and the extended Hamiltonian methods are the more accurate, yielding similar results. However, the Dirac bracket approach is less computationally expensive. Prompted by this observation, we employ here only the Dirac bracket approach, which is enhanced though by the imposition of two additional regularization terms in the loss function minimized during the HDNN training. 

The total loss function is given by \eqref{loss_gen} with $\textbf{z}(t)=(x(t),y(t),p_x(t),p_y(t))^T$ and the Hamilton-Dirac equation residual in the first term of \eqref{loss_gen} is determined by Eqs. \eqref{dx_pendulum}--\eqref{dpy_pednulum}. The regularization term is constructed using the total Hamiltonian \eqref{Ht_pendulum} and the constraints \eqref{prim_constr_1} and \eqref{secondary_constr_1}. {The HDNN is trained for various values of the  pendulum mass $m$, i.e. $m$ is the parameter $\omega$ in \eqref{loss_gen}. For this experiment the HDNN learns the dynamics for masses within the range $[m_1,m_2]=[1.0,2.0]$ and for times within the interval $[0,30 T]$, where $T$ is the period of the non-linear pendulum.}

\begin{figure}[ht!]
    \centering
    \includegraphics[scale=0.3]{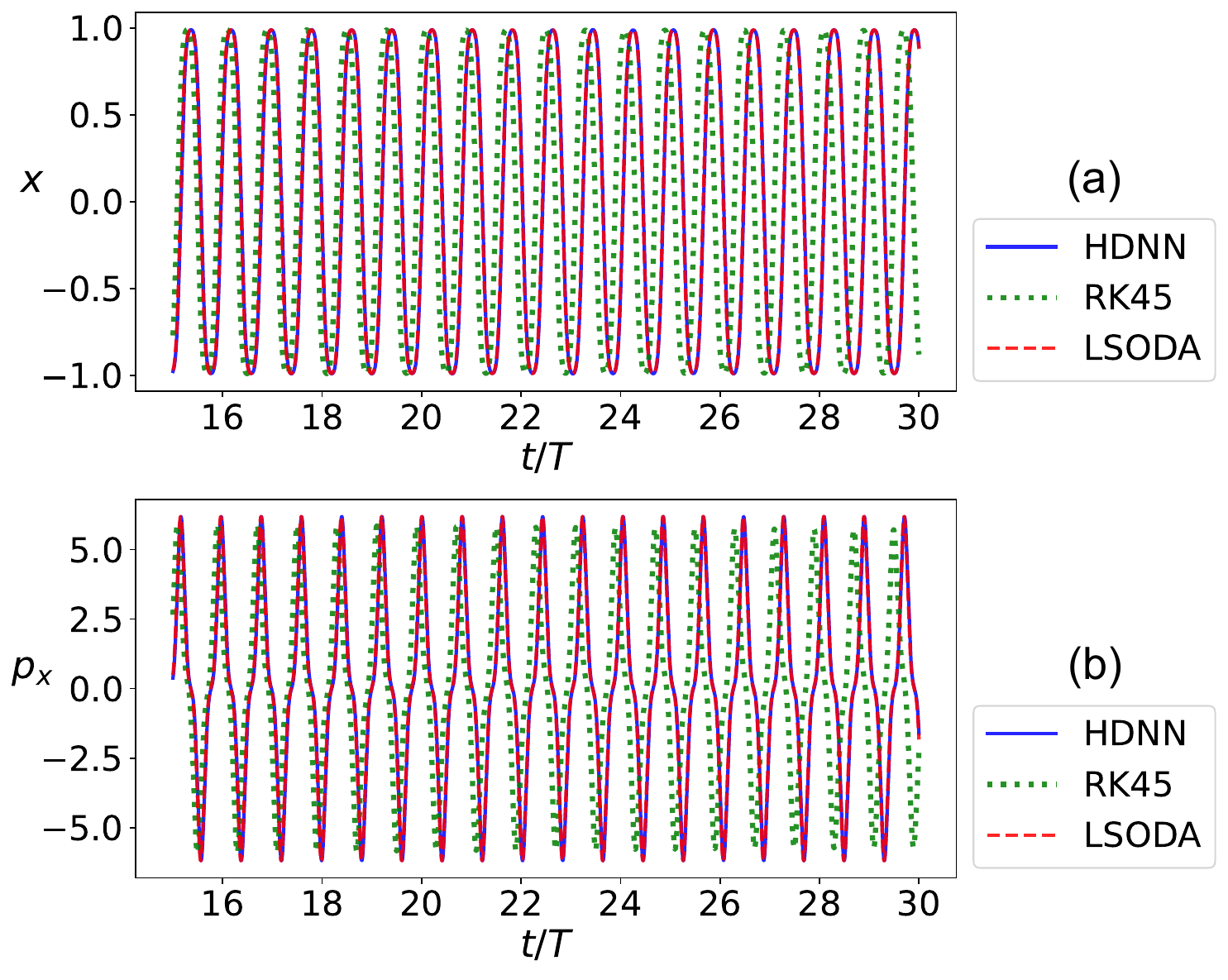}
    \includegraphics[scale=0.3]{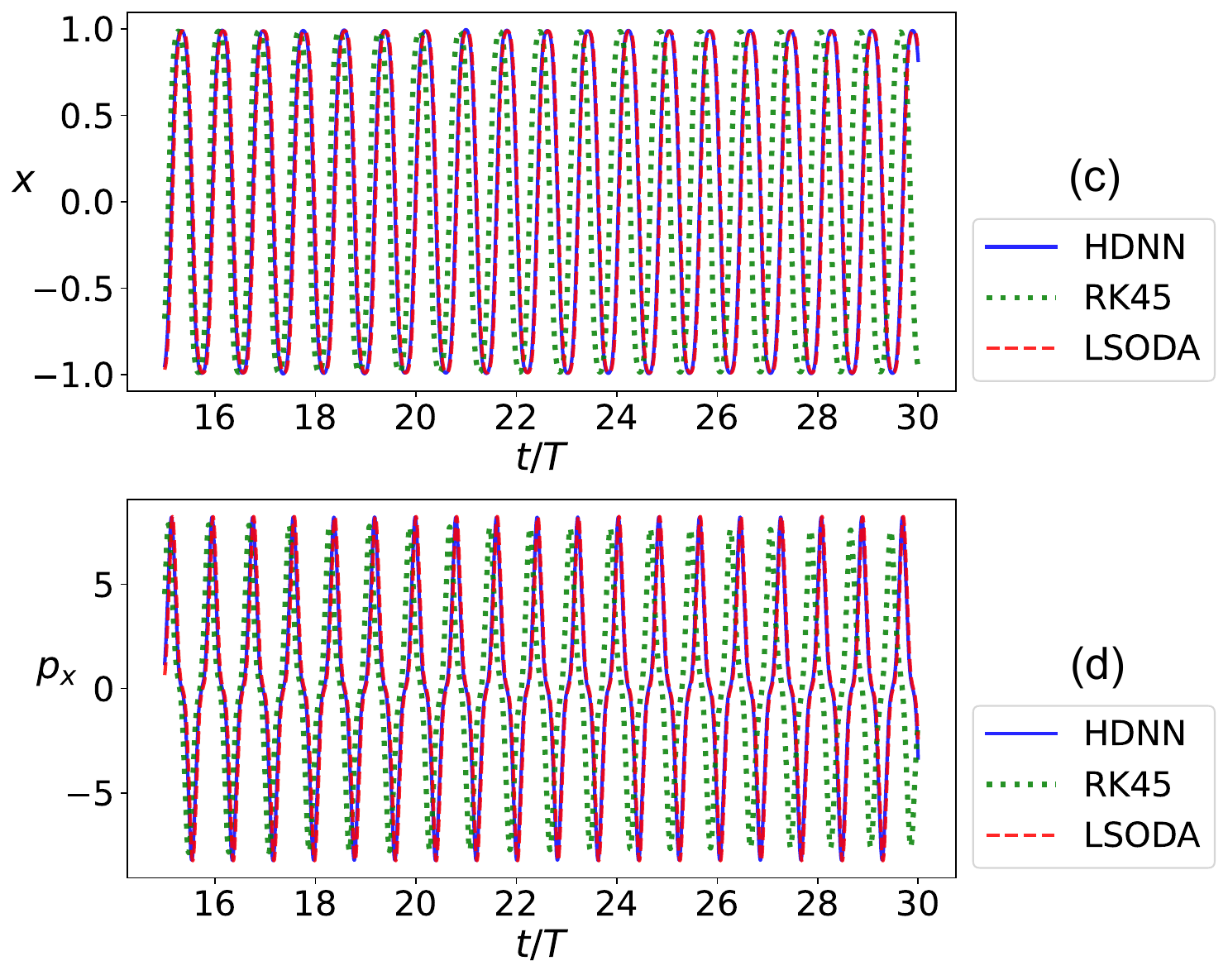}
    \caption[font=small,labelfont=bf]{Time series for the canonical variables $x,p_x$ produced by the HDNN solution \eqref{z_net} (blue solid line), compared with the ``exact'' LSODA solution (dashed red line) and the RK45 method (green dotted line) in the time window $t\in[15T,30T]$. Panels (a) and (b) show predictions for $m=1.5$ while (c) and (d) display the corresponding series for $m=2.0$. The RK45 solution exhibits energy dissipation, with the  momentum amplitude gradually decreasing and a slight phase drift. In contrast, the HDNN solution closely reproduces the exact motion and preserves the oscillation amplitude.}
    \label{fig_pend_timeseries}
\end{figure}

In Fig.~\ref{fig_pend_timeseries} we plotted the time series for $x,p_x$ obtained by the neural network solution \eqref{z_net} for the median value $m=1.5$ and the maximum seen value $m=2.0$, versus a high precision numerical solution, informally referred to as ``exact solution''. This is obtained by solving the Hamilton-Dirac equations with the LSODA algorithm\footnote{{ LSODA is an Adams/BDF numerical method with automatic stiffness detection and automatic switching \cite{Hindmarsh1983}.}}  using a dense grid. To evaluate the constrained neural network method, we also integrated Eqs.~\eqref{dx_pendulum}--\eqref{dpy_pednulum} using the fifth-order Runge-Kutta (RK45) method. The HDNN prediction closely reproduces the exact motion for both values $m=1.5$ and $m=2.0$. In both cases, it is evident that the HDNN solution conserves the amplitude of the oscillation and momentum and consequently the energy and the constraints are  conserved. In contrast, the RK45 method shows energy dissipation and a slight phase drift.

\begin{figure}[ht!]
    \centering
    \includegraphics[scale=0.28]{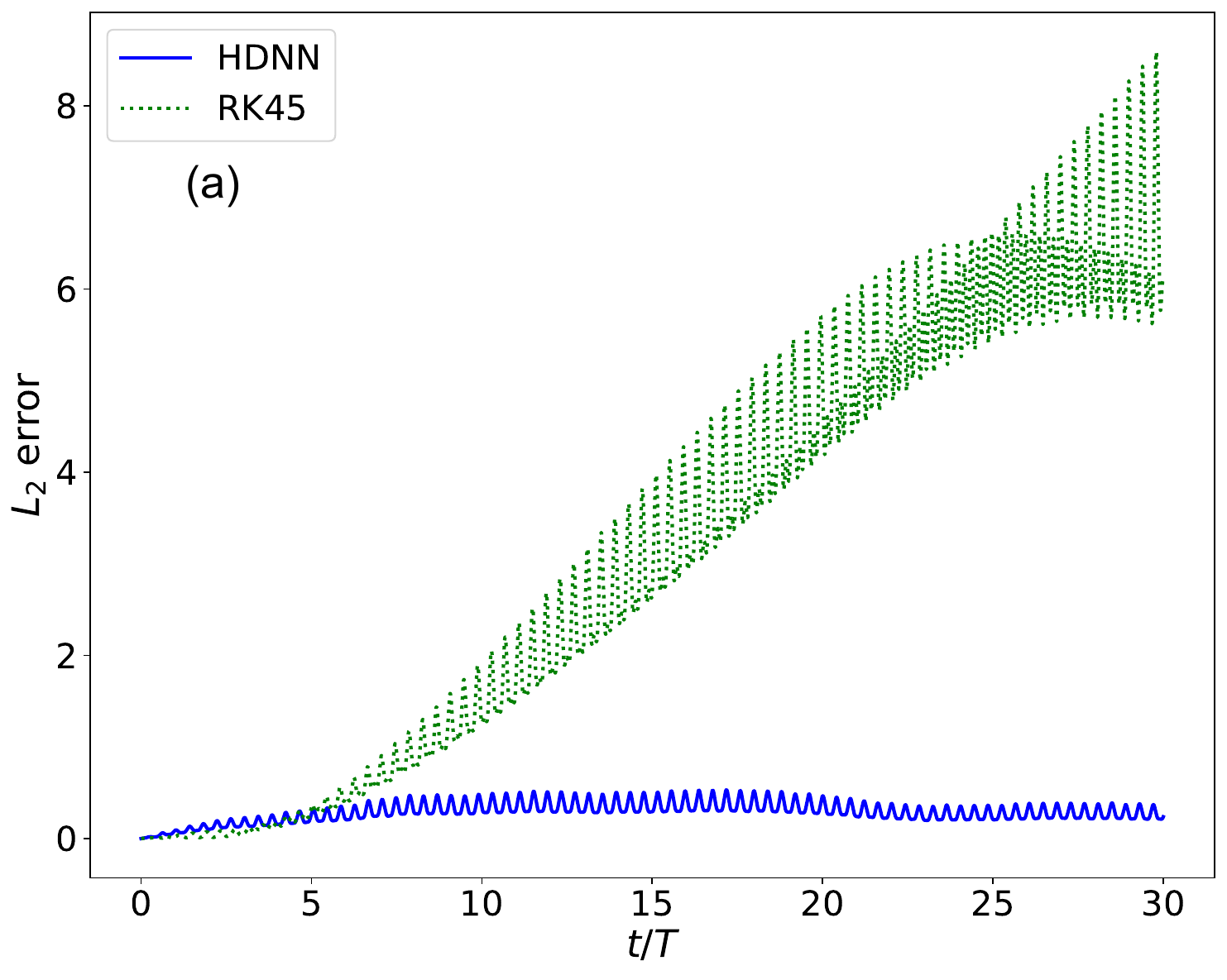}
    \includegraphics[scale=0.28]{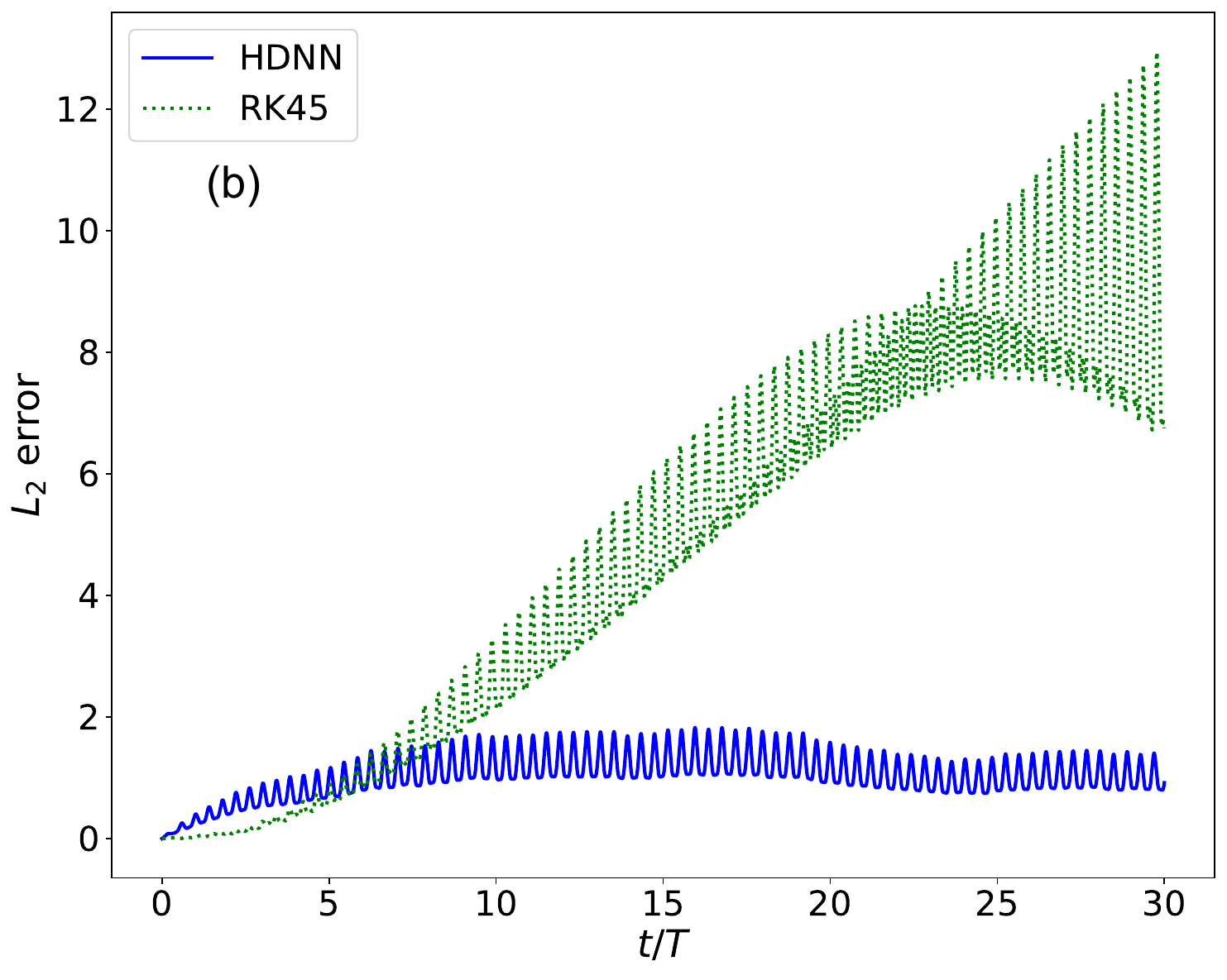}
    \caption[font=small,labelfont=bf]{The $L_2$ error of the predicted solutions (blue solid line) is compared to the RK45 method (green dotted line) for $m=1.5$ in panel (a) and $m=2.0$ in panel (b). In (a) the HDNN prediction is significantly more accurate than the RK45 method. In (b) the HDNN prediction is less accurate for small times but maintains this accuracy over the entire time domain $t$ while the RK45 error increases with time.}
    \label{fig_pend_timeseries_error}
\end{figure}

{ To illustrate these effects, Fig.~\ref{fig_pend_timeseries_error} shows the $L_2$ error of both solutions, computed with respect to the ``exact solution'' over time. At each time $t$, this error is defined as $L_2 = \sqrt{\sum_{i=1}^{2N}(z^i-\tilde{z}^i)^2}$, where $\tilde{\textbf{z}}$ is the LSODA solution. For the median mass value $m=1.5$, the HDNN prediction is significantly more accurate than the RK45 solution. For $m=2.0$, the HDNN prediction is initially less accurate than the RK45 solution, but becomes more accurate at larger times. Thus, once trained, the HDNN can produce all trajectories with better accuracy than RK45, for pendulum masses ranging from $m=1.0$ to $m=2.0$ via a simple forward pass through the network, requiring only inference time. This capability provides a substantial advantage over traditional solvers, which must be invoked repeatedly for multi-query problems.}

\begin{figure}[ht!]
    \centering
    \includegraphics[scale=0.3]{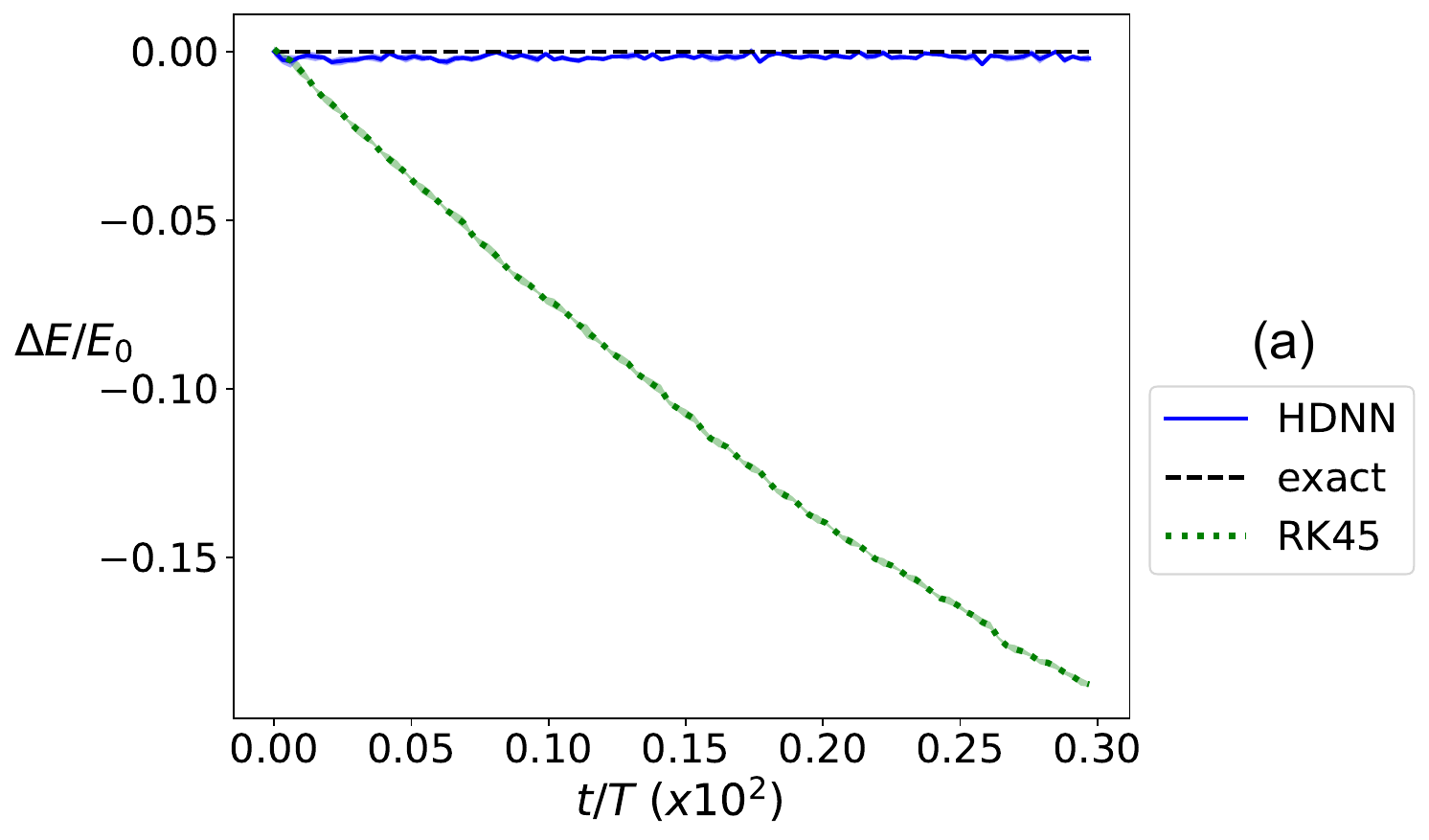}
    \includegraphics[scale=0.3]{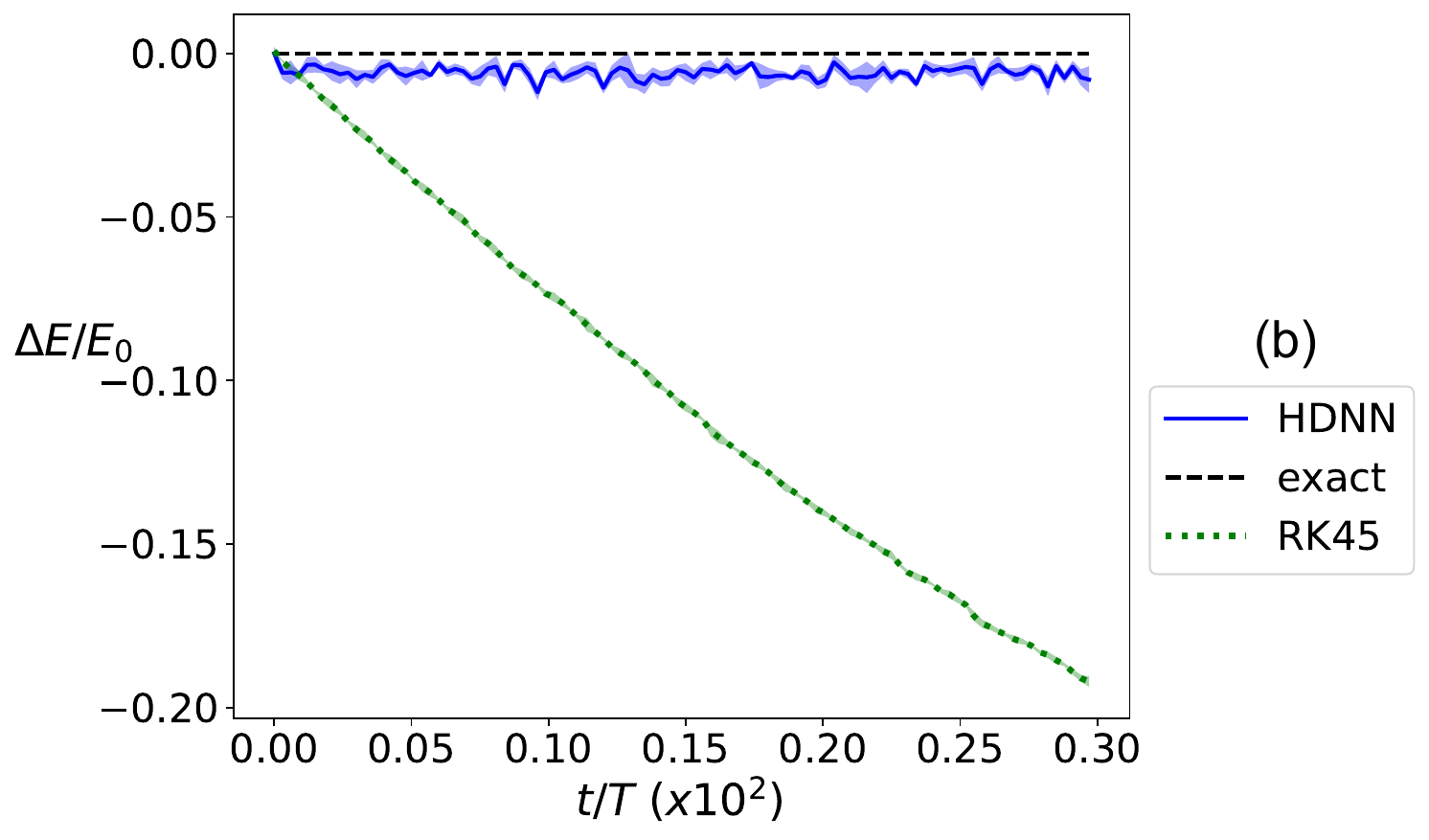}
    \caption[font=small,labelfont=bf]{Energy drift diagrams of the HDNN predictions (solid blue line) and the RK45 solutions (dotted green line) for (a) $m=1.5$ and (b) $m=2.0$.  The energy drift is quantified by the quantity $(E(t)-E_0)/E_0$. The lines represent the mean values of this quantity computed in batches of 100 time steps and the shaded regions represent the standard deviation in these batches. Evidently, the HDNN predictions conserve the energy, while RK45 produces spurious dissipation.}
    \label{fig_pend_energy}
\end{figure}

\begin{figure}[ht!]
    \centering
    \includegraphics[width=74mm,height=80mm]{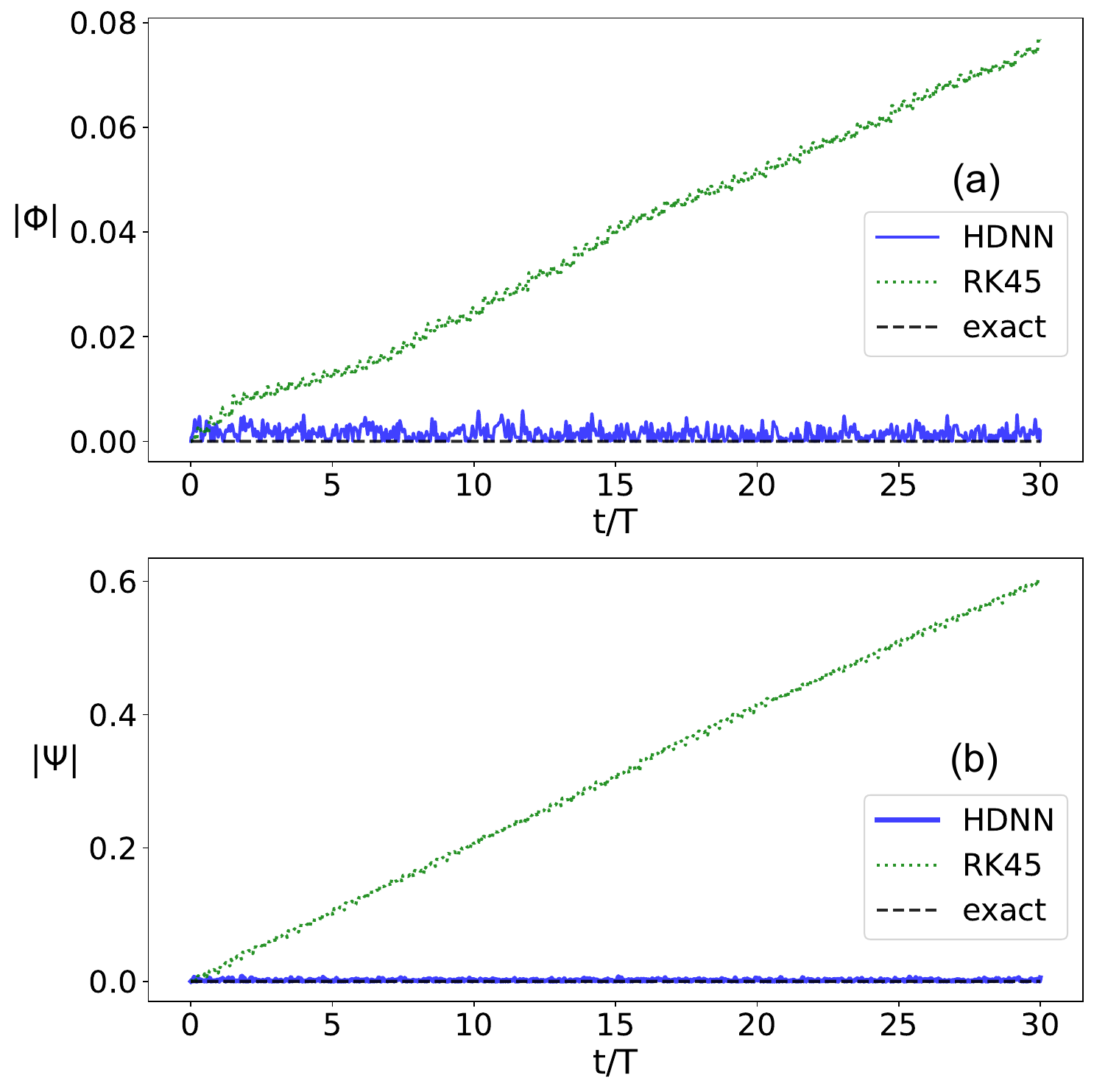}
    \includegraphics[width=74mm,height=80mm]{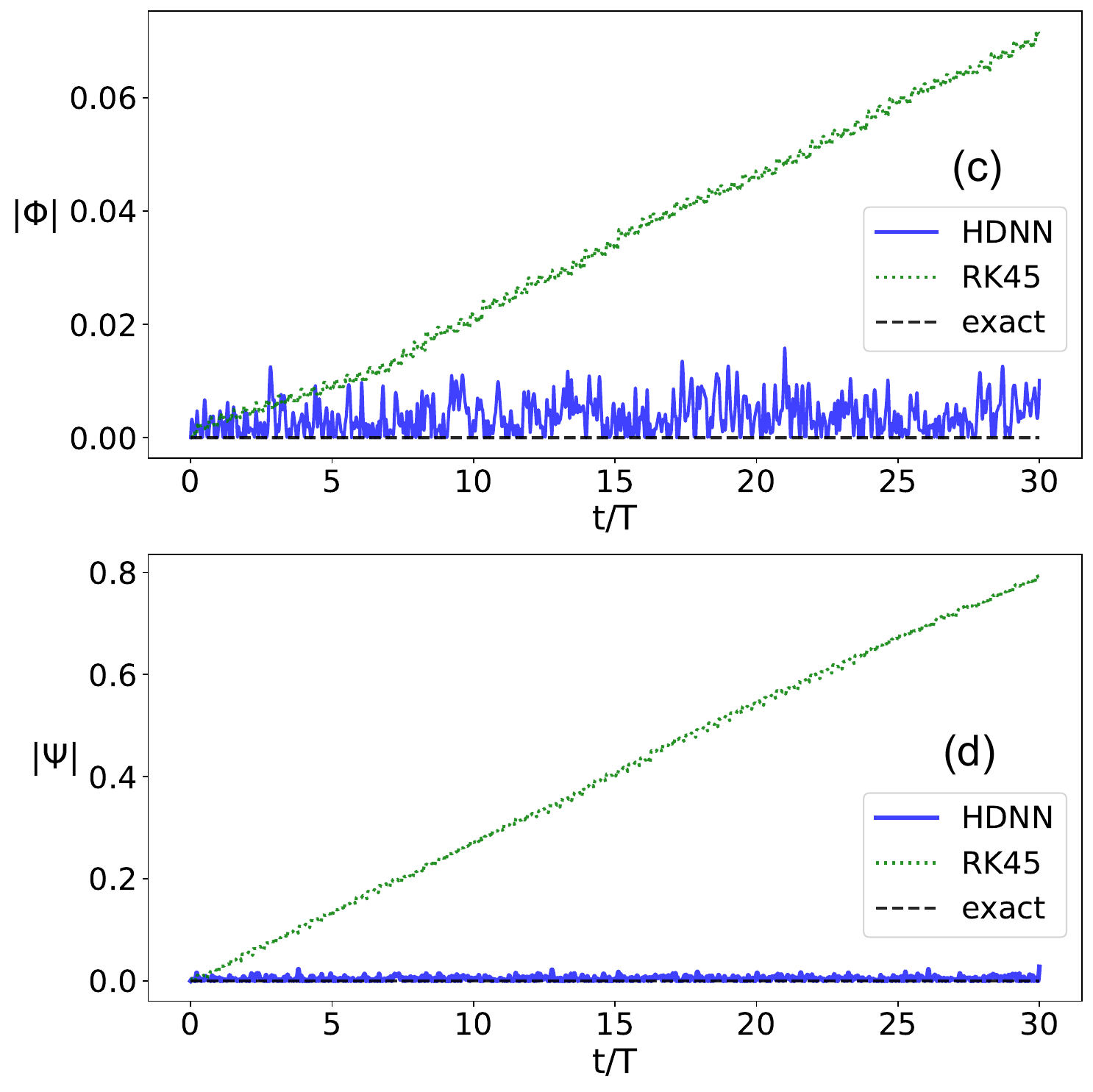}
    \caption[font=small,labelfont=bf]{Constraint violation diagrams for the planar pendulum comparing HDNN predictions and RK45 solutions. Panels (a) and (b) show the evolution of $|\Phi|$ and $|\Psi|$, respectively, for $m=1.5$ while panels (c) and (d) display the corresponding results for $m=2.0$. Clearly, the HDNN predictions preserve the constraints over time, whereas the RK45 solution deviates from the constraint manifold.}
    \label{fig_pend_constrs}
\end{figure}

\begin{figure}[ht!]
    \centering
    \includegraphics[scale=0.3]{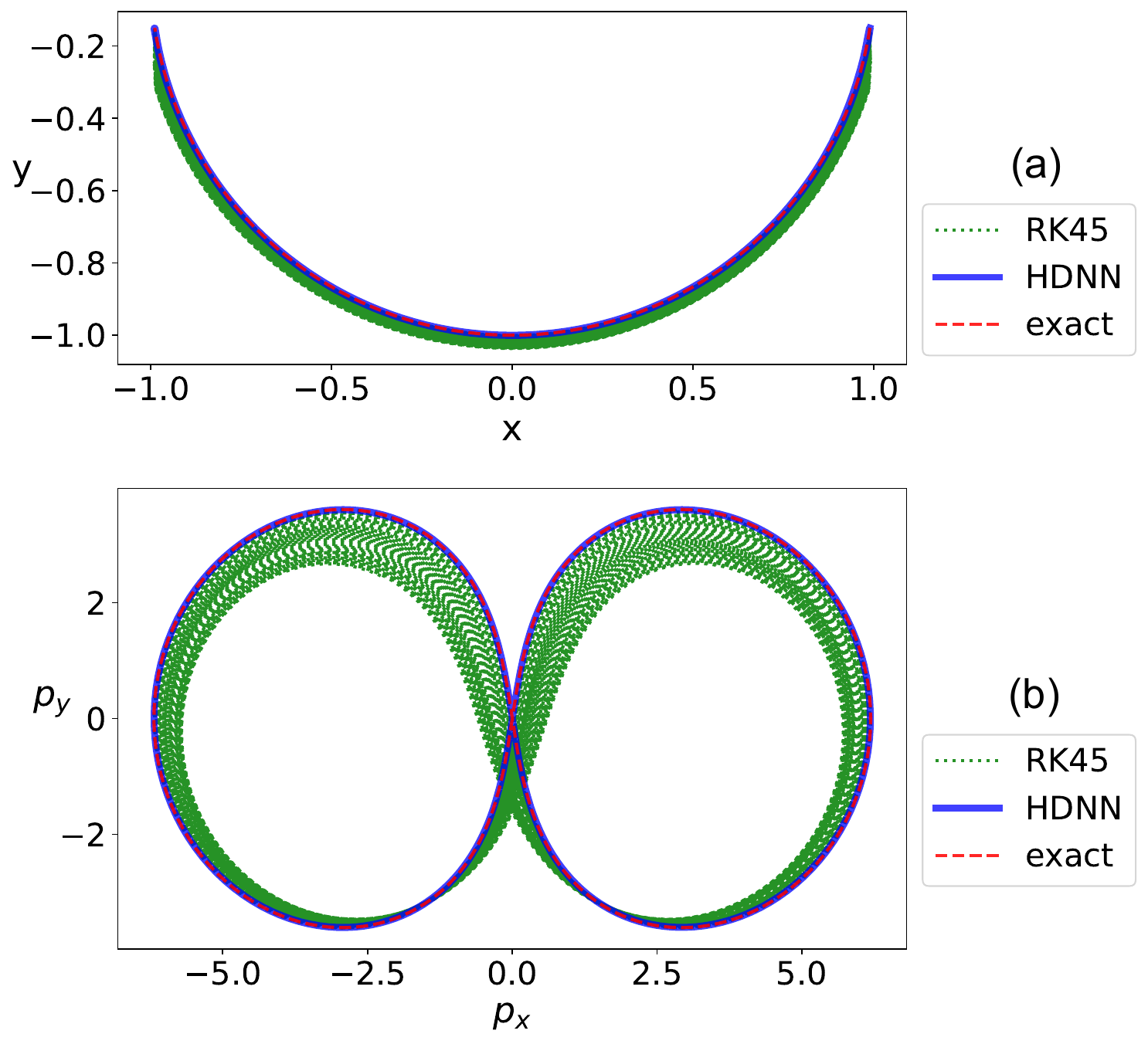}
        \includegraphics[scale=0.3]{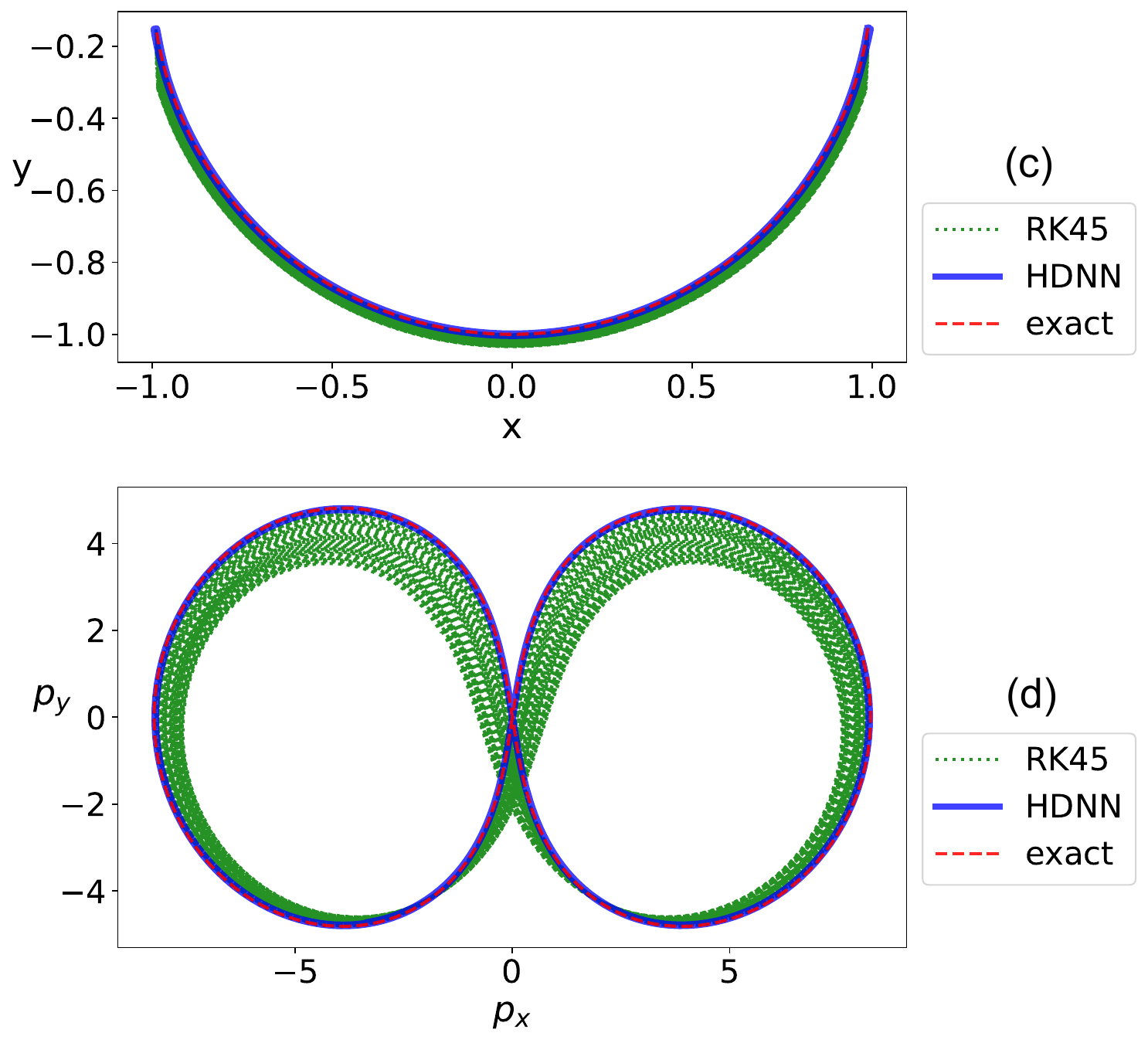}
    \caption[font=small,labelfont=bf]{Panels (a) and (b) display the HDNN-predicted trajectories of the pendulum (blue solid line), along with the LSODA (red dashed line) and RK45 (green dotted line) solutions, in the physical space $(x,y)$ and the $(p_x,p_y)$ section of phase space, respectively, for $m=1.5$. Panels (c) and (d) show the corresponding trajectories for $m=2.0$. The HDNN solution remains on the exact trajectory throughout the time evolution, while the RK45 solution drifts away due to energy dissipation.}
    \label{fig_pend_trajectories}
\end{figure}

\begin{figure}[ht!]
    \centering
    \includegraphics[scale=0.32]{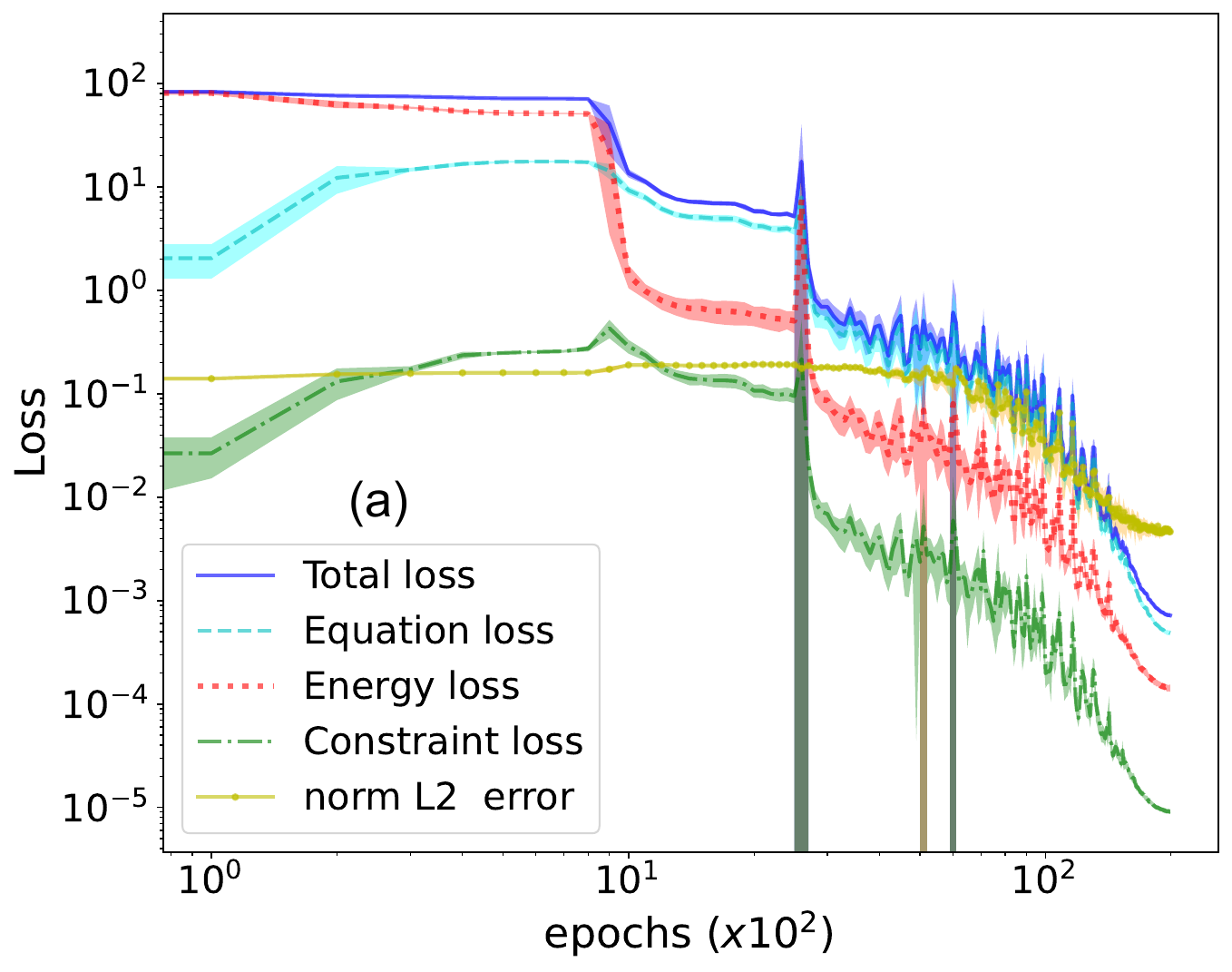}
     \includegraphics[scale=0.32]{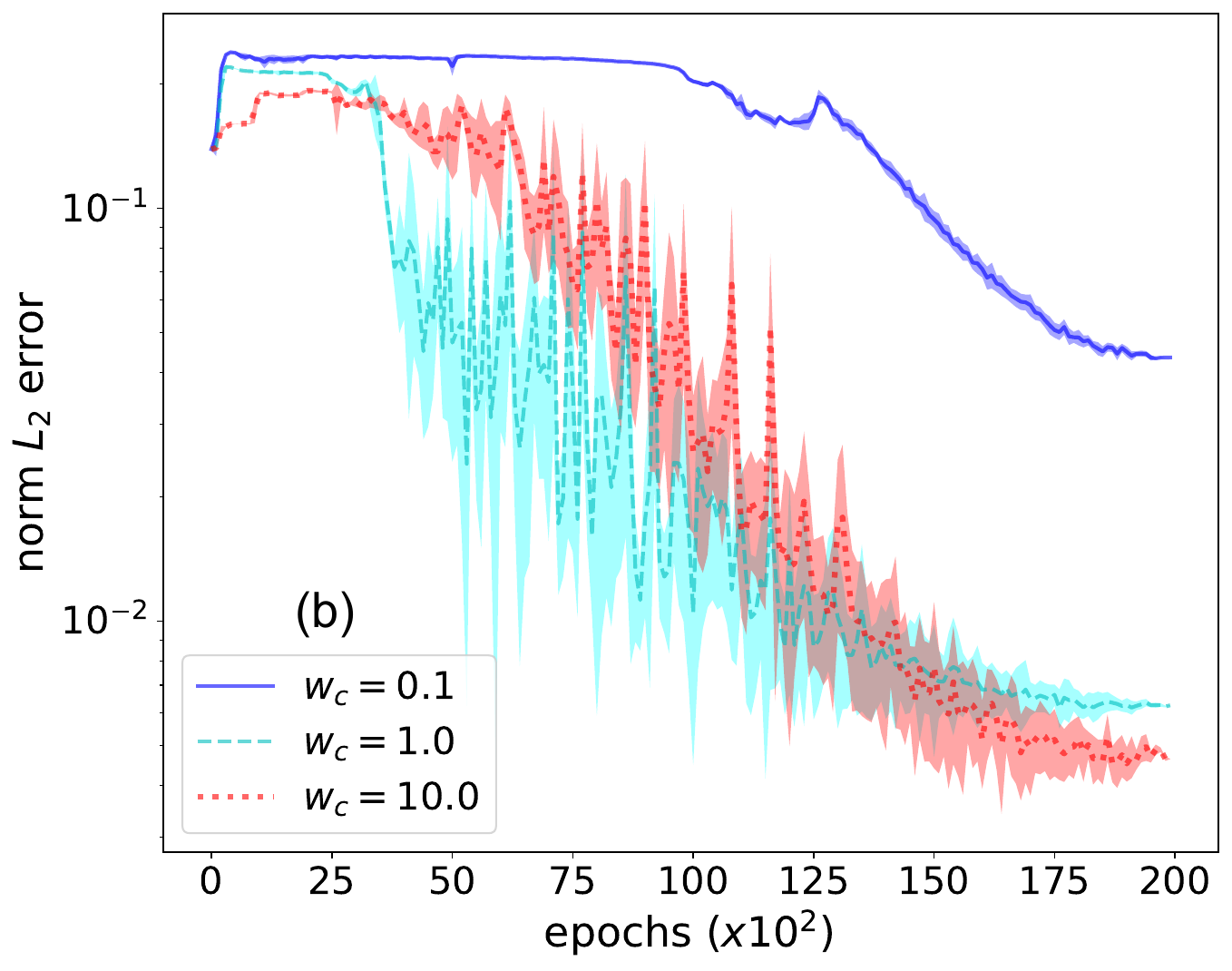}
    \caption[font=small,labelfont=bf]{Panel (a) shows the loss history for the constrained optimization  and. Here the total loss is displayed along with the energy and the constraint terms of the loss function. The normalized $L_2$ error is also displayed for the median value $m=1.5$ calculated on $n$ $t$-grid points. This is defined as $L_2=(1/n)\sqrt{\sum_{i=1}^{2N}\sum_{j=1}^n(z^i(t_j)-\tilde{z}^i(t_j))^2}$. Panel (b) illustrates the impact of the Dirac constraints on the accuracy of the predicted solutions. In both figures, the lines indicate the mean values and the shaded regions represent the standard deviations calculated over batches of 100 epochs. As shown in (b), increasing the weight of the Dirac constraints in the total loss \eqref{loss_gen} from $0.1$ to $1.0$, improves the accuracy of the predicted solution. It has been corroborated that accurate solutions can be obtained also for $w_c\rightarrow 0$ but they require more training epochs and collocation points.}
    \label{fig_pend_losses}
\end{figure}

The energy and constraint-conserving properties of the HDNN, are demonstrated by plotting the normalized energy difference $ (E(t)-E_0)/E_0$ and the absolute values of the constraints $|\Phi|$ and $|\Psi|$, over time. This is done in Figs.~\ref{fig_pend_energy} and \ref{fig_pend_constrs}, respectively where the HDNN predictions are compared with the corresponnding RK45 numerical solutions. {  The HDNN predictions are shown for $m=1.5$ and $m=2.0$. In both cases, the HDNN conserves energy and the Dirac constraints with high precision, in contrast to the RK45 method, which exhibits energy dissipation and constraint violation.} We also plot the pendulum trajectory on the section planes $x-y$ and $p_x-p_y$ of the phase-space in Fig.~\ref{fig_pend_trajectories}. Clearly, both HDNN predictions follow the correct path of the pendulum motion, while the RK45 solution drifts away due to artificial energy dissipation.

For this, and the subsequent examples we have used an HDNN architecture with 4 hidden layers and 160 neurons per layer. The network has been trained for 20000 training epochs using 1000 discrete time points in the $[0,30T]$ time domain.  Regarding the performance of the NN algorithm, the training time was 9.7 minutes on an Nvidia RTX-3060 GPU, whereas the RK45 solver took 0.13 seconds using the same time resolution. Clearly, the training of the HDNNs is far from efficient in terms of computational time. {Nevertheless, the HDNN's ability to predict trajectories for various pendulum masses with higher precision than RK45 at inference time, makes it ultimately more efficient. This method can be generalized by incorporating pre-trained networks on different patches of the $m$ domain, combined into a meta-network capable of predicting trajectories for a much wider range of masses. This is the GPT-PINN concept \cite{Chen2024} which will be utilized elsewhere.}

{Finally, to assess the convergence of the algorithm, we plot the history of the total loss function \eqref{loss_gen} and its different terms in Fig.~\ref{fig_pend_losses} (a). The first term in \eqref{loss_gen} quantifies the residual of the equations of motion, while the two terms comprising $\Lc_{reg}$ in \eqref{loss_reg} quantify the energy drift and the violation of Dirac constraints. The normalized $L_2$ error computed for the median mass value $m=1.5$ is also displayed. In the panel (b) of Fig.~\ref{fig_pend_losses} we present the impact of the term $\Lc_c$ on the accuracy of the predicted solutions. It is evident that increasing the weight $w_c$ of the constraint term from $0.1$ to $1.0$, improves the accuracy of the solution. Accurate solutions can be obtained also for $w_c\rightarrow 0$ but require more training epochs and collocation points. Initialization with appropriately pre-trained networks can also improve the convergence.}


\subsection{Elliptically restricted 2D harmonic oscillator}
\label{subsec_4.2}
The real advantage of the Dirac method becomes evident when considering  holonomic constraints expressed in the form $f(q_1,...,q_N) = 0$, for which there is no suitable coordinate system available to effectively reduce the dimensionality of the problem. In this subsection, we will consider the case of an elliptically restricted 2D harmonic oscillator, wherein a mass $m=1$ moves within a potential of the form:
\begin{eqnarray}
    V(x,y) = \frac{\alpha}{2} x^2 + \frac{\beta}{2} y^2\,,
\end{eqnarray}
and its motion is constrained on the ellipse
\begin{eqnarray}
    x^2 +\frac{y^2}{1-\epsilon^2} - a^2 =0\,,
\end{eqnarray}
where $a$ is the semi-major axis and $\epsilon$ is the eccentricity of the ellipse. 

Although elliptic coordinates could be used to study this problem, such an approach would be more cumbersome. The Dirac method, by contrast, offers greater generality, as it can restrict motion to any shape defined by $f(x,y)=0$. This subsection highlights the clear advantage of HDNNs over traditional numerical solvers, as both LSODA and RK45 fail to confine motion on the constraint manifold, even with a dense time grid. In contrast, HDNN predictions maintain the orbits on the constraint manifold over long times. { Furthermore, once trained, the HDNN predicts accurate orbits across a range of eccentricity values $\epsilon$ at inference time, offering a significant advantage over traditional solvers.}

\begin{figure}[ht!]
    \centering
    \includegraphics[scale=0.3]{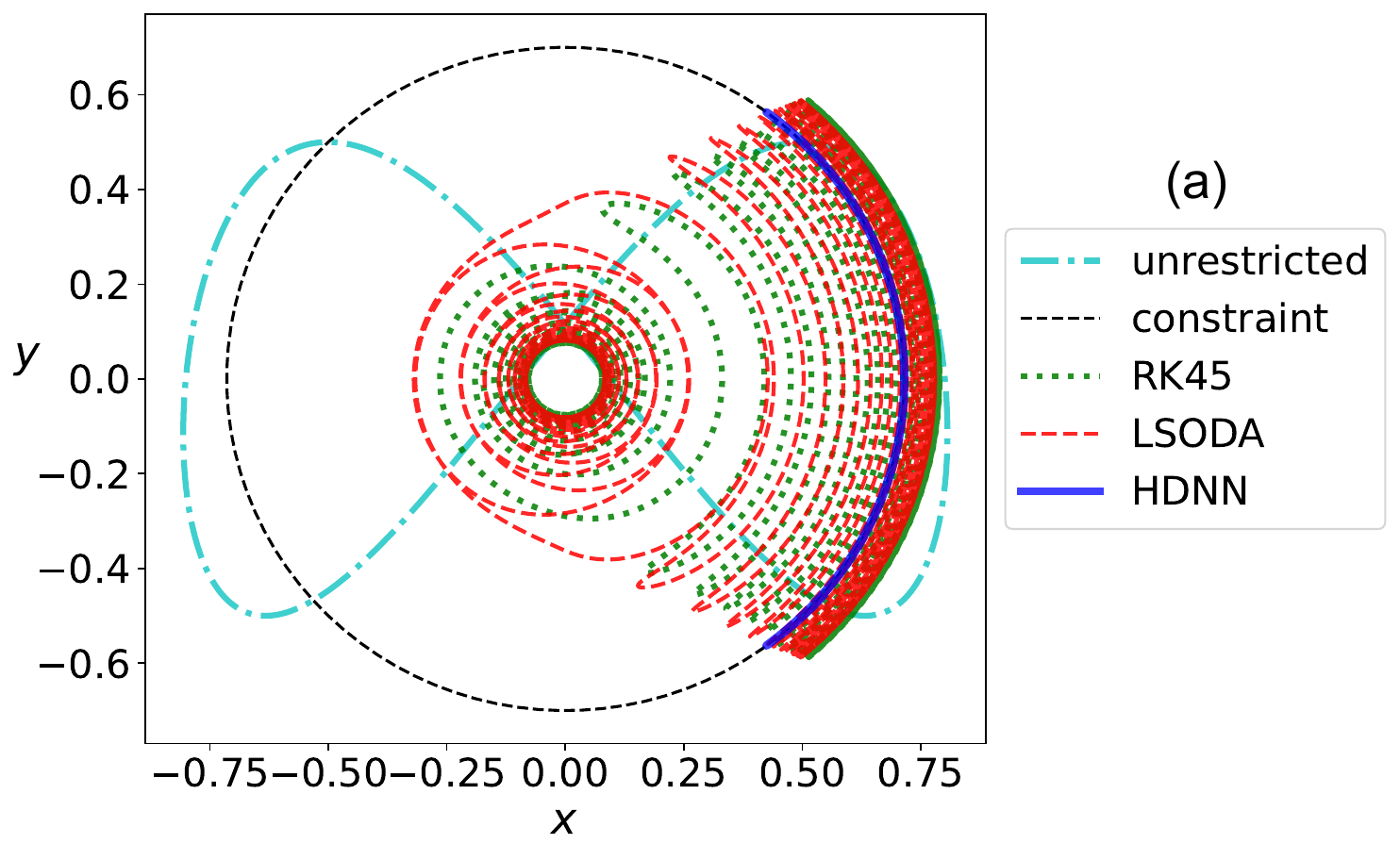}
    \includegraphics[scale=0.3]{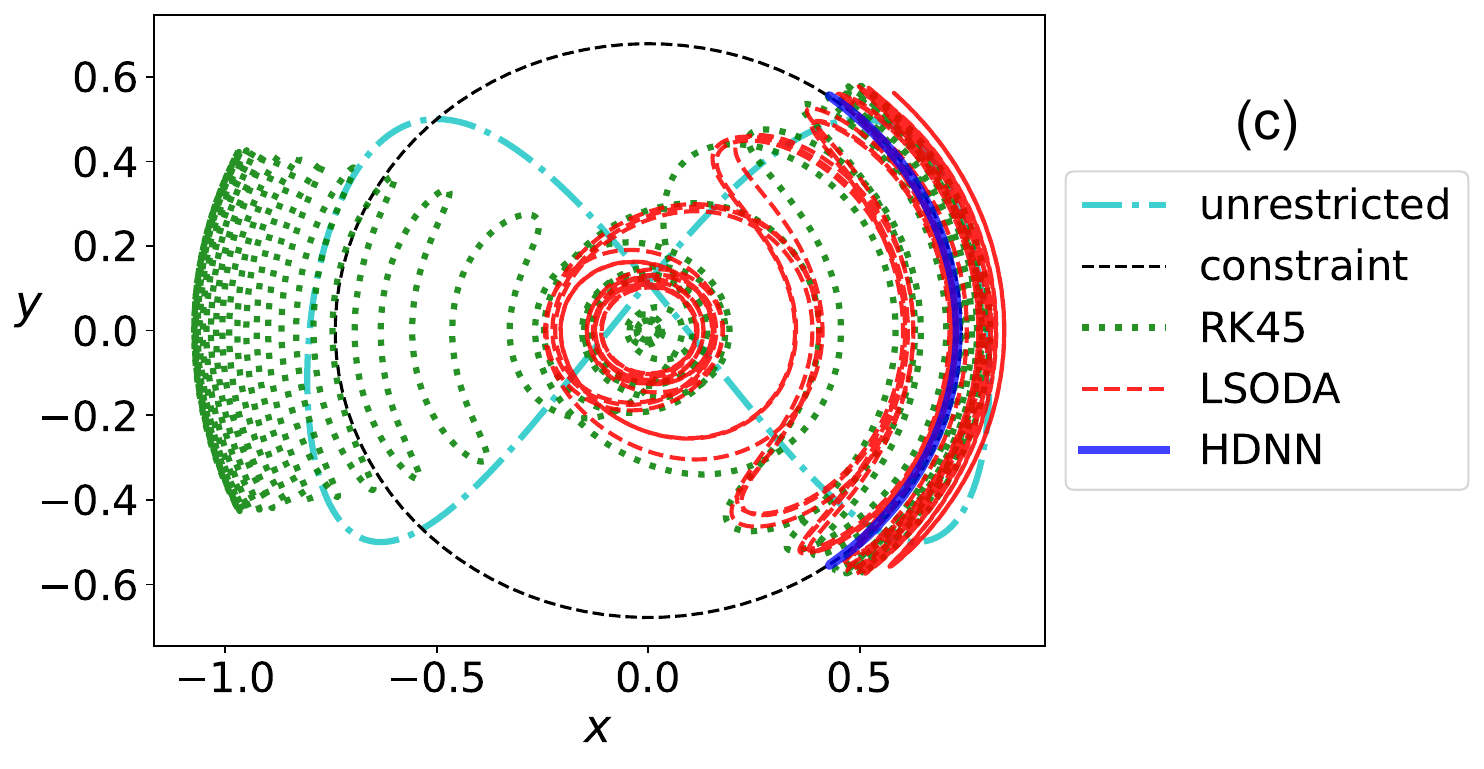}
    \includegraphics[scale=0.3]{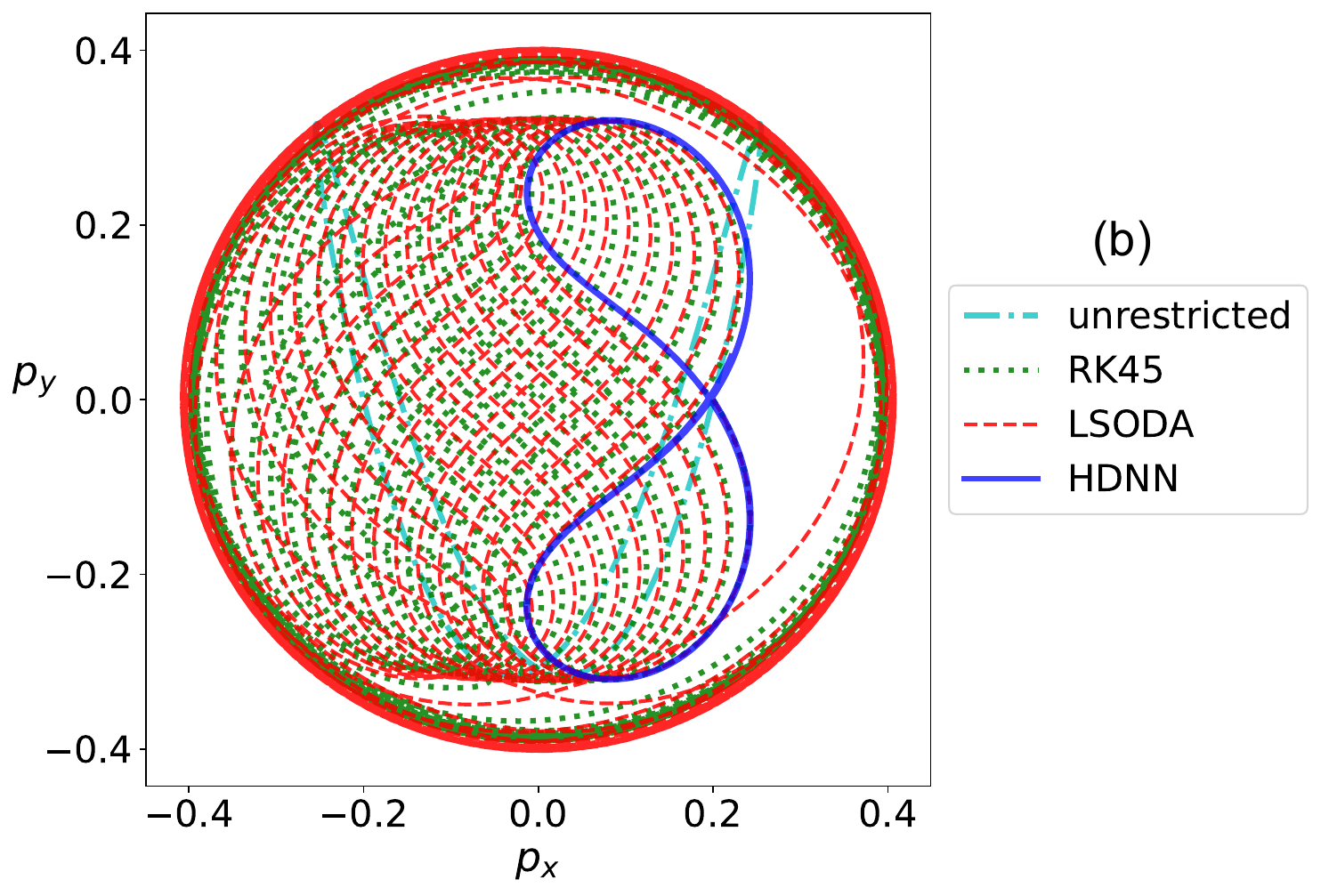}
    \includegraphics[scale=0.3]{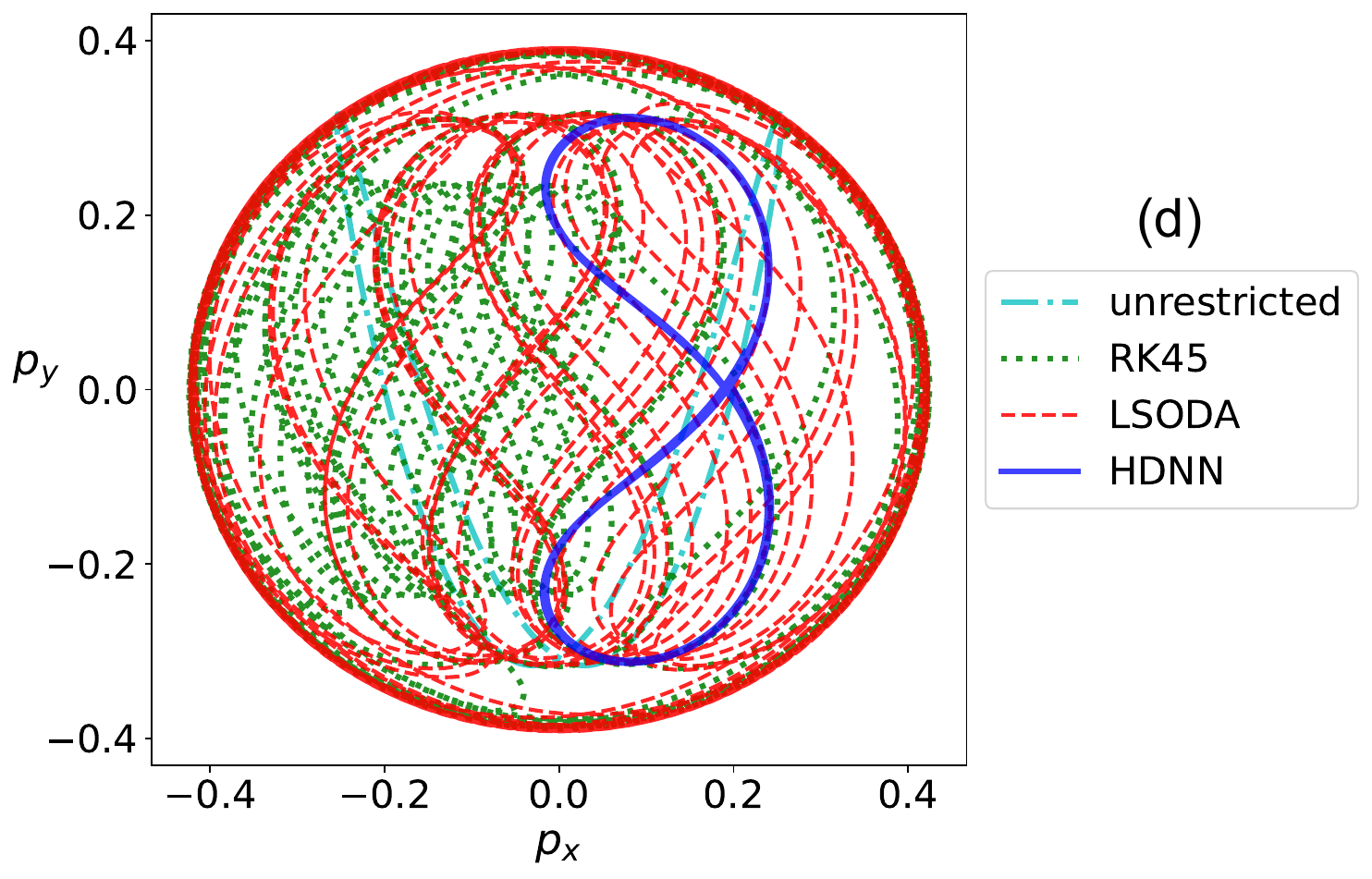}
    \caption[font=small,labelfont=bf]{Panels (a) and (b) display the trajectories of the elliptically restricted harmonic oscillator predicted by the HDNN and computed using the LSODA and RK45 on the $x-y$ and $p_x-p_y$ planes, respectively, for $\epsilon=0.2$. Panels (c) and (d) show the corresponding trajectories for $\epsilon=0.4$. The HDNN predictions consistently adhere to the elliptical constraint, whereas the standard numerical solutions fail to reproduce the true motion.} 
    \label{fig_erho_trajectories}
\end{figure}

In this particular example the primary constraint function is:
\begin{eqnarray}
    \Phi(x,y) = \frac{1}{2}\left(x^2 + \frac{y^2}{1-\epsilon^2} -a^2\right)\,. \label{erho_Phi}
\end{eqnarray}
Repeating the analysis of the previous subsection, we find the following secondary Pfaffian constraint and total Hamiltonian:
\begin{eqnarray}
    \Psi &=& xp_x +\frac{y p_y}{1-\epsilon^2}\,, \label{erho_Psi}\\
    H_t &=& \frac{p_x^2}{2} + \frac{p_y^2}{2} + V(x,y) - \frac{\lambda}{2}\left(x^2 + \frac{y^2}{1-\epsilon^2} -a^2\right)\,, \label{erho_Ht}
\end{eqnarray}
where 
\begin{eqnarray}
    \lambda = -\frac{p_x^2 + \frac{p_y^2}{1-\epsilon^2}-\alpha x^2 -\frac{\beta y^2}{1-\epsilon^2}}{x^2 + \frac{y^2}{(1-\epsilon^2)^2}}\,,
\end{eqnarray}
is the Lagrange multiplier determined by the consistency condition $\{\Psi,H_t\}=0$. The Dirac bracket is:
\begin{eqnarray}
    \{F,G\}_* = \{F,G\} - \frac{1}{x^2 + \frac{y^2}{(1-\epsilon^2)^2}}\left(\{F,\Psi\}\{\Phi,G\}- \{F,\Phi\}\{\Psi,G\} \right)\,,
\end{eqnarray}
and the Hamilton-Dirac equations of motion read as follows:
\begin{eqnarray}
    \dot{x} &=& \{x,H_c\}_*= p_x + \mu x\,, \label{erho_dxdt}\\
    \dot{y} &=& \{y,H_c\}_*= p_y + \mu y\\
    \dot{p}_x &=& \{p_x,H_c\}_*= -\alpha x -\mu p_x + \lambda x\,,\\
    \dot{p}_y &=& \{p_y,H_c\}_*= -\beta x - \mu p_y + \lambda y\,, \label{erho_dpydt}
\end{eqnarray}
where
\begin{eqnarray}
    \mu = -\frac{xp_x+\frac{yp_y}{1-\epsilon^2}}{x^2 + \frac{y^2}{(1-\epsilon^2)^2}}\,.
\end{eqnarray}
We train an HDNN so that \eqref{z_net} satisfies Eqs.~\eqref{erho_dxdt}--\eqref{erho_dpydt}, while it preserves the constraint functions $\Phi$, $\Psi$ and the total Hamiltonian $H_t$ (Eqs.~\eqref{erho_Phi}, \eqref{erho_Psi} and \eqref{erho_Ht}, respectively) by minimizing the loss function as defined in Eq.~\eqref{loss_gen}. {The HDNN is trained for eccentricities of the elliptical constraint \eqref{erho_Phi} ranging from $\epsilon_1=0.0$ to $\epsilon_2=0.4$}. We perform this minimization and solve the system \eqref{erho_dxdt}--\eqref{erho_dpydt} with the RK45 method and the LSODA algorithm, with initial conditions $(x_0,y_0,p_{x0},p_{y0})=(0.5,0.5,0.2,0.0)$ and potential function parameters $\alpha =0.1$, $\beta=0.4$. For this particular choice of parameters the unrestricted motion is in $1:2$ resonance and has a figure-8 trajectory on the $x-y$ plane (see Fig.~\ref{fig_erho_trajectories}). The HDNN solution was trained in the time domain $[0,300]$ on grid  points which are stochastically perturbed in each training epoch, while for the standard numerical methods we used a equidistant grid with 5000 points. In the training procedure we employed transfer learning to enhance efficiency. Initially, the network underwent training for 10000 epochs within the domain $[0,50]$ with 500 points. Subsequently, we gradually extended the domain by $50$ time units until reaching $[0,300]$. After each extension, the network underwent another 10,000 epoch training period, totaling 6 training periods with 10,000 epochs each. The number of points was increased from 500 to 1800 in the last training period, and the entire training lasted 30 minutes.
\begin{figure}[ht!]
    \centering
    \includegraphics[scale=0.3]{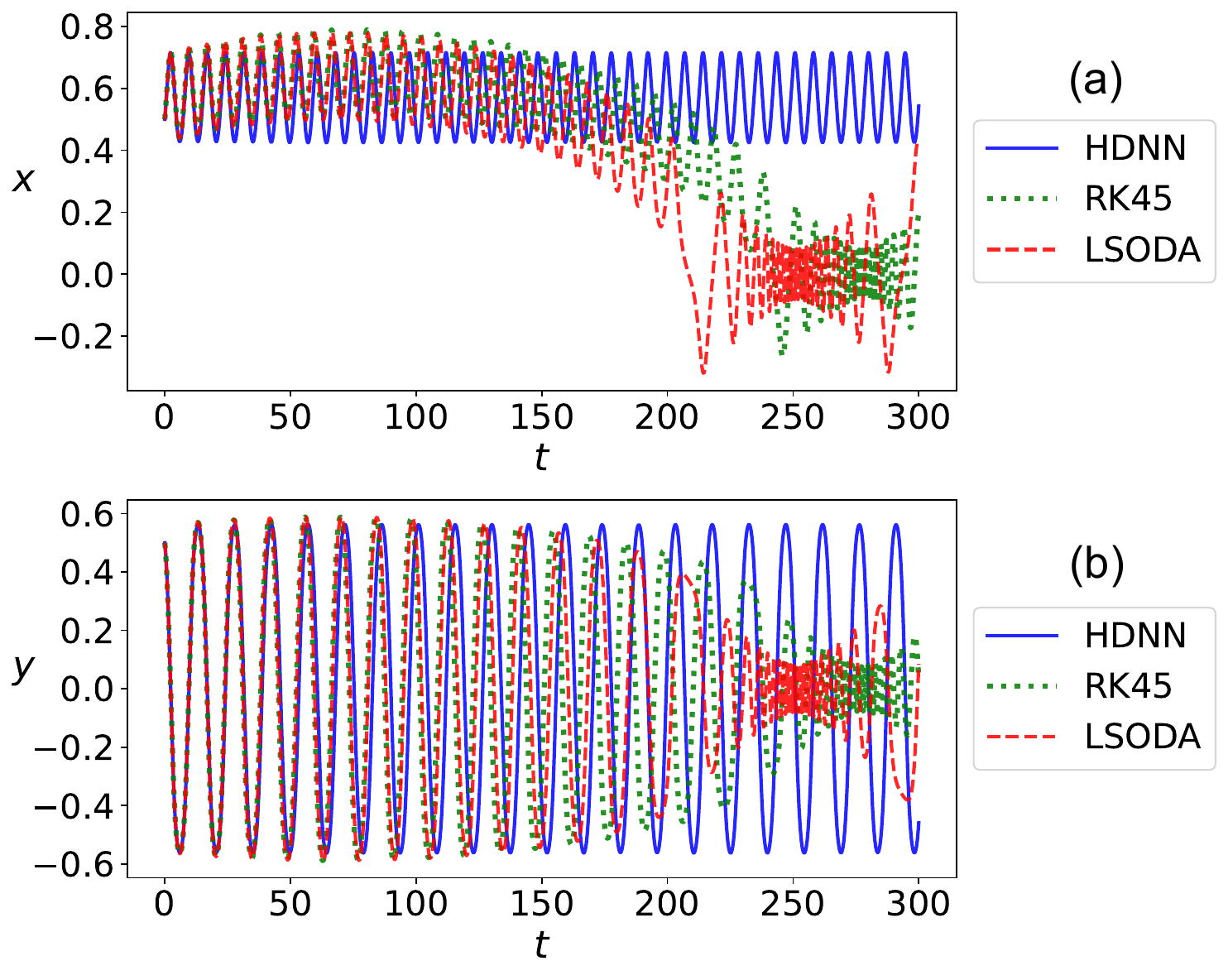}
    \includegraphics[scale=0.3]{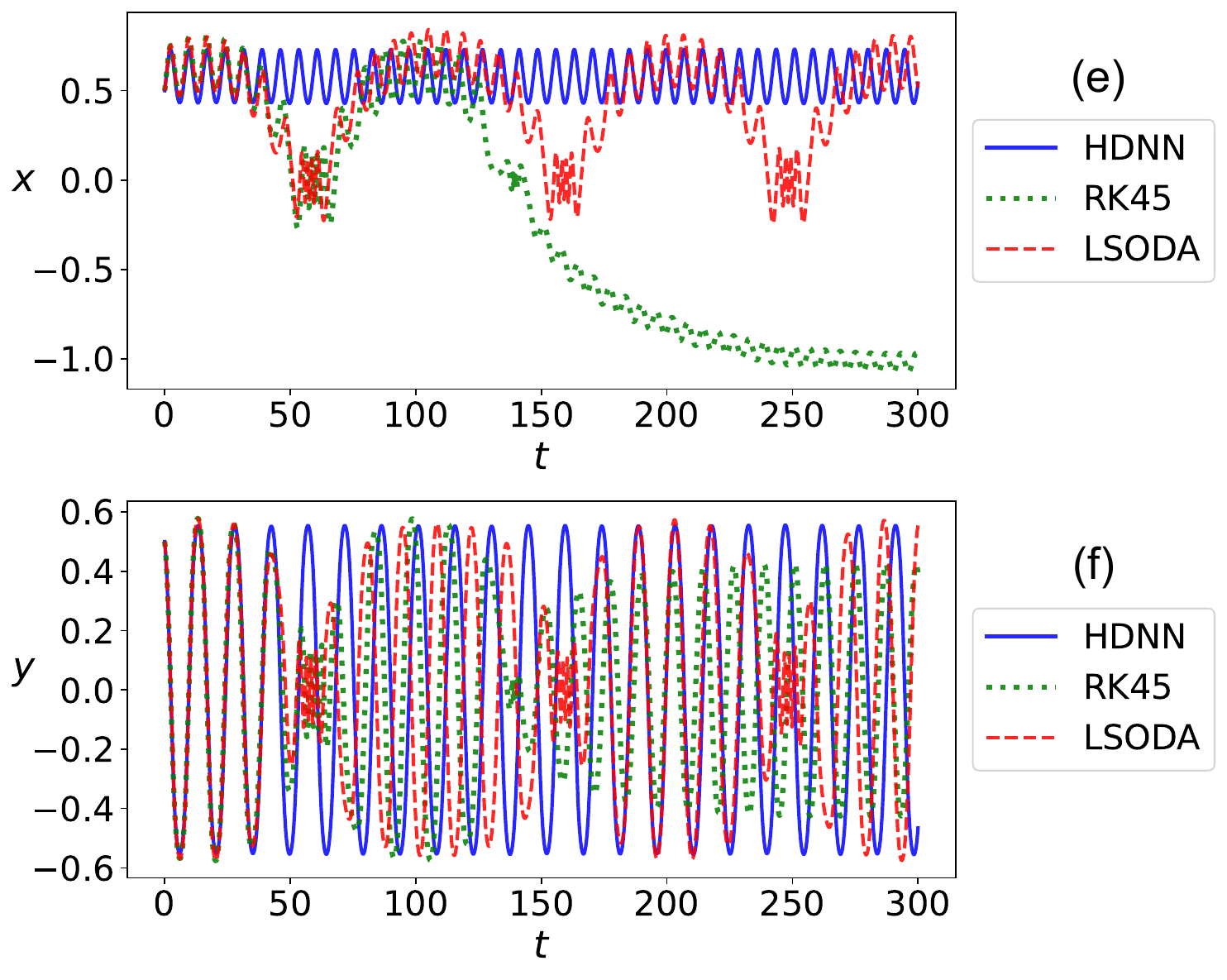}
    \includegraphics[scale=0.3]{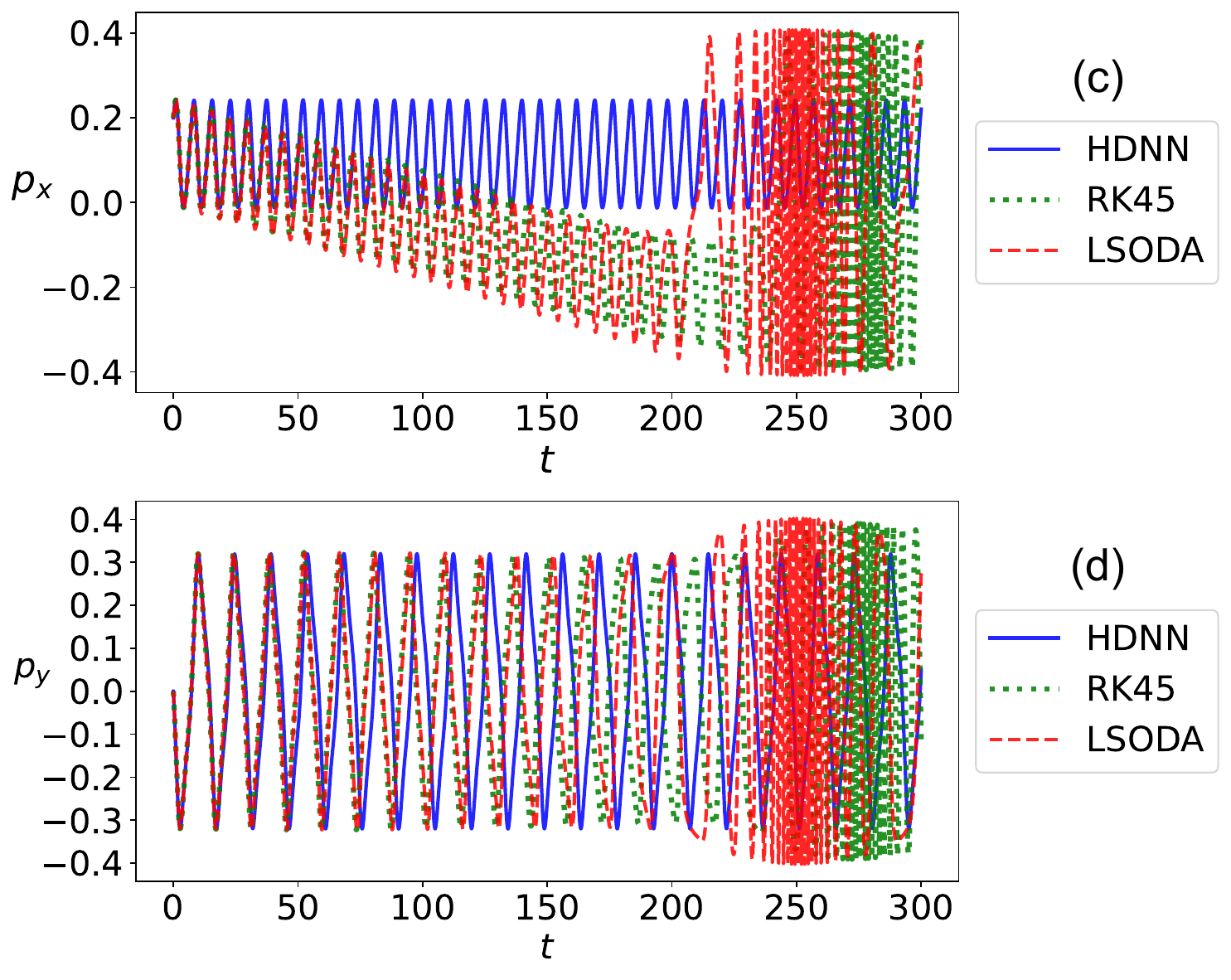}
    \includegraphics[scale=0.3]{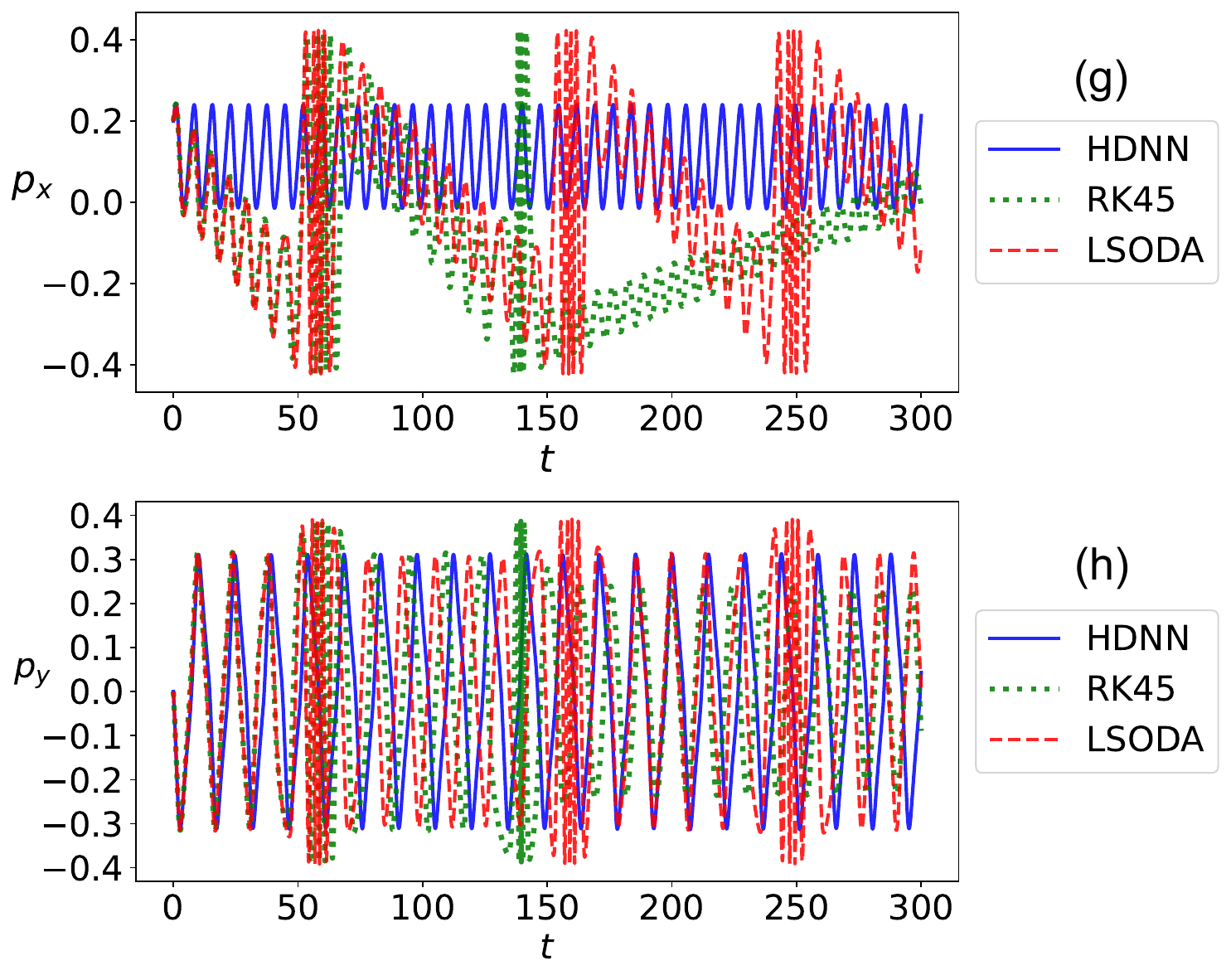}
    \caption[font=small,labelfont=bf]{(a)-(d) Time series for the elliptically restricted harmonic oscillator predicted by the HDNN for $\epsilon=0.20$, compared with the LSODA and RK45 methods. (e)-(h) The corresponding time series for $\epsilon=0.4$. The standard numerical methods exhibit unphysical behavior, while the HDNN predictions remain stable over the entire time domain.}
    \label{fig_erho_timeseries}
\end{figure}

\begin{figure}[ht!]
    \centering
    \includegraphics[scale=0.3]{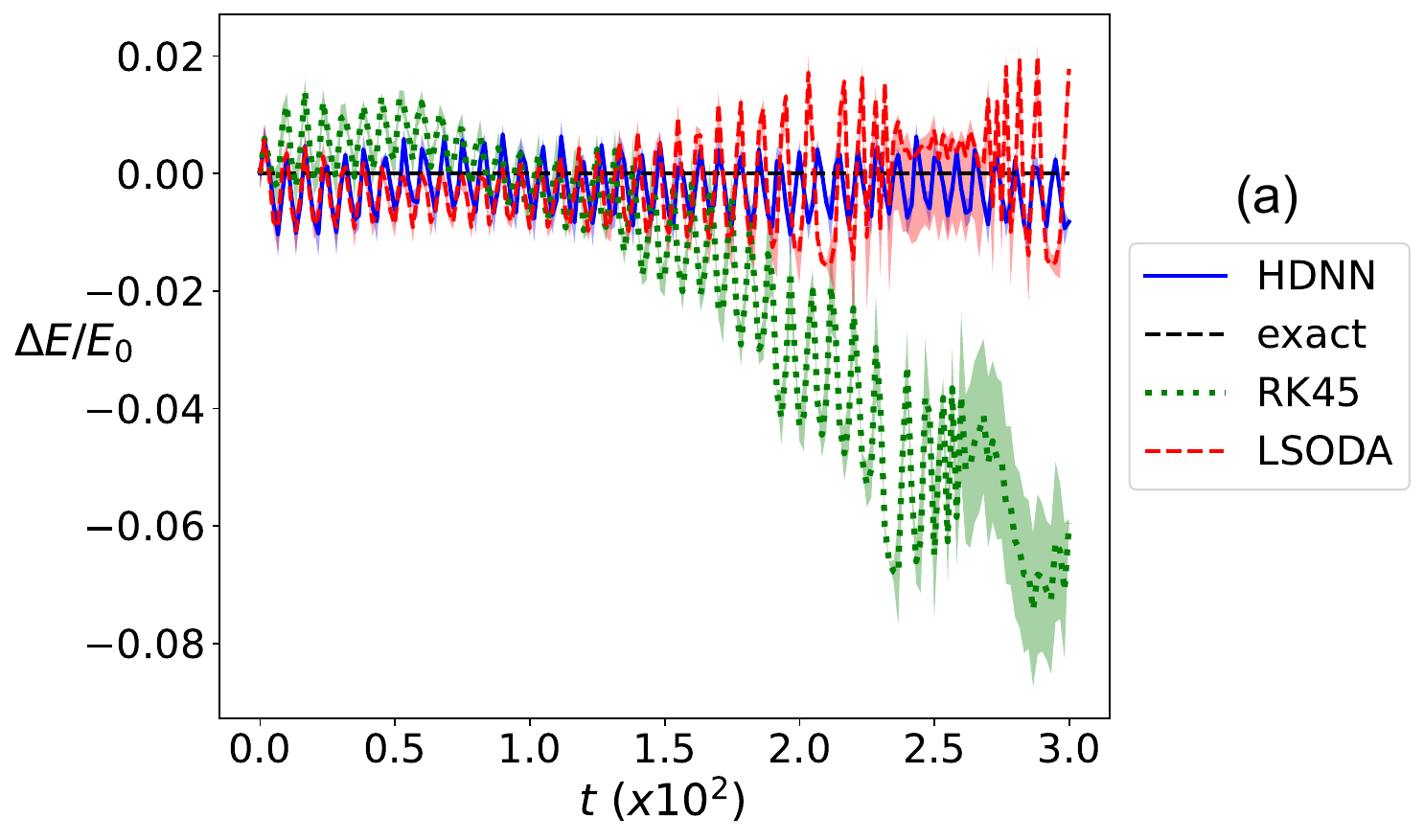}
    \includegraphics[scale=0.3]{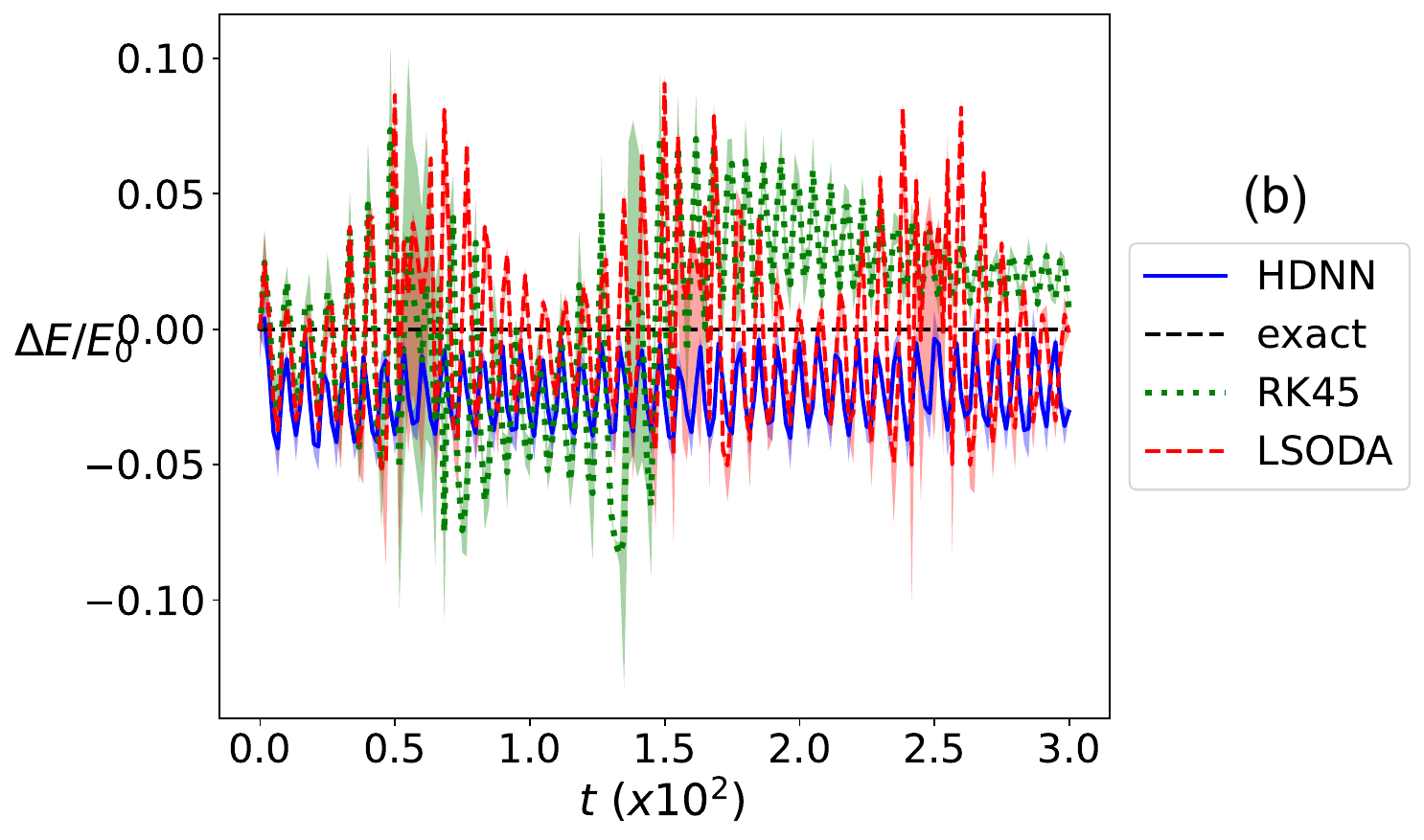}
    \caption[font=small,labelfont=bf]{(a) Time series of the quantity $\Delta E/E_0$, representing the energy drift, for $\epsilon=0.2$. Both the HDNN and the LSODA algorithm maintain the system's energy, while the RK45 solution exhibits significant drift. In panel (b) the corresponding diagrams for $\epsilon=0.4$ are displayed. The diagrams display the mean values computed in batches of 100 grid points, along with their standard deviation delineated by the shaded regions.}
    \label{fig_erho_energy}
\end{figure}

\begin{figure}[ht!]
    \centering
    \includegraphics[scale=0.25]{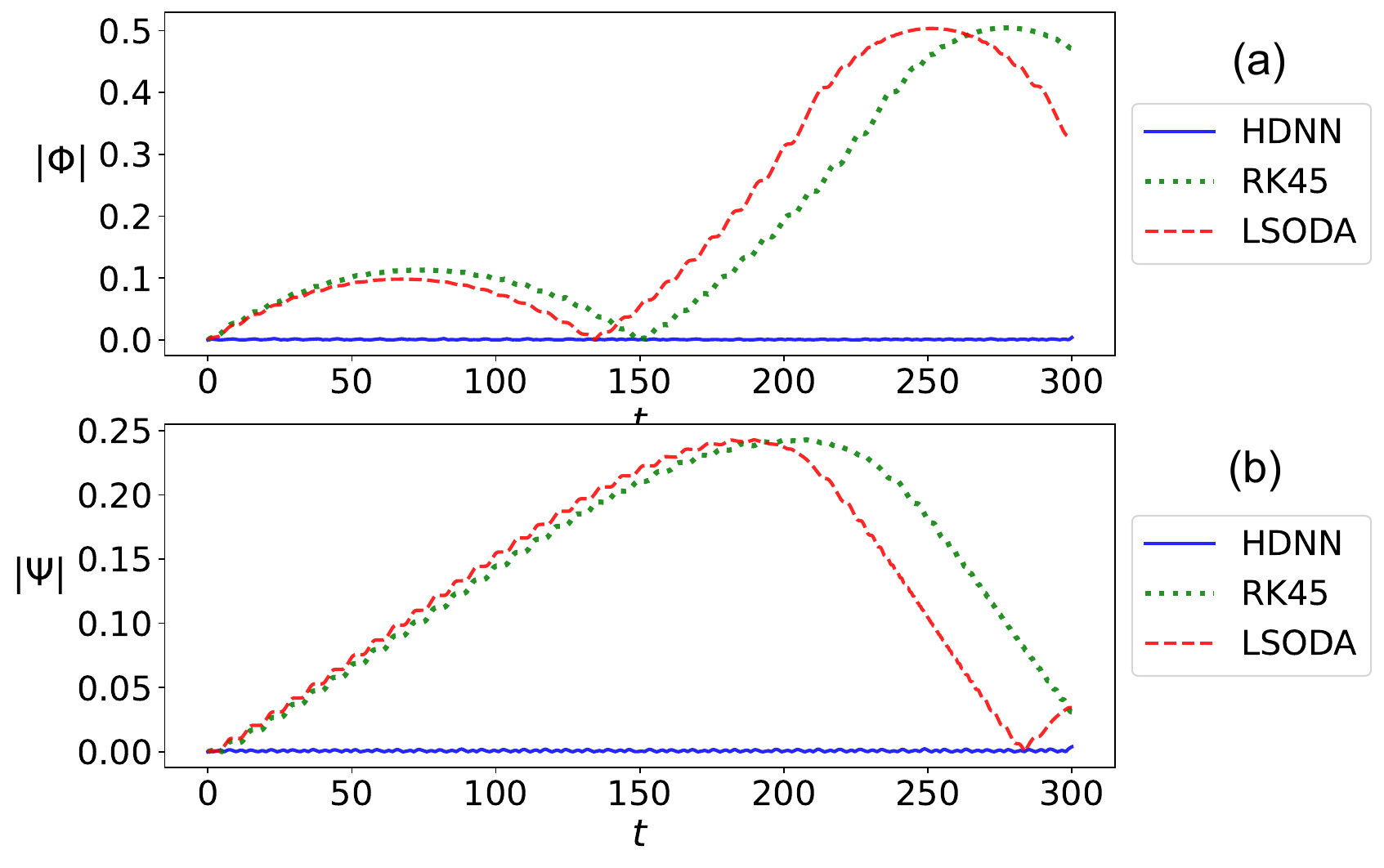}
    \includegraphics[scale=0.25]{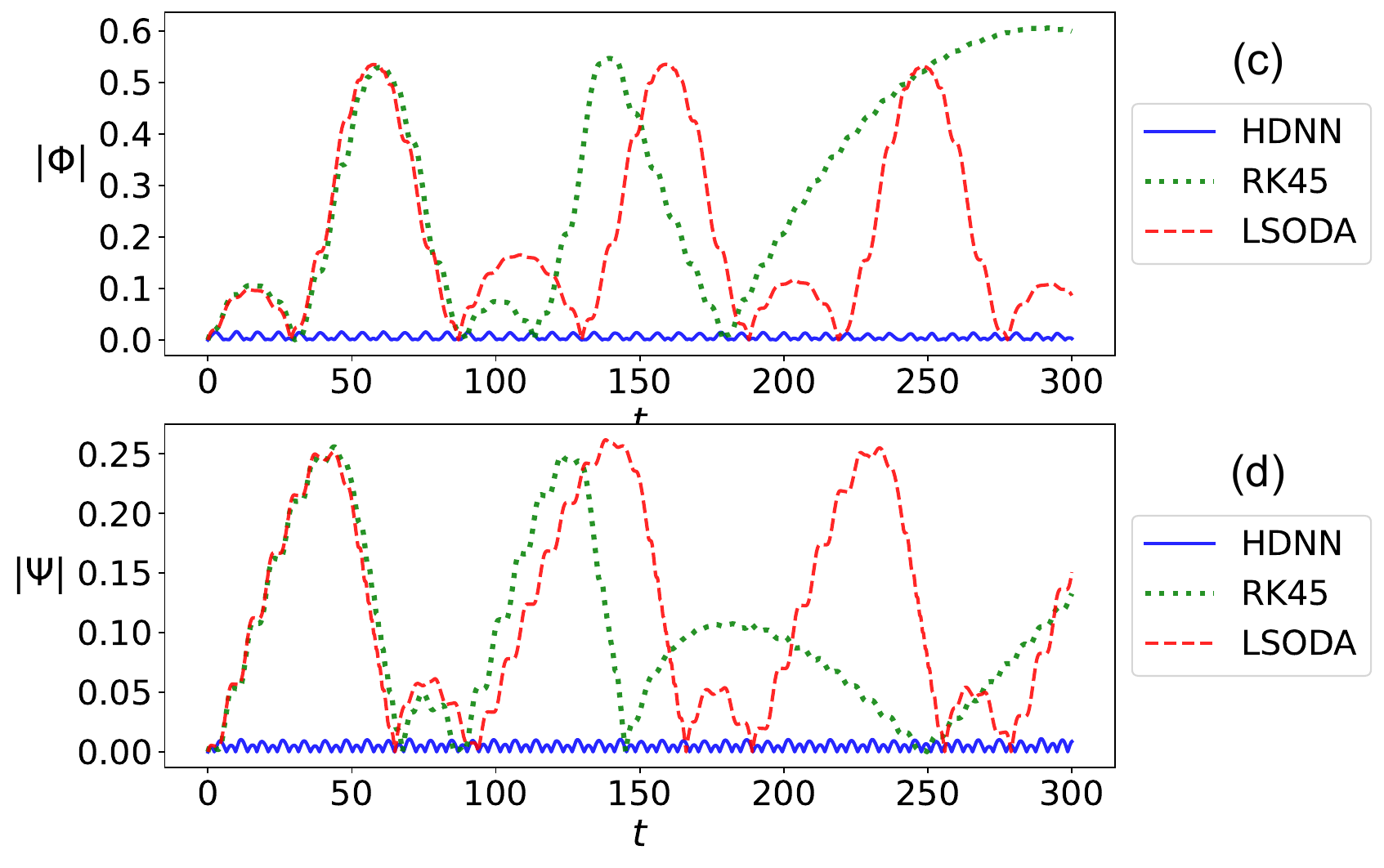}
    \caption[font=small,labelfont=bf]{Evolution of the constraint functions $|\Phi|$ and $|\Psi|$ for HDNN, LSODA, and RK45 solutions. Panels (a) and (b) correspond to $\epsilon=0.2$, while panels (c) and (d) to $\epsilon=0.4$. The HDNN predictions satisfy the Dirac constraints with high precision, in stark contrast to the standard numerical methods.}
    \label{fig_erho_constrs}
\end{figure}

\begin{figure}[ht!]
    \centering
    \includegraphics[scale=0.45]{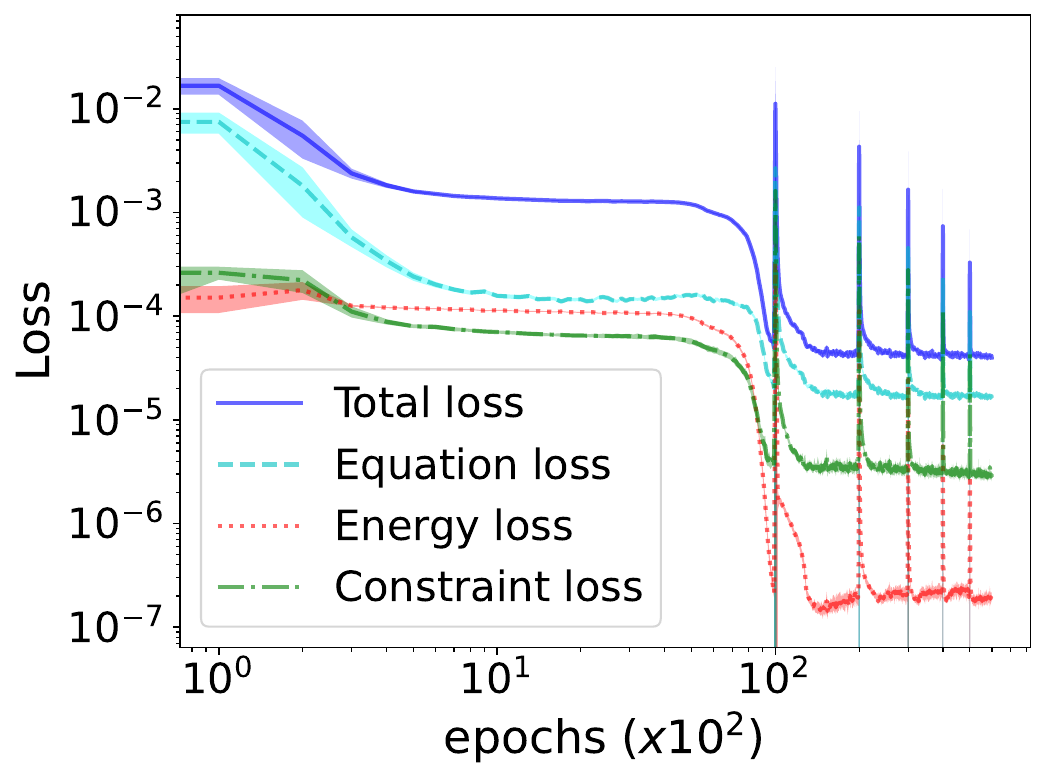}
    \caption[font=small,labelfont=bf]{Training history of the HDNN for the elliptically restricted harmonic oscillator, illustrating the effective minimization of the loss function. The spikes in the various loss function terms after the initial $10^4$ epochs are due to the extensions of the time domain.}
    \label{fig_erho_losses}
\end{figure}

In Fig.~\ref{fig_erho_timeseries}, we compare the HDNN solution with the numerical solutions obtained using LSODA and RK45 methods for the median eccentricity $\epsilon=0.2$ and the highest seen eccentricity value $\epsilon=0.4$. In both cases, the LSODA and RK45 solutions exhibit unphysical behavior, while the HDNN predictions remain stable, conserving energy and preserving Dirac constraints. This stability is evident in the trajectories of the oscillator on the $x-y$ and the $p_x-p_y$ sections of the phase space depicted in Fig. \ref{fig_erho_trajectories}, the energy drift time series shown in Fig. \ref{fig_erho_energy}, and the constraint function evolution in Fig. \ref{fig_erho_constrs}. Although training the HDNNs is quite slow, this example demonstrates their superiority for problems like constrained motion on non-trivial curves over standard solvers which fail to constrain the dynamics on the constraint manifold. {Additionally, the ability to accurately predict system trajectories across different eccentricities at inference time after the initial training is an additional advantage of the HDNN method and highlights its applicability in studying classical restricted  problems with circular and elliptical constraints. 

Finally, the convergence of the HDNN training is illustrated in Fig.~\ref{fig_erho_losses}. The spikes in the various loss function terms after the initial $10^4$ epochs are due to the extensions of the time domain. However, these spikes quickly subside, and the values of the loss function terms stabilize after a relatively small number of additional epochs.}

\subsection{Singular Lagrangians: guiding center motion}
\label{subsec_4.3}
In this subsection we study the motion of a guiding center in a static magnetic field $\boldsymbol{B}(\boldsymbol{x})$ of constant direction. Following the normalization of \cite{Qin2008}, the Lagrangian for the guiding center motion \cite{Littlejohn1983} is given by 
\begin{eqnarray}
    L(\bsx,\dot{\bsx},u_\parallel,\dot{u}_\parallel) = [\bsA(\bsx)+u_\parallel \bsb(\bsx)]\cdot\dot{\bsx} - W(\bsx,u_\parallel)\,, \label{littlejohn_lagrangian}
\end{eqnarray}
where the potential function $W(\bsx,u_\parallel)$ is
\begin{eqnarray}
    W(\bsx,u_{\parallel}) = \mu B(\bsx) +\frac{u_{\parallel}^2}{2} +\phi(\bsx)\,.
\end{eqnarray}
Here, $B(\bsx)=|\bsB(\bsx)|$, $\mu$ is the conserved magnetic moment, $\phi(\bsx)$ is the electrostatic potential, and $\bsA(\bsx)$ is the vector potential, $\bsb(\bsx)$ is the unit vector along the magnetic field direction and $u_\parallel$ the parallel component of the guding center velocity. This is a typical example of non-standard, singular Lagrangian where the Dirac method of constraints applies and a Dirac bracket can be derived. Without referring to the Dirac method for the construction of a generalized Hamiltonian system, the guiding-center equations of motion can be derived by invoking the Euler-Lagrange equations yielding
\begin{eqnarray}
    \dot{\bsx} &=& \frac{u_\parallel \bsB^* -\bsb\times \bsE^*}{\bsb\cdot\bsB^*} \,, \label{gc_gen_eom_1}\\
    \dot{u}_\parallel &=& \frac{\bsB^*\cdot\bsE^*}{\bsb\cdot \bsB^*}\,, \label{gc_gen_eom_2}
\end{eqnarray}
where
\begin{eqnarray}
    \bsA^* &=& \bsA + u_\parallel\bsb\,, \\
    \bsE^* &=& \bsE -\mu \nabla B\,, \\
    \bsB^* &=& \nabla \times \bsA^*\,, 
\end{eqnarray}
Several papers have addressed the topic of structure-preserving discretization of these equations so that the conservation properties of the system are retained on the discrete level. However, since the guiding-center Lagrangian is a non-standard Lagrangian, conventional symplectic integration approaches cannot be directly applied. As a result, several authors have turned to the method of Variational Integrators \cite{Marsden2001} to construct symplectic algorithms capable of preserving conservation properties, making them suitable for long-time simulations of guiding center motion (e.g., \cite{Qin2008,Ellison2018,Morrison2017}). On the other hand though, the guiding center motion is an ideal exemplary for applying the Dirac theory of constraints as demonstrated in \cite{Pfirsch1991} and \cite{Zhang2024}. {For a general presentation of the application of the Dirac algorithm in energetically consistent guiding center theories, readers are referred to Appendix \ref{appendix}.} Here, we use the Dirac theory to construct the Dirac bracket and simulate the dynamics of a guiding center with PINNs for the following toy model for the magnetic field:
\begin{eqnarray}
    \bsB = \left[1+\epsilon\left(\frac{x^2}{\kappa}+y^2\right) \right]\hat{z}\,, \label{magn_field}
\end{eqnarray}
assuming that $\phi(\bsx)=0$ as in \cite{Qin2008}. 
In this case the equations of motion \eqref{gc_gen_eom_1}--\eqref{gc_gen_eom_2} become:
\begin{eqnarray}
    \dot{x} = -\frac{\mu}{B(x,y)}\frac{\partial B}{\partial y}\,,\label{gc_eom_1}\\
    \dot{y} = \frac{\mu}{B(x,y)}\frac{\partial B}{\partial x}\,, \\
    \dot{u}_\parallel =0\,, \label{gc_eom_3}
\end{eqnarray}
and  the guiding center motion has an exact closed elliptic orbit.

We can directly integrate these equations with standard numerical methods such as the symplectic Euler method or the 4th order Runge-Kutta (RK4) to achieve higher accuracy. However, for long time simulations the discretization error of these methods piles up, resulting in a significant drift of the simulated trajectory away from the actual guiding center trajectory. To remedy this problem various variational symplectic integrator approaches have been introduced \cite{Marsden2001,Qin2008,Ellison2018} which preserve a noncanonical symplectic structure. These, integrators though, are implicit and nonlinear solvers have to be employed in order to solve the resulting system of equations after discretization. In contrast, symplectic algorithms for canonical Hamiltonian systems can be designed to be explicit and thus significantly less consuming in terms of computational resources and time. Here, we propose an alternative method that utilizes deep neural networks to resolve the guiding center dynamics. This can be achieved directly without employing the Dirac method of constraints to construct the total Hamiltonian, by solving  \eqref{gc_eom_1}--\eqref{gc_eom_3} and imposing only energy conservation as a regularization term.  However, we choose to utilize the Dirac method to employ the same HDNN architecture as in the previous examples, incorporating constraints and the total Hamiltonian in the regularization term of the loss function. In both cases, the network successfully learns the guiding center dynamics, outperforming traditional ODE solvers in terms of accuracy, with the same or significantly fewer discretization points in time $t$. The computational time required for network training though, is significantly longer. This trade-off can be improved on the side of the deep learning approach by utilizing more advanced strategies within the context of equation-driven machine learning.

The Dirac method of constraints is relevant in the context of the guiding center motion since the Lagrangian \eqref{littlejohn_lagrangian}  is linear in the velocities $\dot{x}$,  $\dot{z}$ and $\dot{y}$, while $\dot{u}_\parallel$ does not appear at all (see also Eq. \eqref{gc_lagrangian} in Appendix \ref{appendix}). Therefore, the equations for the corresponding conjugate momenta are:
\begin{eqnarray}
    p_x &=& \frac{\partial L}{\partial \dot{x}} = A_x\,,\\
    p_y &=& \frac{\partial L}{\partial \dot{y}} = A_y\,,\\
    p_z &=& \frac{\partial L}{\partial \dot{z}} = A_z\,,\\
    p_{u_\parallel} &=& \frac{\partial L}{\partial \dot{u}_\parallel} =0\,,
\end{eqnarray}
which are essentially four phase-space constraints as they cannot be solved for the velocities. The constraint functions are $\Phi_1 = p_x - A_x =0$, $\Phi_2 = p_y -A_y =0$, $\Phi_3 = p_z - A_z=0$ and $\Phi_4 = p_{u_\parallel}=0$, while the canonical and the total Hamiltonian are given, respectively, by:
\begin{eqnarray}
    H_c &=& \dot{x} \frac{\partial L}{\partial \dot{x}}+\dot{y} \frac{\partial L}{\partial \dot{y}}+\dot{z} \frac{\partial L}{\partial \dot{z}} - L = W(x,y,z,u_\parallel)\,, \\ 
    H_t &=& W(x,y,z,u_\parallel) + \sum_{\alpha=1}^4\lambda_\alpha \Phi_\alpha\,,
\end{eqnarray}
where $\lambda_\alpha$, $\alpha=1,2,3,4$ are Lagrangian multipliers. Considering the unidirectional field \eqref{magn_field} and employing the consistency conditions:
\begin{eqnarray}
    \{\Phi_\alpha,H_t\}\approx 0 \,,\quad \alpha =1,...,4\,,
\end{eqnarray}
we can determine the multipliers $\lambda_i$:
\begin{eqnarray}
    \lambda_1=  - \frac{1}{B(x,y)}\frac{\partial W}{\partial y}\,,\\
    \lambda_2= \frac{1}{B(x,y)}\frac{\partial W}{\partial x}\,, \\
    \lambda_3 = u_\parallel\,,\quad \lambda_4 =0\,,
\end{eqnarray}
where $W=\mu B$. Thus, the total Hamiltonian reads
\begin{eqnarray}
    H_t = W(x,y) - \frac{(p_x-A_x)}{B(x,y)} \frac{\partial W}{\partial y} + \frac{(p_y-A_y)}{B(x,y)} \frac{\partial W}{\partial x} + u_\parallel (p_z-A_z)\,. \label{Ht_gc}
\end{eqnarray}

Note that the motion on the $x-y$ plane decouples from the uniform parallel dynamics and thus the non-trivial dynamical equations are
\begin{eqnarray}
    \dot{x} &=& \frac{\partial H_c}{\partial p_x} + \lambda_1 \frac{\partial \Phi_1}{\partial p_x} = - \frac{\mu}{B} \frac{\partial B}{\partial y}\,, \label{dx_gc}\\
    \dot{y} &=& \frac{\partial H_c}{\partial p_y} + \lambda_2 \frac{\partial \Phi_2}{\partial p_y}  =  \frac{\mu}{B} \frac{\partial B}{\partial x}\,,\\
    \dot{p}_x &=& -\frac{\partial H_c}{\partial x} - \lambda_1 \frac{\partial \Phi_1}{\partial x} - \lambda_2 \frac{\partial \Phi_2}{\partial x} = -\mu\left(1- \frac{1}{B}\frac{\partial A_y}{\partial x}\right) \frac{\partial B}{\partial x} - \frac{\mu}{B} \frac{\partial A_x}{\partial x}\frac{\partial B}{\partial y} \,,\\
    \dot{p}_y &=& -\frac{\partial H_c}{\partial y} - \lambda_1 \frac{\partial\Phi_1}{\partial y} - \lambda_2 \frac{\partial \Phi_2}{\partial y}  = - \mu\left(1+\frac{1}{B}\frac{\partial A_x}{\partial y}\right)\frac{\partial B}{\partial y} + \frac{\mu}{B}\frac{\partial A_y}{\partial y}\frac{\partial B}{\partial x}\,.\label{dpy_gc}
\end{eqnarray}
By employing the Dirac approach we essentially embed the dynamical system into a higher-dimensional space, the phase-space. The additional dimensions, which correspond to the conjugate momenta $p_x$ and $p_y$, are redundant {in the sense that the dynamics of the canonical momenta do not influence the equations for the coordinates $x$ and $y$, which can be integrated directly to obtain the guiding center trajectory. Coupling of the canonical momenta dynamics with the dynamical equations for the guiding center coordinates can be achieved by imposing the constraints $\Phi_1$ and $\Phi_2$ through the regularization term \eqref{constr_loss}, thus enforcing the trajectories to respect the Hamiltonian dynamics so that phase-space volumes are preserved by the Hamiltonian flow. Alternatively, a direct coupling would be possible if we had considered constraints $g_1$ and $g_2$ that are functions of $\Phi_1=p_x-A_x$ and $\Phi_2=p_y-A_y$. In that case, the canonical momenta $p_x$ and $p_y$ would have appeared in the dynamical equations for $x$ and $y$ and vanish only in the weak sense.  A similar approach was adopted in \cite{Zhang2024} where the authors use the Euler-Lagrange equation $\bsb\cdot \dot{\bsx}=u_{\parallel}$ to eliminate $u_{\parallel}$ from the Lagrangian \eqref{gc_lagrangian}. The new Lagrangian is:
\begin{eqnarray}
    L(\bsx,\dot{\bsx}) = \frac{\left(\bsb\cdot\dot{\bsx}\right)^2}{2}  +\bsA(\bsx)\cdot\dot{\bsx} - \left[\mu B(\bsx) + \phi(\bsx)\right] = \frac{ b_i b_j }{2}\dot{x}_i \dot{x}_j +\bsA(\bsx)\cdot\dot{\bsx} - \left[\mu B(\bsx) + \phi(\bsx)\right]\,. \label{gc_lagrangian_2}
\end{eqnarray}
This is also a singular Lagrangian since $det\left(\partial^2L/\partial \dot{x}_i\partial\dot{x}_j\right)= det(b_ib_j)=0$, but the difference here is that the canonical momenta are: 
\begin{eqnarray}
    p_i = b_i \bsb\cdot\dot{\bsx}+A_i(\bsx)\,, \quad i=1,2,3\,.
\end{eqnarray}
This led the authors of \cite{Zhang2024} to introduce the following Dirac constraints:
\begin{eqnarray}
    g_1=b_1(\bsx)\left[p_2-A_2(\bsx)\right] - b_2(\bsx)\left[p_1-A_1(\bsx)\right]=0\,,\\
    g_2=b_1(\bsx)\left[p_3-A_3(\bsx)\right] - b_3(\bsx)\left[p_1-A_1(\bsx)\right]=0\,.
\end{eqnarray}
Following the procedures described in Section \ref{sec_II} the following general constrained equations of motion are derived in \cite{Zhang2024}:
\begin{eqnarray}
        \dot{\bsx} &=& \bsxi + \frac{\bsZ+\bsb\times\bsT}{\bsxi\cdot(\nabla\times \boldsymbol{b})+B}\,,\\
    \dot{\bsp} &=& \nabla \bsA\cdot \left(\bsxi+ \frac{\bsZ+\bsb\times\bsT}{\bsxi\cdot(\nabla\times \bsb)+B}\right)+\nabla \bsb\cdot \frac{(\xi_1/b_1) \bsZ+\bsxi\times\bsT}{\bsxi\cdot(\nabla\times\bsb)+B}-\bsT\,,
\end{eqnarray}
where
\begin{eqnarray}
    \bsxi:=\bsp-\bsA\,,\nn \\ 
    \bsZ= \bsxi\times\left[(\bsxi\cdot\nabla) \bsb\right]\,, \nn\\
    \bsT=\nabla\phi+\mu\nabla B\,.
\end{eqnarray}
Applying these equations for our specific example, with $\phi=0$ and $\bsB$ given by \eqref{magn_field}, we find the following equations of motion:
\begin{eqnarray}
    \dot{x} &=& (p_x-A_x)-\frac{\mu}{B} \frac{\partial B}{\partial y}\,, \label{dx_gc_2}\\
    \dot{y} &=& (p_y-A_y) + \frac{\mu}{B}\frac{\partial B}{\partial x}\,,\\
     \dot{p}_x &=& \frac{\partial A_x}{\partial x} \left[(p_x-A_x)-\frac{\mu}{B} \frac{\partial B}{\partial y}\right] +\frac{\partial A_y}{\partial x} \left[(p_y-A_y) + \frac{\mu}{B}\frac{\partial B}{\partial x}\right] - \mu \frac{\partial B}{\partial x}\,, \\
     \dot{p}_y &=& \frac{\partial A_x}{\partial y} \left[(p_x-A_x)-\frac{\mu}{B} \frac{\partial B}{\partial y}\right] +\frac{\partial A_y}{\partial y} \left[(p_y-A_y) + \frac{\mu}{B}\frac{\partial B}{\partial x}\right] - \mu \frac{\partial B}{\partial y} \,. \label{dpy_gc_2}
\end{eqnarray}
Note that equations \eqref{dx_gc_2}--\eqref{dpy_gc_2} reduce to equations \eqref{dx_gc}--\eqref{dpy_gc} on the constraint manifold. Both approaches allow for a Hamiltonian description of the dynamics and the use of symplectic HDNNs. 
}

\subsubsection{Unsupervised HDNN learning} 
\label{unsupervised}

\begin{figure}[ht!]
    \centering
    \includegraphics[scale=0.3]{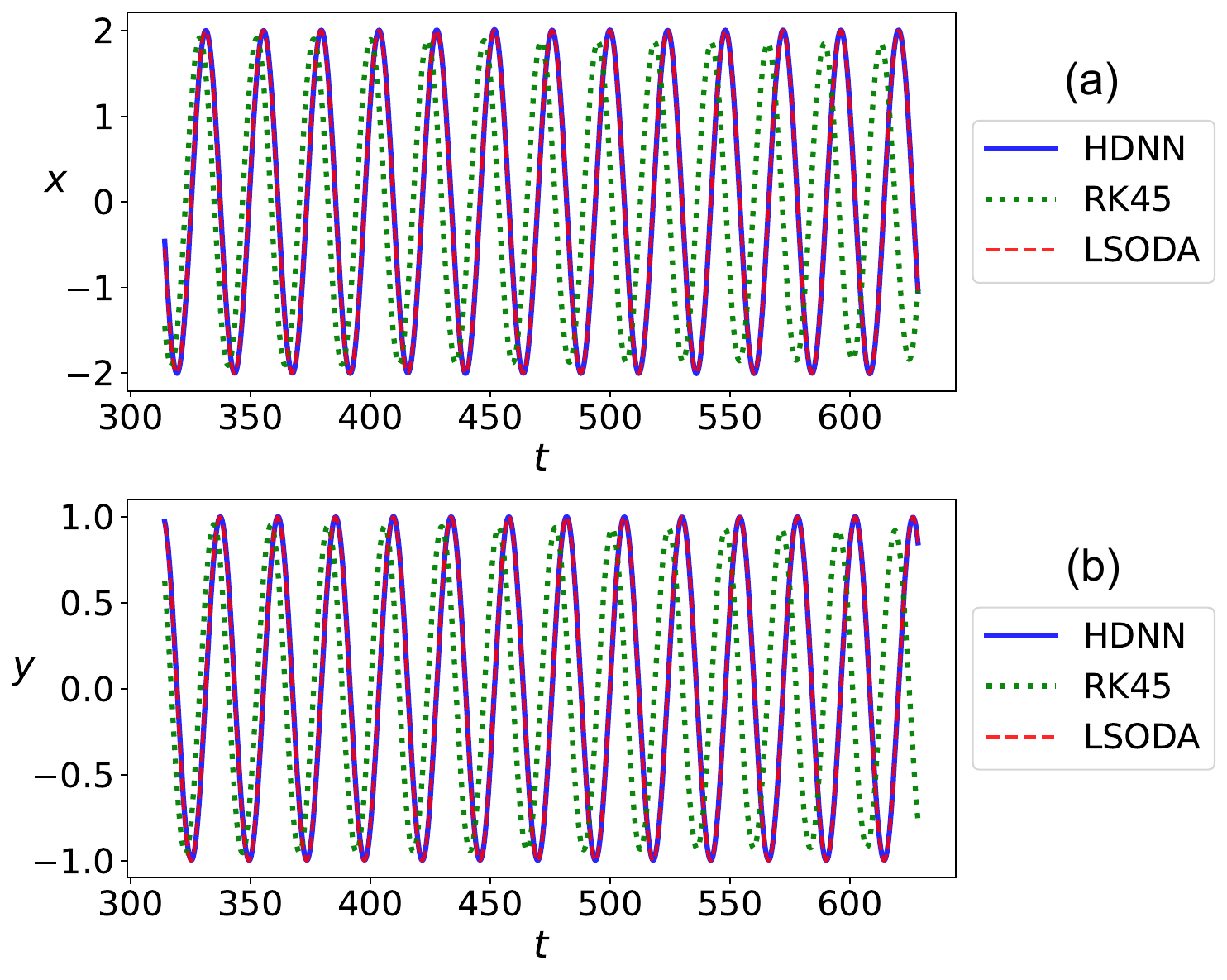}
    \includegraphics[scale=0.3]{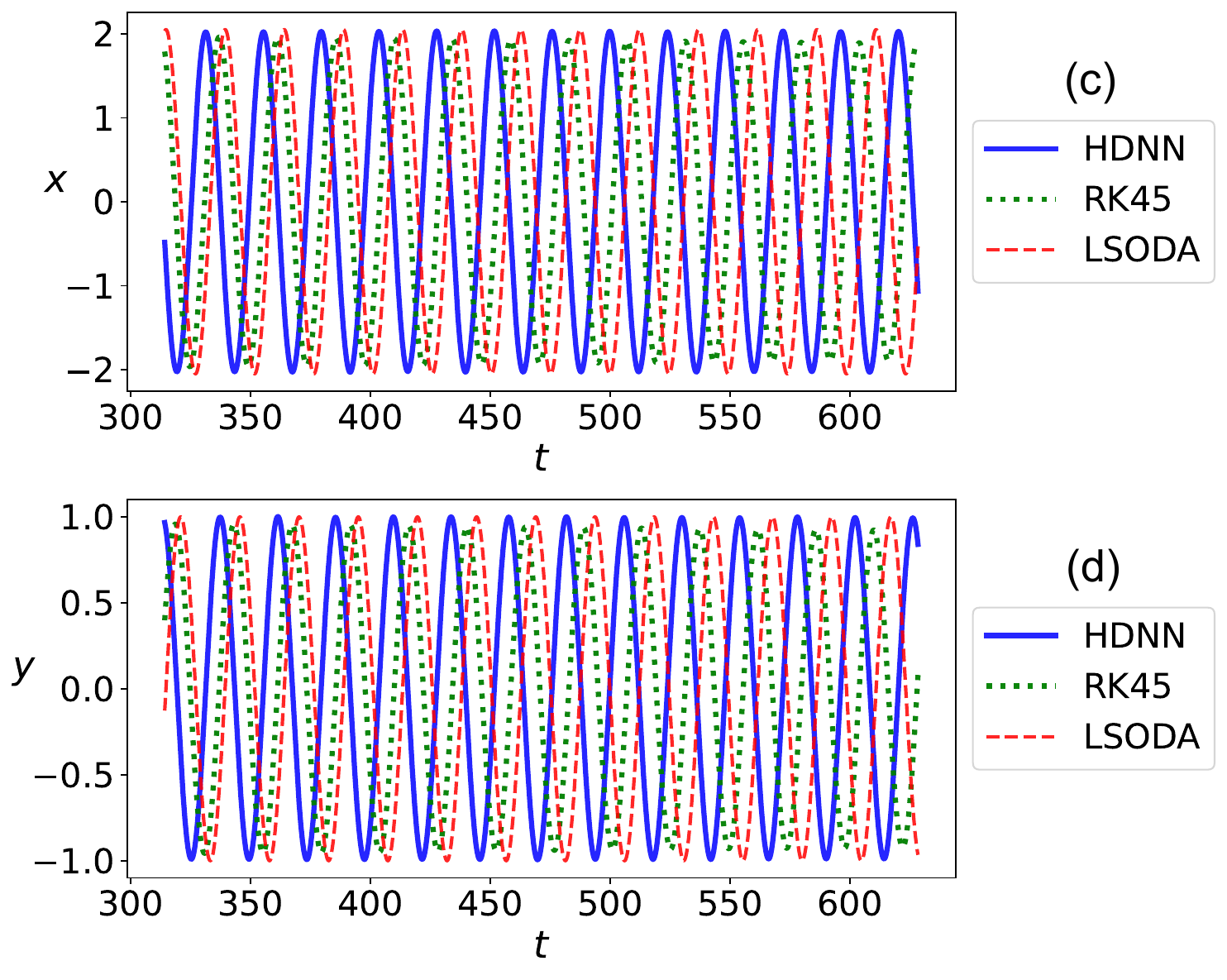}
    \caption[font=small,labelfont=bf]{Panels (a) and (b) present the time series for the $x$ and $y$ coordinates of the guiding center motion, respectively, in the magnetic field \eqref{magn_field}, over the time window $t\in[100\pi,200\pi]$ for $\kappa=4.0$. The HDNN prediction (blue solid line) is compared with the RK45 (green dotted line) and LSODA (red dashed line) solutions. The HDNN is clearly more accurate than RK45 and preserves the amplitude of the guiding center's elliptical motion. Panels (c) and (d) show the corresponding time series for $\kappa=4.2$. While both the HDNN and RK45 solutions exhibit phase drift, the HDNN prediction maintains the amplitude of the motion.} 
    \label{fig_gc_timeseries}
\end{figure}

\begin{figure}[ht!]
    \centering
    \includegraphics[scale=0.3]{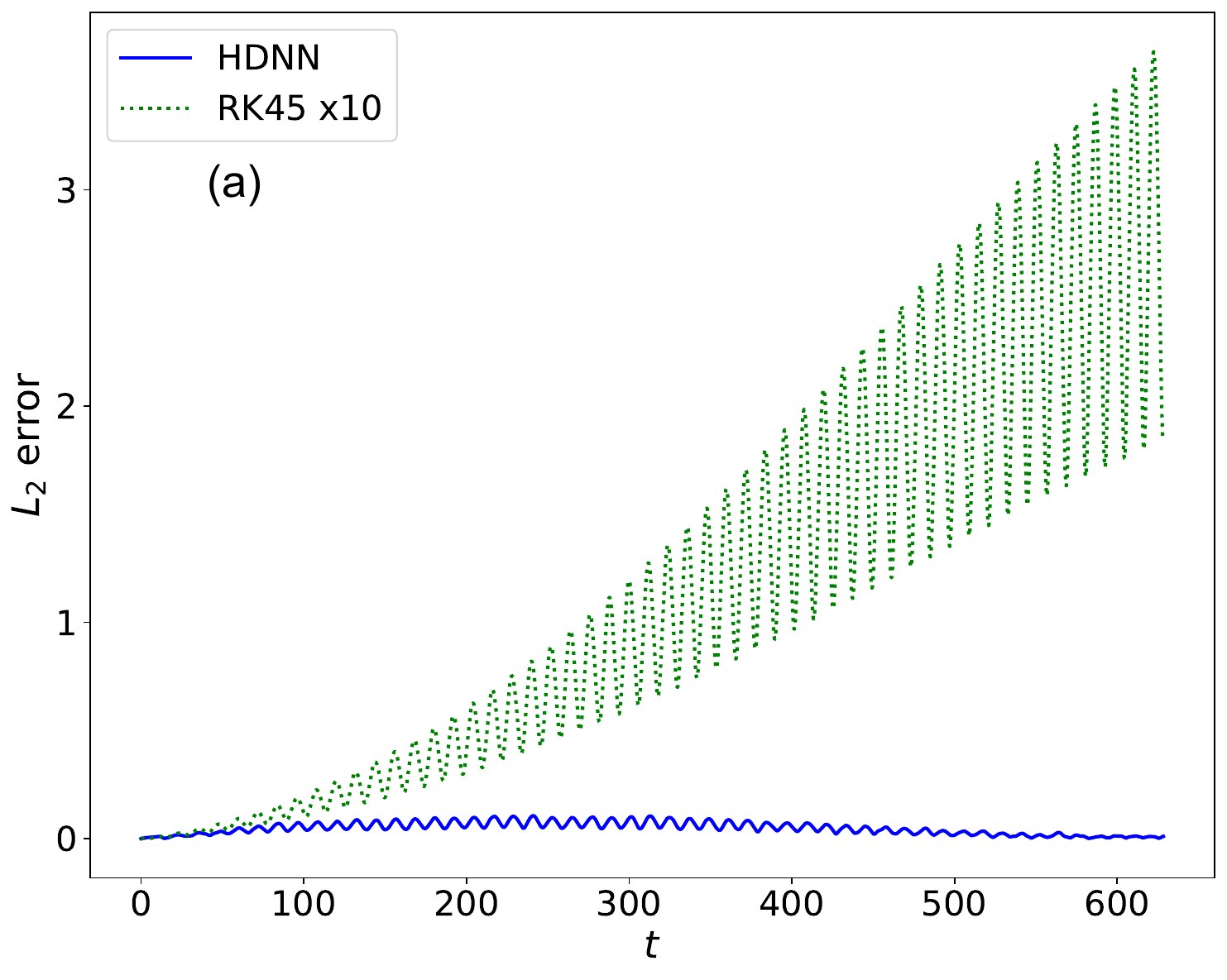}
        \includegraphics[scale=0.3]{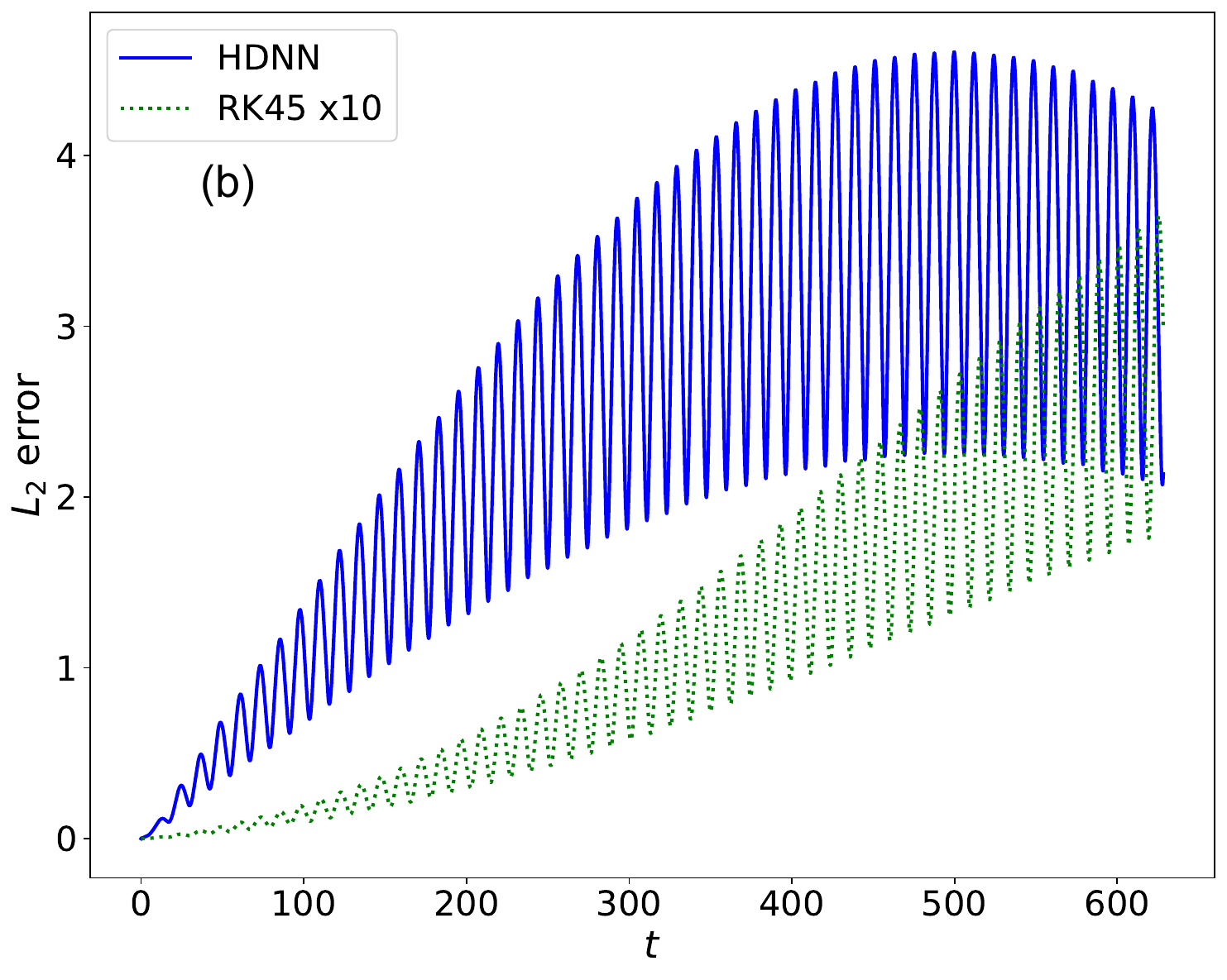}
    \caption[font=small,labelfont=bf]{In panel (a) the evolution of the $L_2$ error of the HDNN prediction and the RK45 solution for $\kappa=4.0$ are displayed. Clearly, the HDNN prediction is much more accurate than the RK45 solution. In panel (b) we present the corresponding plots for $\kappa=4.2$. The $L_2$ error of the HDNN prediction is larger than the RK45 due to the larger phase drift. }
    \label{fig_gc_L2_error}
\end{figure}

\begin{figure}[ht!]
    \centering
    \includegraphics[scale=0.26]{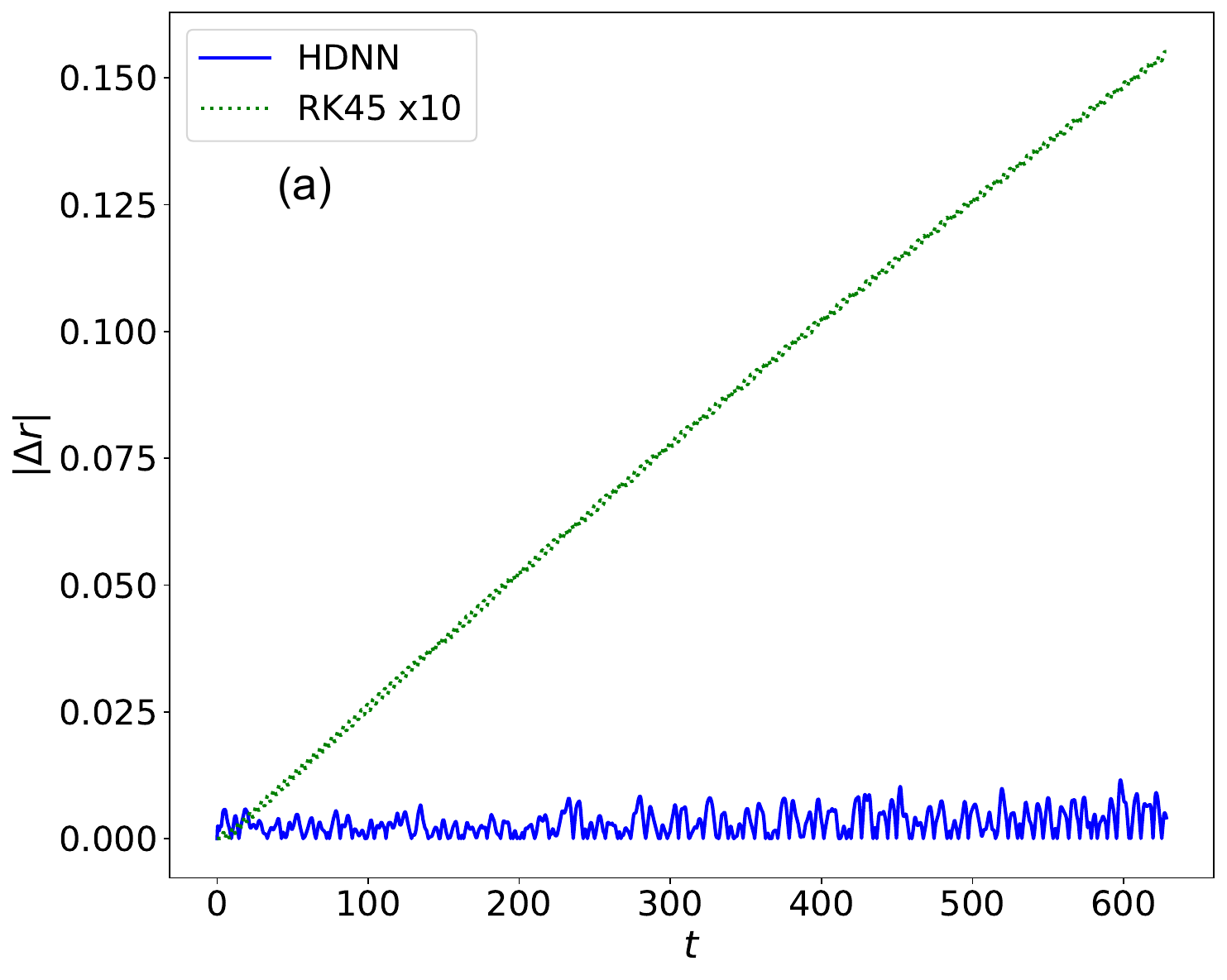}
    \includegraphics[scale=0.26]{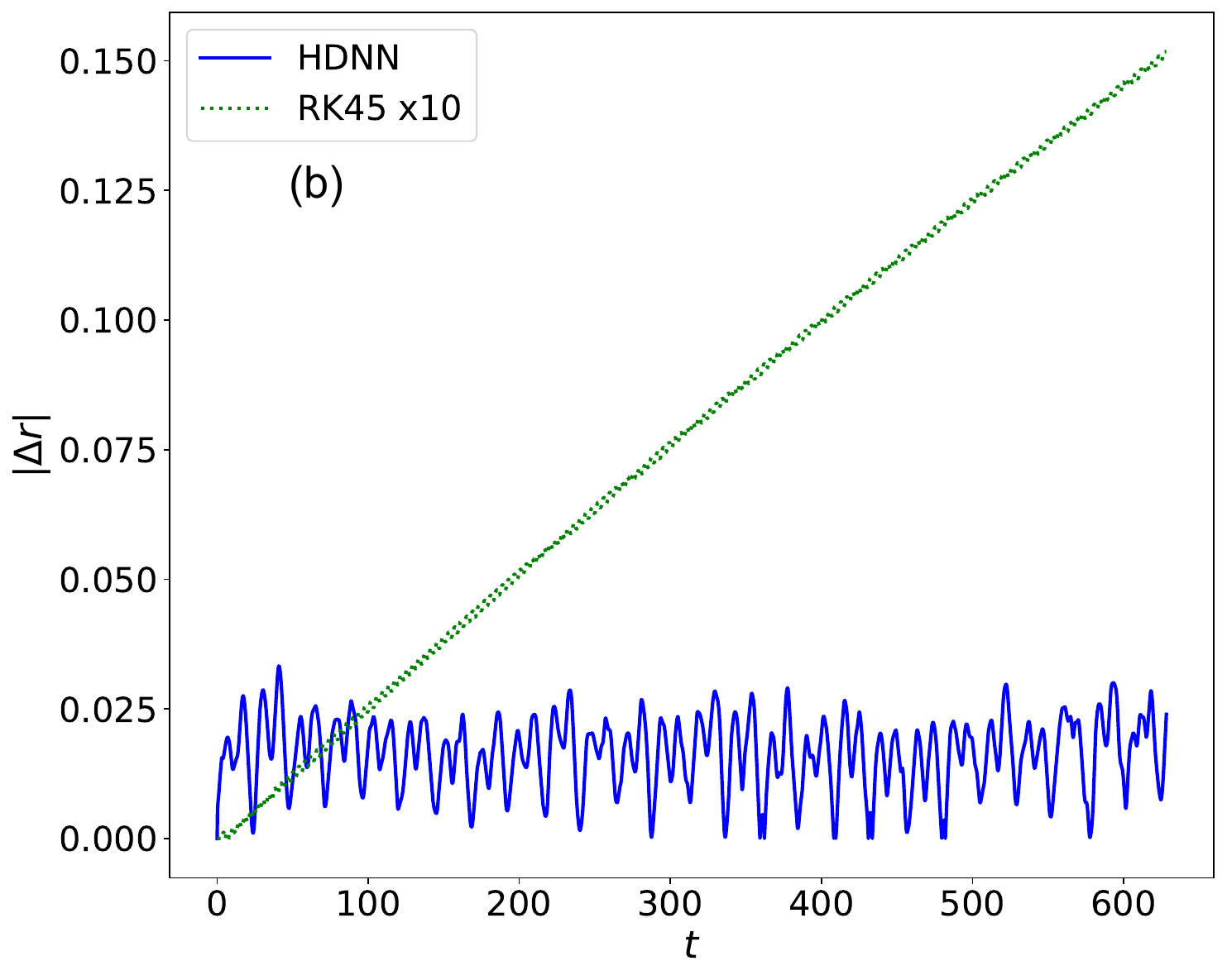}
    \caption[font=small,labelfont=bf]{Panel (a) illustrates the time evolution of the deviation of the HDNN and RK45 trajectories from the true elliptical path of the guiding center for $\kappa=4.0$. The HDNN trajectory remains close to the true path, while the RK45 solution drifts away. Panel (b) presents the corresponding deviations for $\kappa=4.2$, where the HDNN again predicts a more accurate path than the RK45.}
    \label{fig_gc_elliptic_viol}
\end{figure}

\begin{figure}[ht!]
    \centering
    \includegraphics[scale=0.3]{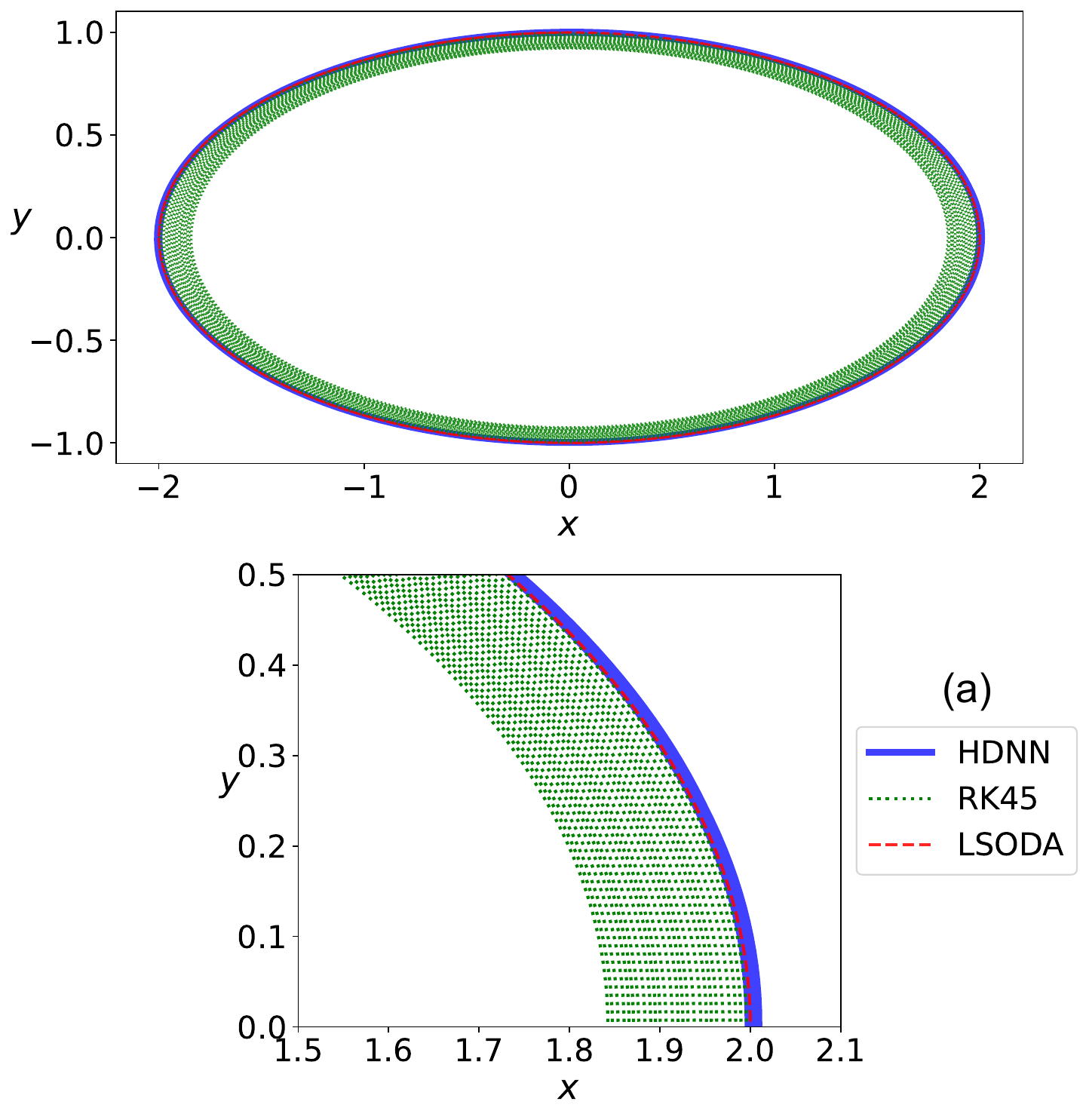}
        \includegraphics[scale=0.3]{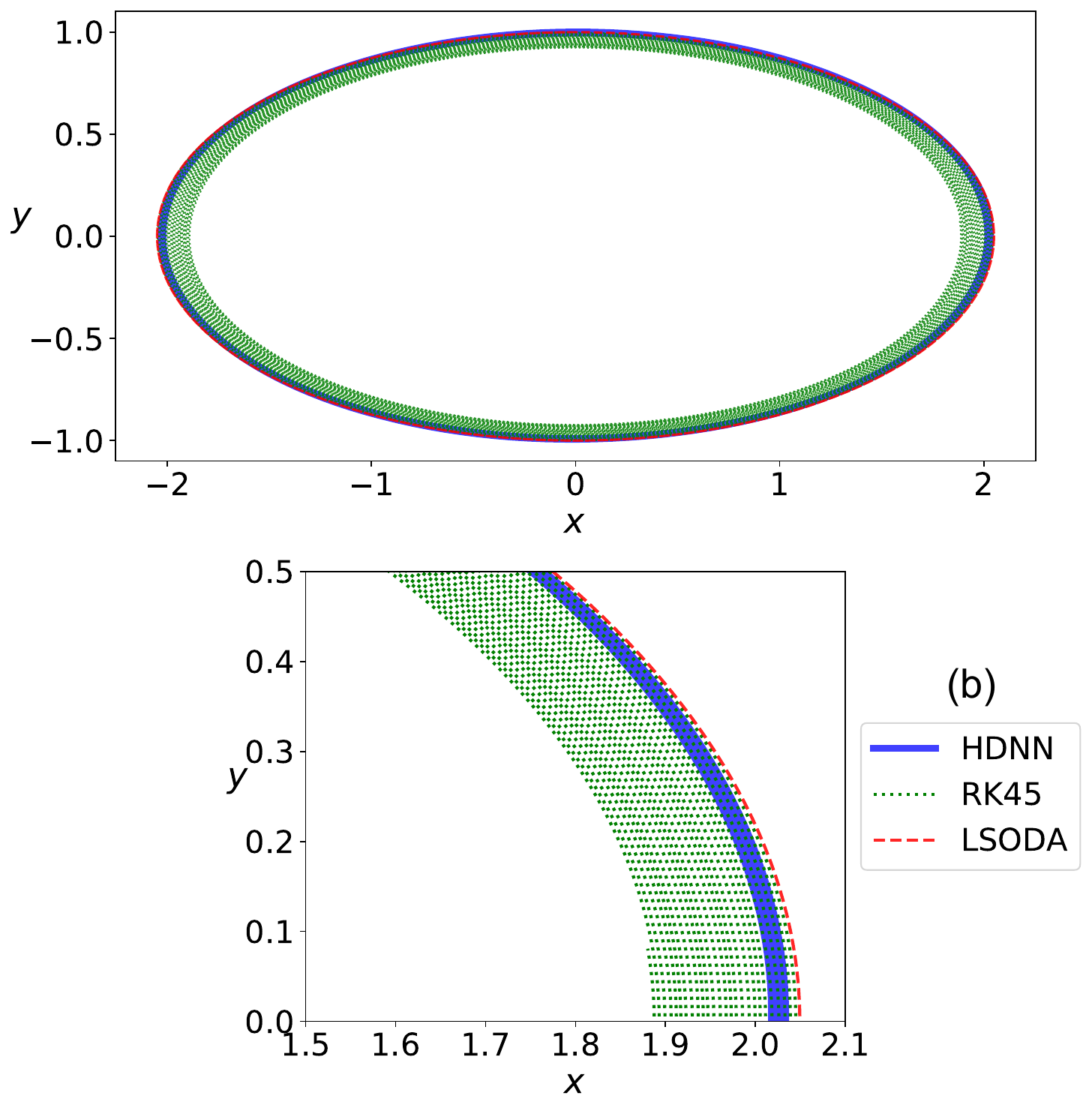}
    \caption[font=small,labelfont=bf]{The trajectory of the guiding center in the magnetic field \eqref{magn_field}, as predicted by the trained HDNN (solid blue line), compared with the RK45 solution (green dotted line) and the LSODA solution (red dashed line) for $\kappa=4.0$ (panel (a)) and $\kappa=4.2$ (panel (b)). In panel (a), the HDNN solution lies exactly on the true elliptical path, while in panel (b), it exhibits a small deviation. In contrast, the RK45 solution drifts toward the center of the elliptical orbit in both cases due to spurious energy dissipation.}
    \label{fig_gc_trajectories}
\end{figure}

\begin{figure}[ht!]
    \centering
    \includegraphics[scale=0.3]{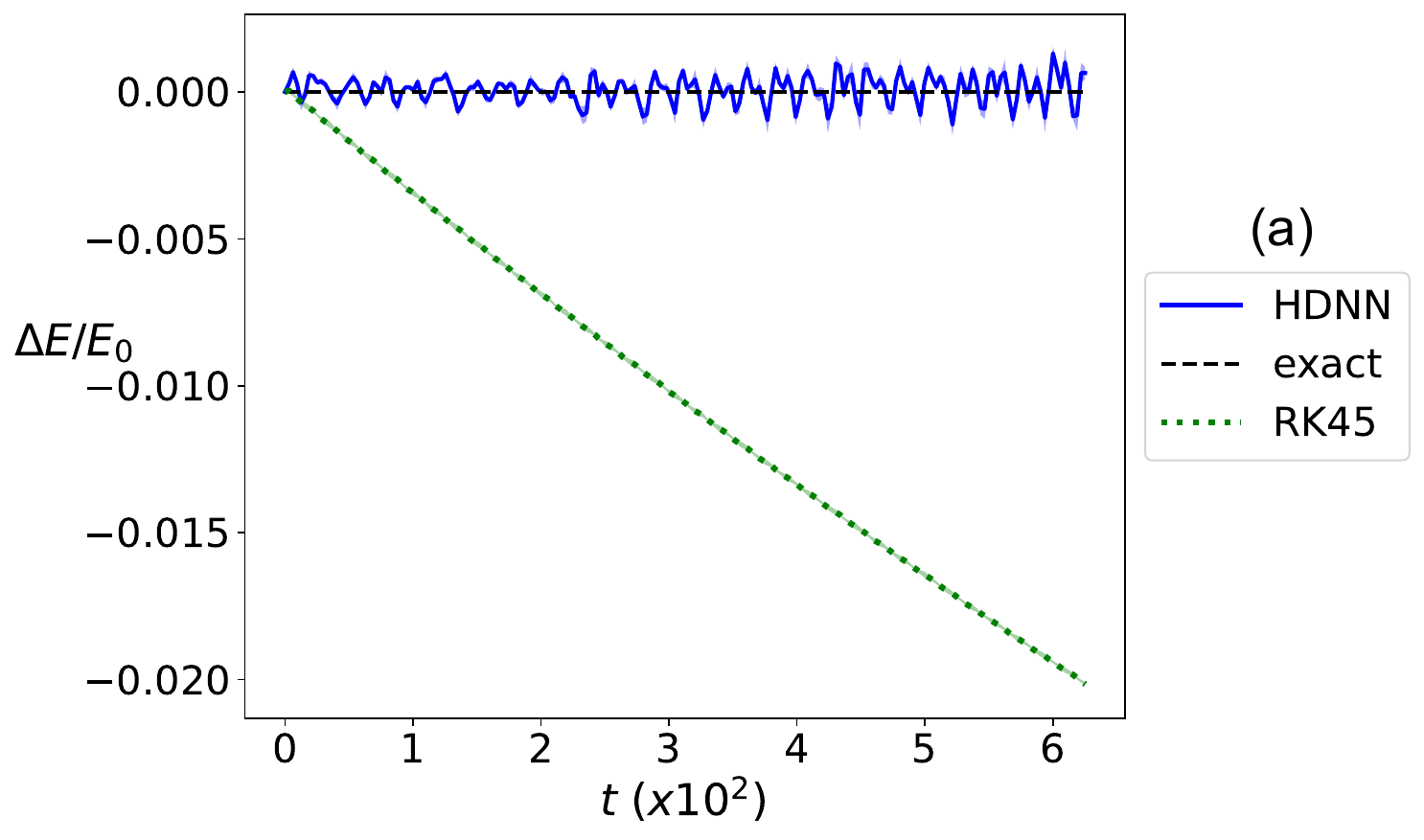}
    \includegraphics[scale=0.3]{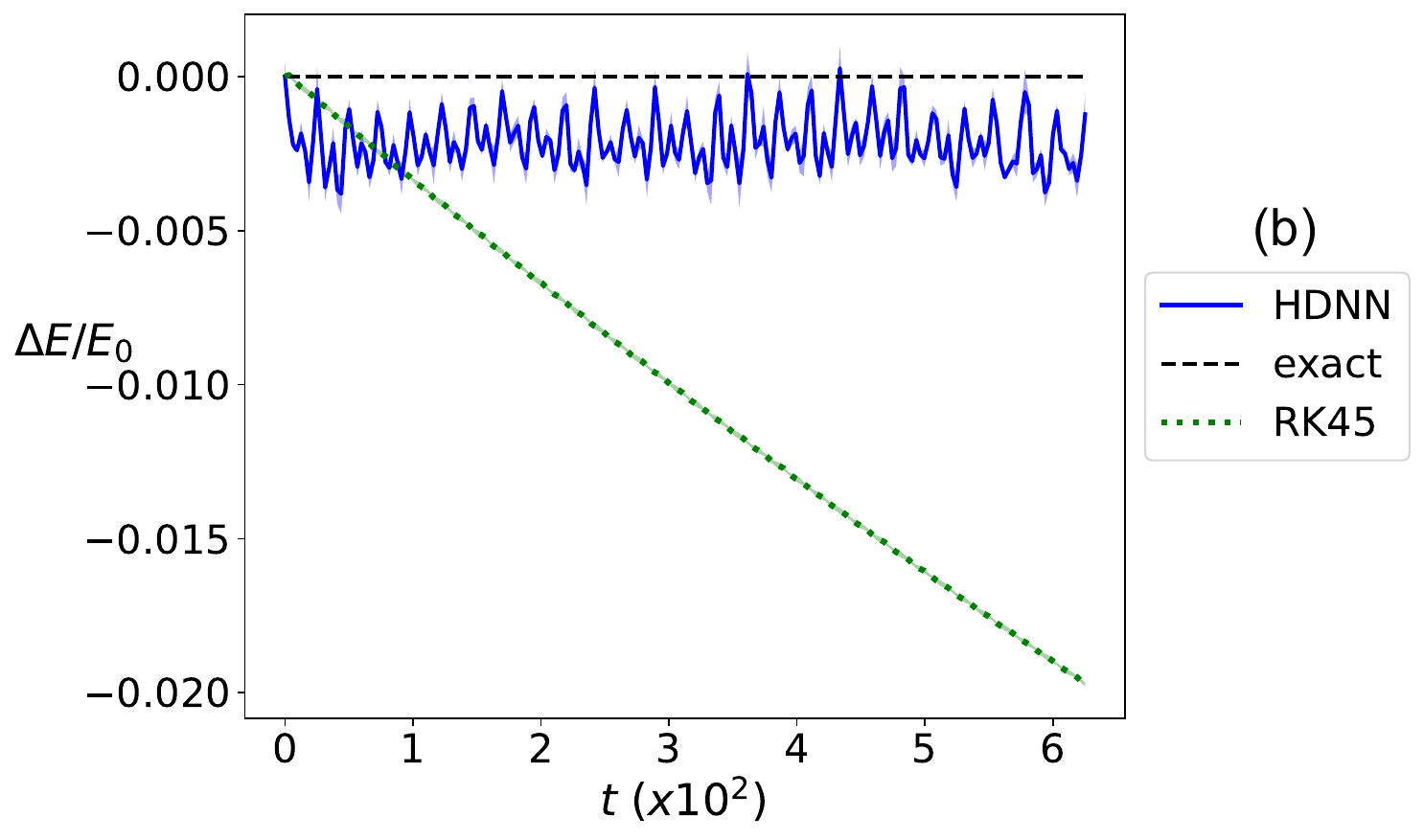}
    \caption[font=small,labelfont=bf]{The time evolution of the mean value of $\Delta E/E_0$ computed in batches of 100 time steps, is shown for (a) $\kappa=4.0$ and (b) $\kappa=4.2$. The HDNN prediction (blue solid line) conserves energy far more accurately over time compared to the RK45 solution (green dotted line), which exhibits unphysical energy dissipation.}
    \label{fig_gc_energy}
\end{figure}

Again, the loss function that is minimized to train the HDNN is given by \eqref{loss_gen} with $\textbf{z}(t) = (x(t),y(t),p_x(t),p_y(t))^T$. The dynamical equations are \eqref{dx_gc_2}--\eqref{dpy_gc_2} and the regularization term $\Lc_{reg}$ contains the total Hamiltonian \eqref{Ht_gc}. For the numerical experiments of this section we used the same HDNN architecture as in the previous sections,{ i.e. 2 input neurons, one for the time variable and one for a problem parameter, 4 hidden layers with 160 neurons per layer and 4 output neurons for the four canonical variables $x,y,p_x,p_y$.  As input parameter we have considered all three characteristic parameters of this problem, i.e. $\epsilon,\kappa$ and $\mu$. Here we present only the case where the network is trained for multiple $\kappa$ values in the range $\kappa\in[3.8,4.2]$. In the other two cases similar results and behavior were obtained. In contrast to the previous examples though, the HDNN seems to struggle to learn the parametric dependencies on $\epsilon,\kappa$ and $\mu$ in the sense that the $L_2$ error of the HDNN prediction becomes larger than the corresponding RK45 error over an extended range of values of the learned parameter.

The network was trained for 10424 epochs, divided in 7 training periods. During the first period of 1100 epochs, the network was trained in the interval $t\in[0,50\pi]$ using 500 training points.  Subsequently, the time interval was extended by $25\pi$ time units in each period while the number of collocation points $n_t$ was increased by 250 per period.  In the final period, lasting 1,948 epochs, the network was trained over the interval $t\in [0,200 \pi]$ with 2000 collocation points. The total training time for the HDNN was $2.8$ minutes, compared to just a fraction of a second required for the RK45 method. 

In Fig.~\ref{fig_gc_timeseries}, we present the time window $[100\pi,200\pi]$ of the $x$ and $y$ time series for the HDNN predictions alongside the LSODA and the RK45 numerical solution for the median value $\kappa=4.0$ and for the maximum seen value $\kappa=4.2$. The network predicts accurately the guiding center trajectory for the median value but exhibits a phase drift for $\kappa=4.2$. This can be seen also in Fig. \ref{fig_gc_L2_error} where the $L_2$ error of the HDNN predictions are compared with the corresponding errors of the RK45 solutions.  For $\kappa=4.0$ the HDNN prediction is clearly more accurate than the RK45 solution, but it has larger error through the entire time domain for $\kappa=4.2$. This is due to the phase drift of the HDNN prediction as seen in Fig. \ref{fig_gc_timeseries}. This behavior was also observed in the other two cases where multiple $\mu$ and  $\epsilon$ values were fed to the network during training. The observed phase drift is likely due to the fact that all three parameters $\epsilon,\kappa$ and $\mu$ affect significantly the period of the guiding center motion. The HDNN appears to face challenges in accurately learning multiple solutions with varying periods.

However, the HDNN effectively captured other characteristics of the guiding center motion across different input parametric values. For instance, all predicted solutions within the range $\kappa\in[3.8,4.2]$ show only minor deviations from the true elliptical path, as illustrated in Fig. \ref{fig_gc_elliptic_viol}. This figure depicts the deviation of the predicted and numerical solutions from the true elliptical trajectory of the guiding center, quantified by the measure $|\Delta r|=|1-(x^2/\kappa + y^2)|$. It is evident that the HDNN predictions remain close to the true path, while the RK45 solution deviates significantly for both $\kappa=4.0$ and $\kappa=4.2$. This can also be corroborated by the plotted trajectories in Fig. \ref{fig_gc_trajectories}. Clearly, the RK45 solution drifts away from the true guiding center path due to energy dissipation as shown in the energy drift diagrams in Fig. \ref{fig_gc_energy}. 

Therefore, while the trained HDNN reproduces accccurately the true trajectory only near the median value of the parameter $\kappa$ within the training range $[\kappa_1,\kappa_2]$, it consistently stays on the true guiding center path and conserves energy more accurately than the RK45 method across the entire parametric range.}

\begin{figure}[ht!]
    \centering
    \includegraphics[scale=0.26]{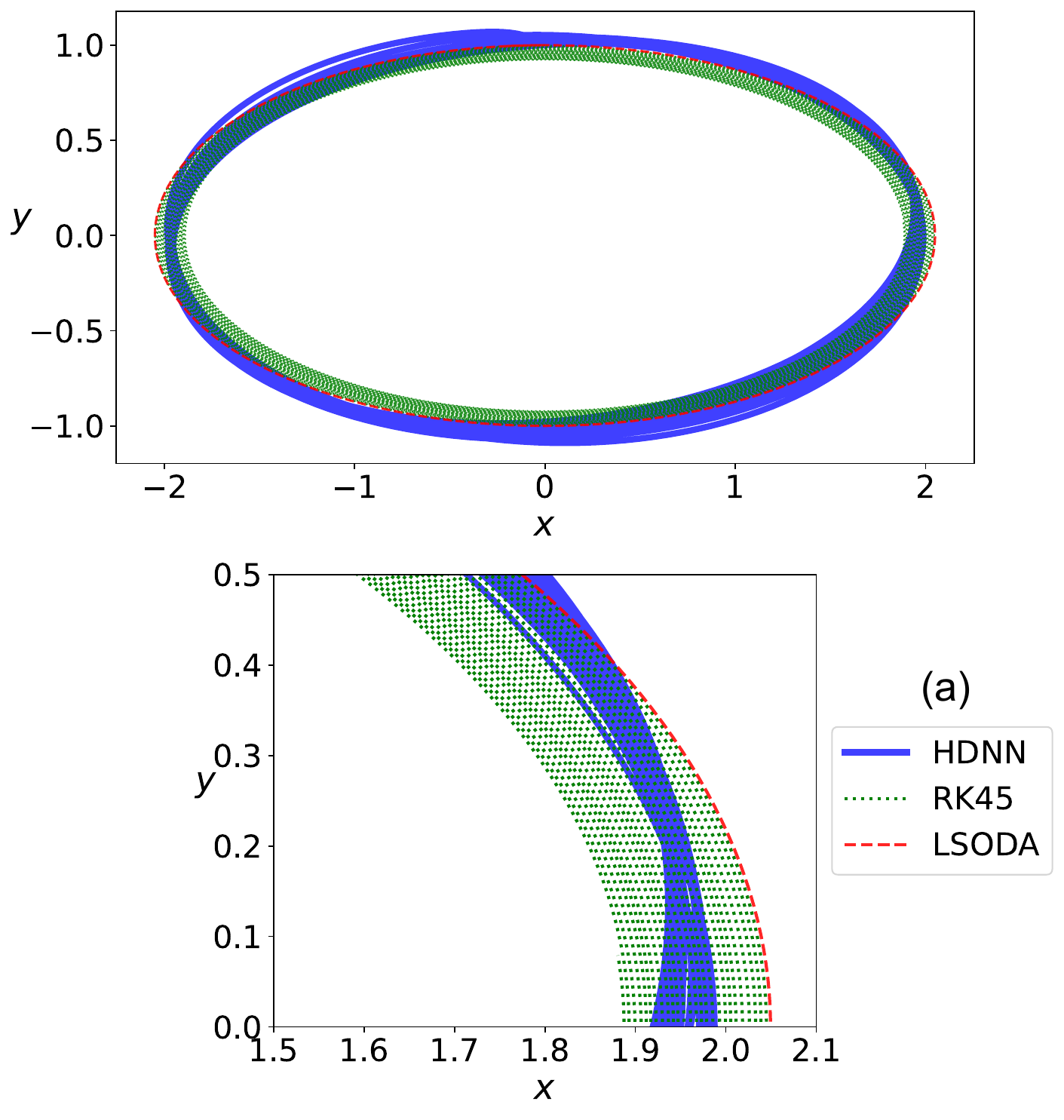}
    \includegraphics[scale=0.26]{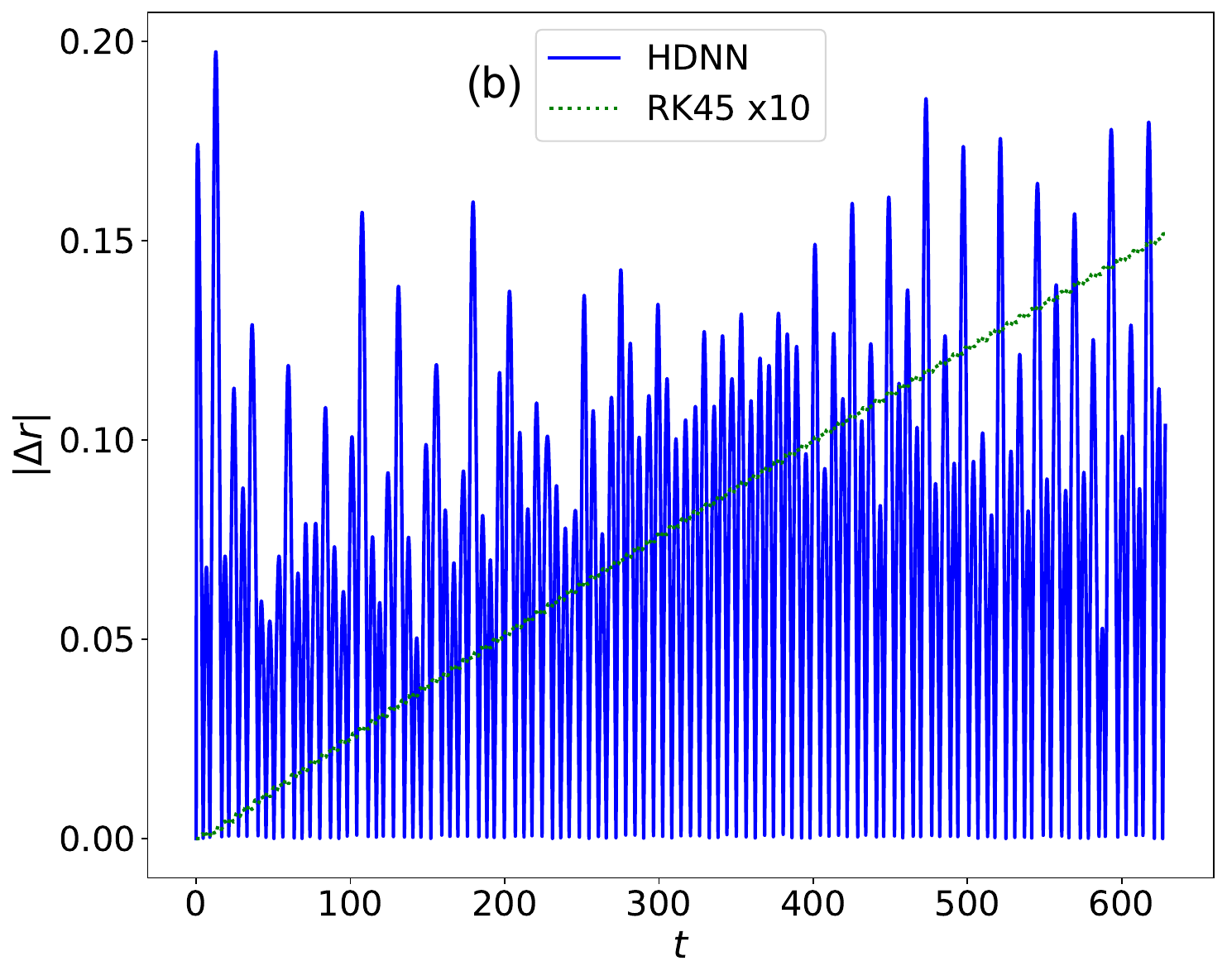}
    \caption[font=small,labelfont=bf]{Panel (a) shows the predicted trajectory for $\kappa=4.2$ from an HDNN trained exclusively with $\kappa=4.0$ along with the corresponding RK45 solution. In (b) we present the time evolution of the quantity $|\Delta r|$ for both the HDNN prediction and the RK45 method. It is evident that the HDNN struggles to reproduce the true path for the unseen value $\kappa=4.2$, whereas as shown in Fig. \ref{fig_gc_trajectories} this was not the case for the HDNN trained over the range $[\kappa_1,\kappa_2]=[3.8,4.2]$.}
    \label{fig_gc_single_seen_value}
\end{figure}

{ Finally, to underscore the importance of training over a range of $\kappa$ values in order to enhance the predictive capability of the HDNN across the entire range, we train an HDNN exclusively for $\kappa=4.0$ and evaluate the acccuracy of the predicted solution for the unseen value $\kappa=4.2$. As shown in Fig.~\ref{fig_gc_single_seen_value}, the predicted trajectory deviates significantly from the true elliptical path and the $|\Delta r|$ of the HDNN prediction is larger than that of the RK45 solution. Hence, training over a range of parameter values, clearly enhances the accuracy of the predictions across the entire range.}

\begin{figure}[ht!]
    \centering
    \includegraphics[scale=0.4]{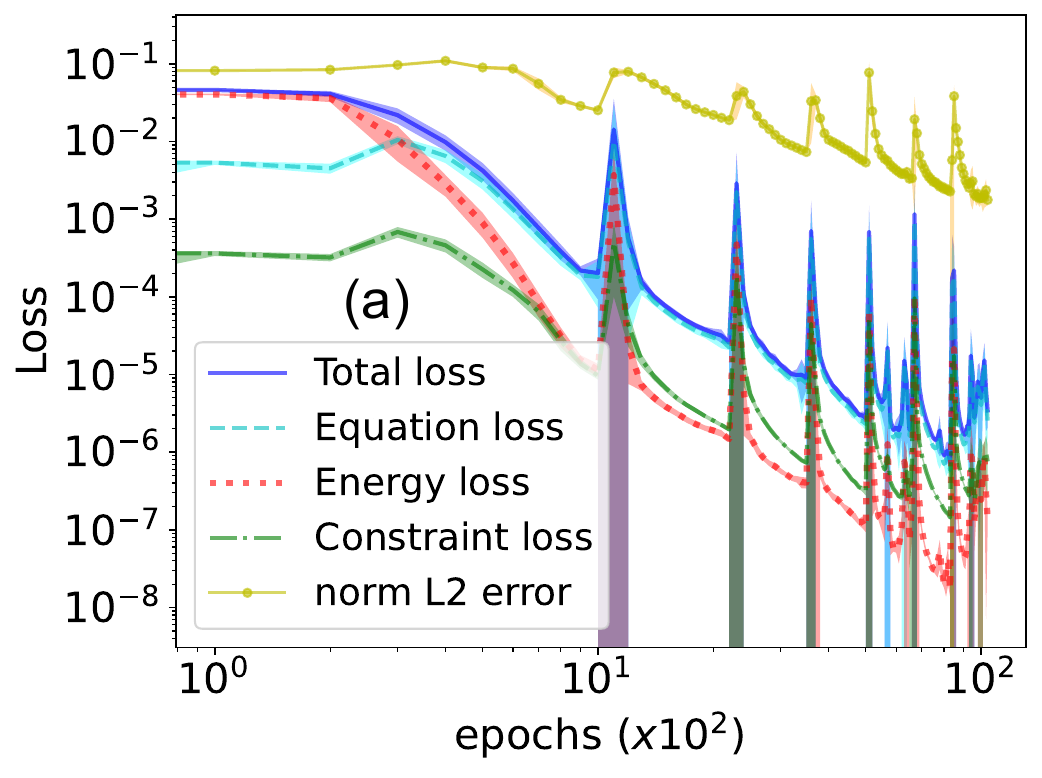}
    \includegraphics[scale=0.4]{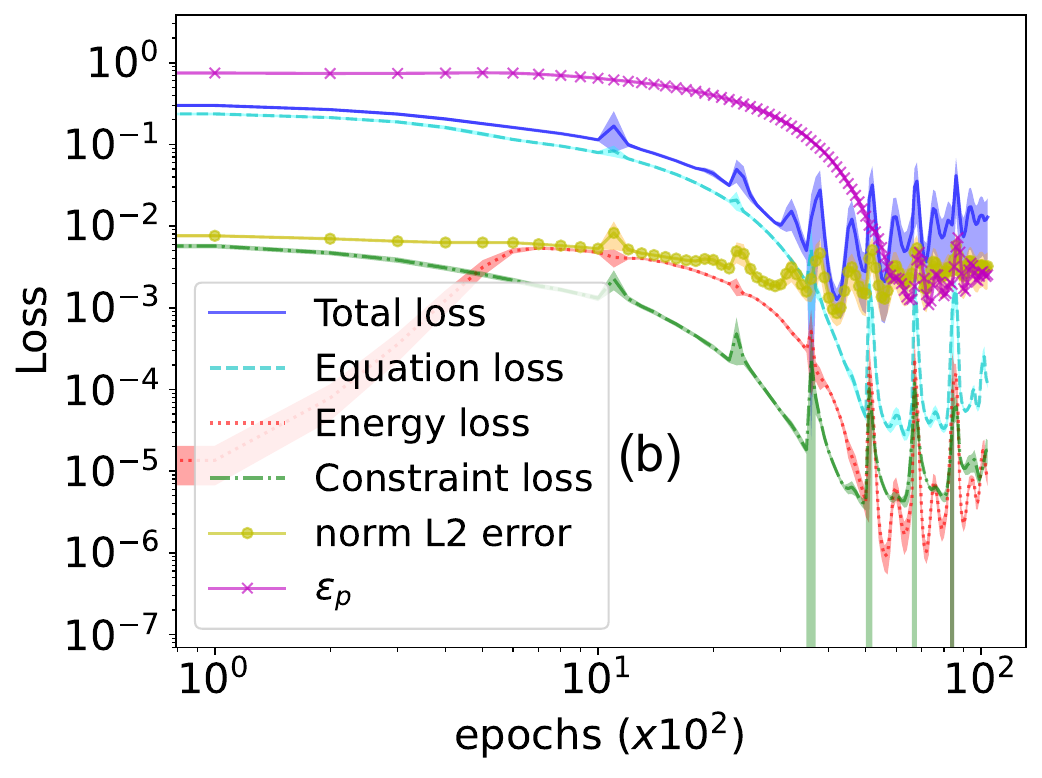}
    \caption[font=small,labelfont=bf]{Panel (a) shows the training history of the loss function versus the number of epochs for the unsupervised training described in section \ref{unsupervised}. The various terms of the total loss function \eqref{loss_gen} are plotted, along with the evolution of the normalized $L_2$ error. In panel (b), we present the corresponding plots for the data-driven, semi-supervised training from subsection \ref{semi_supervised}. The history of the inference error $\varepsilon_p$ is also shown. Towards the end of the training, $\varepsilon_p$ stabilizes at approximately $10^{-3}$, indicating successful inference of the values of $\mu$ and $\kappa$.}
    \label{fig_gc_losses}
\end{figure}

{
\subsubsection{Semi-supervised HDNN learning}
\label{semi_supervised}
As a final example, we address the problem of learning the guiding center dynamics using  a data-driven, semi-supervised setting. In this approach, we utilize a set of ground-truth data $\tilde{\textbf{z}}(t)$ generated by solving Eqs. \eqref{dx_gc}--\eqref{dpy_gc} with the LSODA method on a set $\Tc_d$ of $n_d=500$ equidistant points within the time interval $[0,200 \pi]$, for fixed values of the guiding center magnetic moment $\mu$ and parameter $\kappa$  in \eqref{magn_field}. We denote these fixed values as $\tilde{\mu}$ and $\tilde{\kappa}$, respectively.  The parameter $\epsilon$ is also fixed and equal to $0.15$. The HDNN solution is evaluated on the discrete time points and the mean squared error between the HDNN prediction and the ground-truth data is formed:
\begin{eqnarray}
    \Lc_d = \frac{1}{n_d} \sum_{i=1}^{4}\sum_{t\in \Tc_d} |z^i(t)-\tilde{z}^i(t)|^2\,.
\end{eqnarray}
Now, following the standard data-driven approach for parametric inference (e.g. see \cite{Raissi2019}), we include this data term in the training of the HDNN, forming the following total loss function:
\begin{eqnarray}
    \Lc = \Lc_{hd}+\Lc_{reg}+\Lc_{d}\,, \label{L_tot_data}
\end{eqnarray}
where $\Lc_{hd}$ is the first term in  Eq. \ref{loss_gen}, i.e. the mean squared error of the Hamilton-Dirac equations residual. During the HDNN training, we minimize \eqref{L_tot_data} and two of the trainable parameters are the intrinsic magnetic moment $\mu$ of the guiding center and the parameter $\kappa$ which determines the elongation of the trajectory. Apart from this change the training setting and hyperparameters are the same as in the unsupervised experiment. The algorithm converges without problems to approximate the guiding center dynamics and infers the parametric values with high precision. To quantify this precision we record the error:
$$\varepsilon_p=\frac{\sqrt{(\mu-\tilde{\mu})^2+(\kappa-\tilde{\kappa})^2}}{\sqrt{\tilde{\mu}^2+\tilde{\kappa}^2}}\,,$$
where $\mu$ and $\kappa$ are the inferred values, while the ground truth values are $\tilde{\mu} = 2.0$ and $\tilde{\kappa} = 4.0$. We plot this error along with the training losses in Fig. \ref{fig_gc_losses}. We observe that, by the end of the training, this error reaches the order of  $10^{-3}$, indicating successful parametric inference.}

\section{Conclusion}
\label{sec_V}
In this study, we applied the Dirac method of constraints to enforce holonomic constraints in Hamiltonian systems. The constrained dynamics were computed by solving the Hamilton-Dirac equations using physics-informed neural networks, referred to as HDNNs (Hamilton-Dirac Neural Networks. Additionally, we employed the Dirac algorithm to address a problem with a singular Lagrangian, specifically examining the motion of a guiding center in a magnetic field.

The holonomically constrained systems examined in this study include the nonlinear planar pendulum in Cartesian coordinates and a two-dimensional, elliptically restricted harmonic oscillator. In both cases, HDNNs demonstrated superior accuracy compared to standard explicit algorithms, such as the RK45 method. {The HDNNs effectively learned one parametric dependence of each problem over a range of parameter values.} Moreover, the predicted solutions conserved energy and preserved the Dirac constraints, unlike the standard explicit RK45 method. These preservation properties were achieved by incorporating appropriate regularization terms into the total loss function, which was minimized during the training process.

In the case of guiding center motion, enforcing energy conservation proved crucial for computing stable and accurate orbits, offering a significant advantage over standard explicit numerical solvers. Traditionally, ensuring energy and phase-space volume conservation relies on structure-preserving discretization of the Lagrangian density or the action functional, leading to implicit numerical solvers with large computational times. { By embedding the guiding center motion system in a higher-dimensional space (i.e., the phase space) offers the possibility to use the same HDNN approach to impose additional holonomic or phase-space constraints via the Dirac method. By leveraging the Dirac theory of constraints, the method extends beyond conventional symplectic integration approaches, enabling the application of PINNs to systems with non-standard Lagrangians or phase-space constraints, which are characterized by energy and phase-space-volume conservation.

While the computational time for training the networks exceeded that of explicit methods by many orders of magnitude, the HDNN approach was superior in terms of conserving the energy and the Dirac constraints and also yielded accurate solutions even in cases where the standard numerical methods failed entirely, as shown in section \ref{subsec_4.3}. Moreover, the efficiency of neural network algorithms can be significantly improved through parallel computing and the use of pre-trained networks, such as generative pretrained models, which can help reduce training time. Future research will focus on utilizing alternative and advanced network architectures and types, such as echo state networks and generative pre-trained models, for constrained systems.  Another possibility, related to GPT-PINNs,  is the one-shot transfer learning of PINNs introduced in \cite{Desai2022}. In this method, linear combinations of PINNs, which are pre-trained on families of differential equations, can be re-used to solve new differential equations. The optimal linear weights required to satisfy an unseen instance of a differential equation can be computed in a single optimization step. We aspire that these enhancements will enable us} to explore also chaotic dynamics and further investigate the potential significance of equation-driven machine learning in systems with singular Lagrangians.

{
\section*{Code and data availability}
The code and data supporting the findings of this article are available at \cite{dakalts_git}.
}

\section*{Acknowledgements}
The author would like to thank Prof. George Throumoulopoulos for valuable discussions and Prof. Jian Liu for providing  useful information regarding the derivations in Ref. \cite{Zhang2024}. He is also grateful to the anonymous referees for their valuable comments, which improved the quality of the paper.

\begin{appendices}
\section{Dirac constraints in energetically consistent guiding center theories}
\label{appendix}
Let us examine an example where Eqs.~\eqref{second_constraints} can either be used to determine the multipliers $\zeta^\alpha$ or can also lead to tertiary constraints, considering the Lagrangian
\begin{eqnarray}
    L = \dot{q}^i G_i(q,t) -W(q)\,,\quad i=1,...,N\,, \label{gc_lagrangian}
\end{eqnarray}
which emerges in the context of energetically consistent guiding center theories, describing the motion of charged particle guiding centers in strong magnetic fields while satisfying fundamental conservation laws (e.g. see \cite{Littlejohn1983,Wimmel1983,Pfirsch1984,Wimmel1984}). The conjugate momenta corresponding to the Lagrangian \eqref{gc_lagrangian} are:
\begin{eqnarray}
    p_i = \frac{\partial L}{\partial \dot{q}^i } = G_i(q)\,,
\end{eqnarray}
thus we have the following primary constraints:
\begin{eqnarray}
    \Phi_k = p_k - G_k(q)=0\,, \quad k=1,...,N\,
\end{eqnarray}
and the total Hamiltonian is given by:
\begin{eqnarray}
    H_t = \dot{q}^ip_i - L + \zeta^k[p_k - G_k(q)] = W(q) + \zeta^k[p_k-G_k(q)]\,.
\end{eqnarray}
Following Dirac's algorithm, we check whether the primary constraints are preserved by the dynamics
\begin{eqnarray}
    \dot{\Phi}_k =\{ \Phi_k, H_t\} \approx - \frac{\partial G_k}{\partial q^i} \zeta^i -\left(\frac{\partial W}{\partial q^i}-\zeta^j \frac{\partial G_j}{\partial q^i}\right) \delta^{i}_{k} = -\frac{\partial W}{\partial q^k} - \left( \zeta^i \frac{\partial G_k}{\partial q^i} - \zeta^j \frac{\partial G_j}{\partial q^k} \right)\,,
\end{eqnarray}
Then, the equations
\begin{eqnarray}
    \Psi_k = -\frac{\partial W}{\partial q^k} - \left( \zeta^i \frac{\partial G_k}{\partial q^i} - \zeta^j \frac{\partial G_j}{\partial q^k} \right) =  -\frac{\partial W}{\partial q^k} - \Zc_{ik}\zeta^i \approx 0\,,\quad i,k=1,...,N\,,  \label{lin_syst_zeta}
\end{eqnarray}
can be used to uniquely determine the multipliers $\zeta^k$ if $det(\Zc) \neq 0$, thus concluding the Dirac's algorithm. On the other hand, if the system is not solvable, we end up with a set of secondary constraints. For example, let us consider the case where the second term in \eqref{lin_syst_zeta} vanishes for some $k=\ell$, hence, we end up with a secondary constraint of the form
\begin{eqnarray}
    \Psi_\ell = - \frac{\partial W}{\partial q^\ell}\,, 
\end{eqnarray}
and we have to check its conservation:
\begin{eqnarray}
    \dot{\Psi}_\ell=\{\Psi_\ell,H_t\} \approx -\zeta^k\frac{\partial^2 W}{\partial q^\ell \partial q^k} = -\sum_{m\neq\ell} \zeta^m \frac{\partial^2 W}{\partial q^m \partial q^\ell} - \zeta^\ell \frac{\partial^2 W}{\partial q^\ell \partial q^\ell} \approx 0 \,.
\end{eqnarray}
If $\partial^2 W/\partial q^\ell \partial q^\ell \neq 0$ we have an   equation that determines the multiplier $\zeta^\ell$:
\begin{eqnarray}
    \zeta^\ell = - \frac{1}{\partial^2 W/\partial q^\ell \partial q^\ell} \sum_{m\neq \ell} \zeta^m \frac{\partial^2 W }{\partial q^m \partial q^\ell}\,.
\end{eqnarray}

\end{appendices}
\printbibliography

\end{document}